\documentclass[aps,superscriptaddress,twocolumn,nolinenumbers, nofootinbib]{revtex4-2}
\usepackage{lipsum, amsmath, amsthm, amsfonts, physics, adjustbox, bm, graphicx, dsfont, wrapfig, enumerate}
\usepackage[english]{babel}
\usepackage{xcolor}
\usepackage{appendix}
\usepackage{comment}
\usepackage{tikz}
\usepackage{thmtools,thm-restate}
\usepackage{xr-hyper}
\usepackage[hidelinks,hypertexnames=false]{hyperref}
\usepackage[capitalize]{cleveref}
\usepackage{bbm}
\usetikzlibrary{quantikz2}
\usepackage{tcolorbox}
\usepackage[ruled, vlined]{algorithm2e}

\usepackage{bbold} % Identity operator = \mathbb{1}
\usepackage{placeins} % allows use of \FloatBarrier
\usepackage{booktabs}
\newcommand{\jyv}{Faculty of Information Technology, University of Jyväskylä, Finland}

\newcommand{\ethz}{Institute for Theoretical Physics, ETH Zürich, Switzerland}
\newcommand{\qumami}{Quantum Mathematics, Algorithms and Machine Intelligence (QUMAMI), UPM, Spain}
\newcommand{\upm}{Universidad Politécnica de Madrid, Spain}

\newtcolorbox{boxA}[1]{colback=blue!5!white,
colframe=black!75!black,fonttitle=\bfseries,
title={#1}}

\makeatletter
\newcommand*{\addFileDependency}[1]{
\typeout{(#1)}
\@addtofilelist{#1}
\IfFileExists{#1}{}{\typeout{No file #1.}}
}\makeatother

\begin{document}
\title{\ProjectTitle}
\newcommand{\ProjectTitle}{A Basis-Dependent Kernel for Learning Quantum Phases of Matter}

\author{Clemens Lindner}
\email[Email: ]{lindnerclemens@outlook.com}
\affiliation{\jyv}
\author{Vin\'icius Mohr}
\affiliation{\ethz}
\author{Adrián Pérez-Salinas}
\affiliation{\ethz}
\affiliation{\qumami}
\affiliation{\upm}

\begin{abstract}
Identifying the phases of a quantum many-body system as its parameters vary is a central problem in condensed-matter physics. Data-driven approaches address this task by comparing quantum states, for instance through their fidelity, and locating transitions where these comparisons change abruptly. In this work, we introduce the inverse participation ratio (IPR) kernel, a kernel method built for learning quantum phases of matter from IPRs in physically motivated bases. Unlike fidelity-based kernels, the IPR kernel can identify states with physically meaningful similarity even when they are orthogonal. We test the method on ground states of the transverse-field Ising (TFIM) and Heisenberg XXZ models. For TFIM Gibbs states, we further show that the IPR kernel retains phase-relevant structure over a range of nonzero temperatures where the phase signal of fidelity-based kernels is already degraded. More broadly, our results suggest that quantum kernels benefit more from feature maps tailored to the physics of the problem than from maximal expressivity. 
\end{abstract}
\maketitle

\section{Introduction}\label{sec:introduction}

Characterizing the different phases of quantum many-body systems as a function of their physical parameters is a central task in quantum physics. These parameters can control phenomena such as ferromagnetic order, metallic or insulating behavior, and superconductivity \cite{sachdev_quantum_phase_2011,vojta_quantum_phase_2003}. At zero temperature, these phases are typically encoded in the properties of the ground state.

Finding this ground state is challenging even when the Hamiltonian governing the system is fully known \cite{kempe_complexity_local_2005}, unless additional structure can be exploited, such as restricted entanglement enabling efficient matrix-product-state (MPS) descriptions in gapped one-dimensional systems \cite{hastings_area_law_2007,verstraete_matrix_product_2006,landau_polynomial_time_algorithm_2013,schuch_classifying_quantum_2011}, or integrability permitting exact treatment in particular models \cite{caux_remarks_notion_2011}. A complementary route is to classify phases directly from data using machine learning \cite{carrasquilla_machine_learning_2020, carrasquilla_machine_learning_2017, zhang_interpretable_machine_2019,scarato_hybrid_quantum_classical_2026}. Within this data-driven setting, ground-state properties can be predicted efficiently from examples drawn from the same phase, whereas such guarantees need not extend across a phase transition \cite{huang_provably_efficient_2022}.

The data-driven approach still relies on either the states themselves or state-derived features that remain difficult to compute classically. Quantum processors can naturally supply both. For example, variational algorithms prepare approximate ground states on quantum hardware \cite{peruzzo_variational_eigenvalue_2014,cerezo_variational_quantum_2021}, entanglement has been measured on trapped-ion and superconducting devices \cite{linke_measuring_renyi_2018,yang_probing_entanglement_2026,koh_measurement_induced_entanglement_2023}, and short-depth circuits have been proposed to estimate coherence and discrete Berry phases \cite{tan_probing_quantum_2025}. In particular, learning phases of matter from prepared quantum states or their measurement outcomes is an active area of research \cite{bespalova_benchmarking_quantum_2026,schoulepnikoff_discovering_quantum_2026}. Quantum kernels address this task by comparing states directly \cite{sancho_lorente_quantum_kernels_2022,wu_quantum_phase_2023,ali_finite_size_resource_2026,khosrojerdi_learning_classify_2024}, and for specially constructed learning problems they provably outperform all classical learners \cite{muser_provable_advantages_2024,liu_rigorous_robust_2021}.

The inverse participation ratio (IPR) measures how concentrated a probability distribution is in a given basis \cite{wegner_inverse_participation_1980, murphy_generalized_inverse_2011}. Primarily studied in the context of Anderson transitions \cite{evers_anderson_transitions_2008,evers_fluctuations_inverse_2000,rossignolo_localization_transitions_2019} and many-body localization \cite{alet_many_body_localization_2018,bera_many_body_localization_2015}, the IPR has seen repeated use in detecting phase transitions \cite{pal_many_body_localization_2010, baena_entropies_ipr_2022, zhang_interpretable_machine_2019}. Furthermore, it appears in contexts ranging from ergodicity breaking and the eigenstate thermalization hypothesis \cite{de_luca_ergodicity_2013, brown_quantum_2008, graf_quantum_2025, alam_onset_ergodicity_2026a} to quantum-classical computational separations \cite{dalzell_random_quantum_2022, mari_counting_collisions_2024, boixo_characterizing_quantum_2018, bremner_average_case_complexity_2016, aaronson_computational_complexity_2010}, where the IPR is understood as a collision probability or as a controlling measure of anti-concentration over circuit output probabilities.

In this work, we design an IPR-based kernel that assists learning algorithms in classifying phases of matter. The kernel compares quantum states by how concentrated their probability distributions are in a few chosen bases. The bases must be chosen beforehand, guided by the physics deep inside the phases, where ground states are easier to compute than near a transition. We build our kernel on the observation that the ground state of a given phase may appear highly localized in the basis that diagonalizes the dominant term of the Hamiltonian while being strongly delocalized in a competing basis. That is, the corresponding IPRs attain opposite extrema over broad parameter ranges. At the phase transition, the competing influence of two or more terms induces delocalization in all of these bases \cite{perez_salinas_multiple_basis_representation_2025}. 

Our algorithm assumes access to prepared quantum states, with many copies of each. For every state, the IPR is evaluated in a set of bases chosen for the problem at hand, and the resulting values are collected into a feature vector (see \cref{fig:scheme}). The kernel is then defined as the inner product between these feature vectors, yielding a positive-semidefinite similarity measure that can be used to classify the phase of a previously unseen state. A key difference of this kernel with respect to fidelity-based proposals (e.g., \cite{sancho_lorente_quantum_kernels_2022}), which use the fidelity as a phase-transition indicator \cite{zanardi_ground_state_2006,damski_fidelity_approach_2015,gu_fidelity_approach_2010,rossini_ground_state_fidelity_2018}, is that two orthogonal quantum states can still be assigned a high similarity. This builds a physically motivated inductive bias into the kernel. 
As an example, consider the ground and first excited states corresponding to a phase of matter far away from a phase transition. A fidelity-based kernel would signal no similarity, although most of the phase-relevant physical properties of the two states may be nearly identical (see \cref{fig:ising_example}). 

We test the utility of this bias by benchmarking the IPR kernel on ground states of the transverse-field Ising model (TFIM) and its corresponding temperature-varying Gibbs states for up to $L=100$ spins. For thermal states our kernel shows the most promise over fidelity-based approaches. Finite sampling, however, limits the accuracy close to the transition, where all IPRs become exponentially small. Another test on the Heisenberg XXZ chain up to $L=24$ resolves both phases but sees the learned boundary shifted from the infinite-size prediction.

The paper is organized as follows. \Cref{sec:background} reviews IPRs, multiple-basis representations, and kernel methods. \Cref{sec:ipr_kernel} describes the IPR kernel and the classification algorithm for phases of matter. \Cref{sec:results} applies it to TFIM and XXZ ground states and to TFIM Gibbs states, where we compare it with fidelity-based kernels. \Cref{sec:discussion} discusses the results and open directions. The appendices give details on data generation, further numerical results, and quantum circuits for estimating the kernel. 

\section{Background}\label{sec:background}

\subsection{Inverse participation ratios}\label{sec:ipr}

The inverse participation ratio (IPR) is a simple quantity that captures how concentrated a probability distribution is across its domain. Let a pure $L$-qubit state $\ket{\psi}$ be written in an arbitrary basis, for instance, the computational basis, as
\begin{equation}
\ket{\psi}
=
\sum_{x \in \{0,1\}^L} c_x  \ket{x}.
\end{equation}
Then, the IPR of order $q > 1$, measured in the computational basis, is defined as
\begin{equation}
I_q^{(Z)}(\psi)
=
\sum_{x \in \{0,1\}^L} |c_x|^{2q},
\end{equation}
and thus $I_q^{(Z)} \in [2^{-L(q - 1)}, 1]$. At the extrema, a localized, sparse state maximizes the IPR near 1, whereas a completely delocalized, uniform state minimizes it toward its lower bound. The IPR is therefore a measure of effective sparsity of the state. Throughout the remainder of this work, we will keep $q = 2$ unless explicitly stated otherwise.

As apparent from its definition, we can compute the IPR with respect to any basis, a property that we use to investigate the state as seen through different lenses. 
Consider a collection of unitaries $\{U_s\}_{s=1}^{M}$, each of which defines a new basis by applying $U_s$ to the elements of the $Z$ basis,
\begin{equation}\label{eq:define_bases}
    \big\{\, U_s \ket{x} \;\big|\; x \in \{0,1\}^L \big\}.
\end{equation}
Let $U_s$ denote one such transformation. Expressing $\ket{\psi}$ in this basis means writing
\begin{equation}
    \ket{\psi}
    =
    \sum_{x \in \{0,1\}^L} c_x^{(s)} \, U_s \ket{x},
\end{equation}
where the coefficients in basis $s$ are $c_x^{(s)}=\bra{x} U_s^{\dagger} \ket{\psi}$. The IPR of $\ket{\psi}$ in the new basis $s$ is then defined as in the previous case, with the updated coefficients, 
\begin{equation}\label{eq:ipr}
    I^{(s)}(\psi)
    =
    \sum_{x \in \{0,1\}^L}
    \left|
    \bra{x} U_s^\dagger \ket{\psi}
    \right|^{4}.
\end{equation}
It quantifies the effective sparsity of $\ket{\psi}$ when expressed in the basis generated by $U_s$. By evaluating this quantity in several physically motivated bases, we obtain a basis-dependent localization profile of the state.

In a similar fashion it is possible to compute the IPR of mixed states. Let $\rho^{(s)} := U_s^\dagger \rho U_s$ be the density matrix $\rho$ written in basis $s$. Its elements are thus $\rho^{(s)}_{xy}:=\bra{x} U_s^\dagger \rho U_s \ket{y}$. We define the IPR of a density matrix $\rho$ in basis $s$ as
\begin{equation} \label{eq:mixed_ipr}
    I^{(s)}(\rho)
    =
    \sum_{x \in \{0,1\}^L}
    \left(
    \bra{x} U_s^\dagger \rho\, U_s \ket{x}
    \right)^2
    =
    \sum_{x \in \{0,1\}^L}
    (
    \rho^{(s)}_{xx}
    )^2.
\end{equation}
Note that for a pure state $\rho=\ketbra{\psi}{\psi}$, we recover the standard pure-state IPR: $I^{(s)}(\rho)=I^{(s)}(\psi)$.

\begin{figure*}
    \centering
    \includegraphics{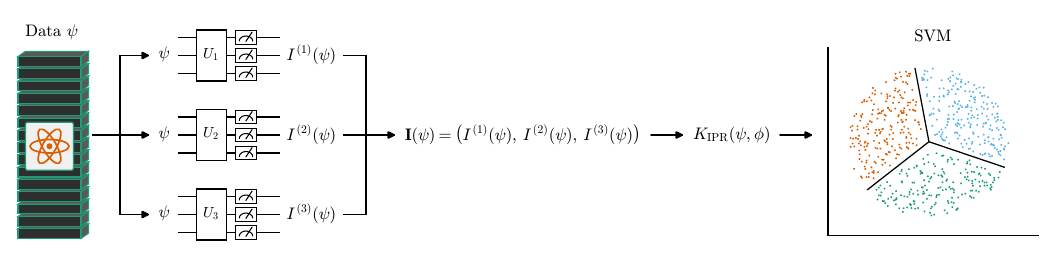}
    \caption{IPR-kernel algorithm. Given a state $\psi$, we calculate its IPR in various bases and concatenate the results into a feature vector. The resulting kernel compares the IPR profiles of two states, and a support vector machine (SVM) uses it to classify their phase.}\label{fig:scheme}
\end{figure*}

\subsection{Multiple-basis representation}\label{sec:mbr}

Quantum states are typically represented by their coordinates in one specified orthonormal basis, following the tradition of linear algebra. However, there is no inherent reason to restrict ourselves to representing a state in a single such basis. One can, for instance, explore multiple-basis representations (MBR) of a state \cite{perez_salinas_multiple_basis_representation_2025}, following the description 
\begin{equation}
\ket{\psi}
=
\sum_{s} \alpha_s U_s
\left(
\sum_{x_s\in \mathcal{J}_s} c_{x_s}^{(s)} \ket{x_s}
\right),
\qquad \mathcal{J}_s \subset \{0, 1\}^L.  
\end{equation}
Here, the unitaries $\{U_s\}$ define different bases of the Hilbert space, as in \cref{eq:define_bases}, while each subset $\mathcal{J}_s$ specifies the support of the corresponding basis-specific sparse representation.

Moderate sizes of the subsets $\mathcal{J}_s$ lead to classically efficient descriptions of the quantum states. As an illustrative example, consider the state
\begin{equation}
\ket{\psi_{0+}}
=
\frac{1}{\sqrt{2(1+2^{-L/2})}}
\left(
\ket{0}^{\otimes L}+\ket{+}^{\otimes L}
\right).
\end{equation}
Although both the computational and Hadamard bases (as well as any other) can represent this particular state, this combination allows one to do it with only two coefficients. Among all possible combinations of bases, this representation becomes of special interest if $\{ U_s \}$ form a set of mutually unbiased bases (MUB) \cite{durt_mutually_unbiased_2010}. The reason is that a basis $s$ demands exponentially large $\mathcal{J}_s$ to capture a constant fraction of a state which is sparse in another basis $s'$, provided that both bases are MUB. The choice of MUBs offers a natural way to split the state into several components. 

The MBR idea has been previously explored as a classical surrogate to approximate ground states \cite{perez_salinas_multiple_basis_representation_2025}. This surrogate method chooses $\{ U_s\}$ as the bases (approximately) diagonalizing the Hamiltonian deep in each phase and selects $\mathcal{J}_s$ heuristically with a size constraint of $\mathcal O(\operatorname{poly}(L))$. Under these conditions, the approximation to the ground state decreases in quality when approaching the phase transition. 

Hence, previous observations suggest that the MBR can be used as an approximation to ground states, although with an inherent limitation at the phase transition. This work aims to turn this limitation into a witness of the phase transition region itself. Any MBR with an efficient description will have a large, hence detectable, IPR in the bases used for the description. However, choosing the wrong basis or measuring a state at the border of the phase transition will generally signal IPRs of smaller values. In particular, if the bases of interest are MUBs, then the IPRs measured in these bases will be exponentially vanishing.

\subsection{Kernel methods and support vector machines}\label{sec:kernel_methods}

Kernel methods \cite{scholkopf_learning_2001,ghojogh_reproducing_2021}, in conjunction with support vector machines (SVMs) \cite{boser_training_1992}, provide a way to compare data points and classify them, as widely explored in the machine learning literature \cite{hofmann_kernel_methods_2008,cervantes_comprehensive_survey_2020}. SVMs are classifiers that find an optimal linear separation for a dataset $\{(x_i, y_i)\}_{i = 1}^N$, where $x_i \in \mathbb R^d$ and $y_i \in \{-1, 1\}$. The resulting decision function can be expressed only in terms of inner products between data points and a subset of training points, called support vectors. The search for the support vectors is performed by solving a soft-margin optimization problem which requires only the inner products $\langle x_i, x_j\rangle$. More generally, the data may be mapped through a feature map $\Phi$, replacing these inner products by a kernel function, that is
\begin{align}
    \Phi &: \mathcal X \rightarrow \mathcal F, \\
    [K]_{ij}
    &=
    \langle \Phi(x_i), \Phi(x_j)\rangle_{\mathcal F}.
\end{align}
Here, $[K]_{ij}$ is the Gram matrix encompassing all pairwise similarity measures, i.e., the kernel functions $K(x_i,x_j)$, between data points. For $K$ to define a valid kernel, it must be positive semidefinite. The choice of the feature map, or equivalently of the kernel, decides what properties of the data the classifier sees. We provide more details in \cref{app:supplementary}.

\section{The IPR kernel} \label{sec:ipr_kernel}

The central ingredient of this work is a kernel based on computing or estimating the IPR of a given state with respect to different bases. The procedure requires access to quantum data and a quantum processing step to estimate the relevant IPR values. These values are then assembled into a classical feature vector and used for phase classification. A schematic overview is provided in \cref{fig:scheme}.

For a given pure quantum state $\ket{\psi}$ we evaluate its IPR across a family of bases $\{U_s\}_{s=1}^{M}$ and define the feature vector
\begin{equation}
    \mathbf{I}(\psi)
    :=
    \big(
    I^{(1)}(\psi),
    I^{(2)}(\psi),
    \ldots,
    I^{(M)}(\psi)
    \big).
\end{equation}
This construction therefore maps our quantum state to a point in an $M$-dimensional feature space which encodes its basis-dependent sparsity structure. Notice that this step requires having access to copies of the input state $\ket\psi$.

Using this specific feature map, we define the IPR kernel between two states $\ket{\psi}$ and $\ket{\phi}$ as the inner product of their respective feature vectors,
\begin{equation}
    K_{\mathrm{IPR}}(\psi,\phi)
    =
    \langle \mathbf{I}(\psi), \mathbf{I}(\phi) \rangle
    =
    \sum_{s=1}^{M}
    I^{(s)}(\psi)\,
    I^{(s)}(\phi),
\end{equation}
which defines a valid kernel and can be generalized to mixed states through \cref{eq:mixed_ipr}. In this form, the kernel quantifies the similarity of the two states in terms of their sparsity properties across the chosen set of bases. To classify phases of matter, the corresponding bases are chosen as representatives of each phase. If two states exhibit a large IPR in the same basis, their kernel value is of order $\mathcal{O}(1)$, and they are regarded as \textit{similar} by the classifier. If their largest IPR values occur in different bases, the IPR kernel will yield smaller values. In the case where the state lies close to a phase transition, we expect all IPRs, and hence the kernel values, to be small.

\begin{figure}[t]
    \centering
    \includegraphics{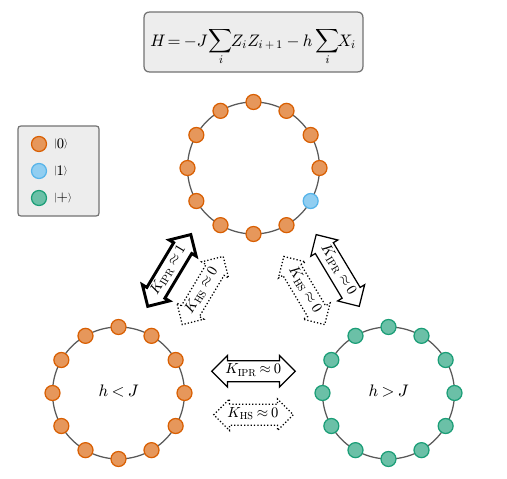}
    \caption{1D TFIM spin configurations and their approximate Hilbert-Schmidt (HS) and IPR kernel values. Bottom: representative ground states for $h<J$ and $h>J$. Top: a single spin flip of the ferromagnetic state.}
    \label{fig:ising_example}
\end{figure}

It is important to highlight some differences of this kernel with respect to other previously used approaches. Quantum kernels are commonly constructed as inner products in a quantum feature space \cite{havlicek_supervised_2019, schuld_supervised_2021}, primarily the Hilbert-Schmidt (HS) inner product between density operators. For pure states, the resulting kernel coincides with the fidelity,
\begin{equation}
K_{\rm HS}(\psi,\phi)=\left|\bra{\psi}\ket{\phi}\right|^2,
\label{eq:hs_kernel}
\end{equation}
while for mixed states the same inner product gives $K_{\rm HS}(\rho,\sigma)=\operatorname{Tr}(\rho\sigma)$. However, in some cases this type of kernel fails to capture certain physical properties. As an example, consider a simple instance of the transverse-field Ising model (TFIM) \cite{pfeuty_one_dimensional_ising_1970, sachdev_quantum_phase_2011}, depicted in \cref{fig:ising_example}, as 
\begin{equation}
H_{\mathrm{TFIM}}=-J\sum_i Z_iZ_{i+1}-h\sum_i X_i,    
\end{equation}
on a regular 1D graph with periodic boundary conditions. Setting $J>0$ and $h\geq0$, we get two phases: ferromagnetic and paramagnetic. The representative ground states of each phase are 
\begin{align}
    0\leq h<J:\quad
    &\ket{\psi_0}
    \xrightarrow{h\to 0}
    \ket{0 0 0 \ldots 0 0},
    \label{eq:tfim_ferromagnetic}\\
    h>J:\quad
    &\ket{\psi_0}
    \xrightarrow{h\to\infty}
    \ket{+ + + \ldots + +},
    \label{eq:tfim_paramagnetic}
\end{align}
up to some trivial degeneracies. To showcase the intuition, consider for $h \ll J$ the state $\ket{0 0 0 \ldots 0 1 0 \ldots 0 0 0}$. This state lives in the subspace of first excited levels. With respect to the order parameter of the TFIM, i.e., magnetization per site $m_z=L^{-1}\sum_i\langle Z_i\rangle$, this state is only $\mathcal O(L^{-1})$ away from the ground state, and hence exhibits similar physics. However, its HS kernel value with respect to the ferromagnetic ground state in \cref{eq:tfim_ferromagnetic} is identically zero, thus perceived as completely different. By contrast, the IPR kernel captures their shared localization in the $Z$ basis and therefore provides a more refined notion of similarity.

The two kernels also differ in their estimation cost. A clear advantage of the HS kernel is the possibility of implementing it easily on a quantum computer by means of a SWAP test \cite{buhrman_quantum_fingerprinting_2001,ekert_direct_estimations_2002}, and its value can be estimated with an accuracy depending on the number of measurements, following a Monte Carlo scaling. To compute the IPR kernel, the most direct approach is to apply the corresponding change of basis, sample the resulting state, and estimate the probabilities $p_x^{(s)}=|\bra{x}U_s^\dagger\ket{\psi}|^2$ entering $I^{(s)}$. However, this is inefficient because it reconstructs many individual probabilities only to combine them into a single quantity. Additionally, for very small values of the IPR, this leads to large accuracy overheads. 

Instead, if two copies are available simultaneously, it is possible to implement modifications of the SWAP test to estimate the IPR, possibly even with quantum amplitude estimation \cite{brassard_quantum_amplitude_2002}. This technique provides an estimate of the IPR with an additive sampling error of $\mathcal O\left(1/\sqrt{N_s}\right)$, with $N_s$ being the number of samples, which amplitude estimation improves to at best $\mathcal O\left(1/{N_s}\right)$. Hence, even with amplitude estimation, we can only expect to estimate IPRs of size $\gtrsim 1 / {N_s}$, and any exponentially vanishing value is beyond our resolution. Thus, the number of the required copies to estimate each IPR to a given precision is determined by the value of the IPR itself. Restricting ourselves to the case without amplitude estimation, we refer the interested reader to \cref{app:circuits} for a quantum-circuit implementation of the IPR kernel.

\subsection{Choice of bases}\label{sec:basis_choice}

The performance of our kernel depends fundamentally on the chosen set of bases and on how well those bases represent the underlying physical problem. Since we are interested in phase detection, we focus on the Hamiltonian encoding the problem at hand and heuristically select the bases that distinctly represent each phase. Many Hamiltonians of interest can be written as a decomposition of two (or a few) competing terms $H_a$ and $H_b$, with $[H_a,H_b] \neq 0$, i.e.
\begin{equation}
    H(\lambda)=H_a+\lambda H_b,
\end{equation}
where each term represents the model deep inside one phase, and $\lambda$ acts as a tuning parameter. Our approach for phase detection would then proceed as follows: we denote the ground state of $H(\lambda)$ by $\ket{\psi_{\mathrm{GS}}(\lambda)}$ and evaluate its IPR in the eigenbases of $H_a$ and $H_b$ separately over a chosen grid of $\lambda$ values. However, the availability of the exact ground states does not constitute a necessary condition for our approach. It is sufficient that their basis-dependent IPR profiles retain information about the underlying phase, as we demonstrate for finite-temperature Gibbs states in \cref{sec:thermal}.

Deep in the regime dominated by $H_a$, the ground state is expected to have support on only one or a few eigenstates of $H_a$. It is therefore highly localized in the eigenbasis of $H_a$, leading to a large IPR in that basis. Conversely, when expressed in the eigenbasis of the competing term $H_b$, the same state is generally spread over many basis states, leading to a smaller IPR. The reverse behavior is expected in the regime where $H_b$ dominates. The IPRs therefore delimit the regimes in which each basis gives a sparse description of the ground state. Between these regimes, both IPRs decay rapidly with system size, and the phase transition as well as the crossing of the IPR curves lie in this range of small IPRs. In specific cases that crossing might coincide with the phase transition point itself, as for the one dimensional TFIM in \cref{sec:tfim} or the SU(2)-symmetric point of the XXZ chain in \cref{app:xxz} viewed through the $X$ and $Z$ bases, where a symmetry or duality relates the two bases. Importantly for us, and since we cannot evaluate exponentially low IPRs in practice, the kernel only needs to consistently separate two states drawn from opposite phases away from the transition, where at least one IPR remains large. 

In some cases, the relevant bases are physically clear and reduce to simple Pauli product bases. This is the case for the transverse-field Ising model discussed in \cref{sec:tfim}, where the two limiting regimes naturally motivate the computational $Z$ and Hadamard $X$ bases. A more challenging situation is encountered when the relevant basis is not a simple product basis, as we will observe in \cref{sec:xxz}.

In practice, however, Hamiltonian terms have degenerate eigenspaces and therefore do not specify a unique basis. Since the IPR is defined with respect to individual basis states, we first have to fix an explicit basis within each subspace. Throughout the numerical applications below, we thus choose the relevant bases explicitly.

This distinction is also reflected in the circuit constructions in \cref{app:circuits}. We provide two approaches. First, if the basis-change unitary is known explicitly, one can apply it directly and estimate the standard IPR in the corresponding complete basis. Second, if the basis is specified only through a Hamiltonian, one can use a phase-estimation-based circuit. In the presence of degeneracies, this second approach naturally resolves entire energy eigenspaces, leading to a generalized block-IPR, see \cref{eq:block_ipr}. This quantity measures localization over energy subspaces instead of localization over individual basis states.

Finally, we note that the IPR kernel can naturally be extended to problems with multiple transition points or control parameters. A systematic treatment of this extension is left for future research. Apart from the XXZ chain in a longitudinal field in \cref{sec:xxz}, we focus on a single transition point and kernels built from two physically motivated bases.

\subsection{Supervised phase classification}\label{sec:classification_protocol}

Before turning to the results, we briefly summarize the phase-classification workflow. Given a Hamiltonian $H(\lambda)$ depending on a tuning parameter $\lambda$, we assume access to a set of labeled quantum states,
\begin{equation}
\left(\ket{\psi(\lambda_i)}, y_i\right), \qquad y_i\in\{-1,+1\},
\end{equation}
where the labels indicate the corresponding phase. In the applications below these states are ground states or Gibbs states of the model, but the construction only requires that the training states can be prepared and, more importantly, labeled reliably.

For the labeled training data, we evaluate the IPR kernel Gram matrix, by classical simulation as in this work or with the quantum circuits described in \cref{app:circuits}, and employ it as a precomputed kernel in a classical SVM. The resulting decision function defines a maximum-margin classifier whose zero crossing can, in the ideal case, be interpreted as an estimate of the phase boundary. Once trained, the classifier assigns phase labels to new quantum states from their kernel values with respect to the support vectors.

The finite-shot estimation of the IPR values constitutes the main practical limitation of our kernel. For an $L$-qubit system, IPR values lie in the range $[2^{-L},1]$, so delocalized states can yield exponentially small contributions. In the two-basis setting this is compensated deep inside either phase, where one basis IPR remains large. Close to a phase transition, however, both IPRs can become small, making the relevant kernel values difficult to resolve with a finite sampling budget $N_s$. Hence, with finite $N_s$, we expect reliable classification deep inside each phase, while states close to the transition become harder to distinguish as $L$ increases.

Finally, it is important to note that our IPR kernel effectively limits the number of data features to which the SVM is sensitive. The HS kernel, in contrast, is sensitive to all differences between states, which can lead to overfitting and generalization issues \cite{thanasilp_exponential_2024}. This problem is mitigated in the IPR approach, at the expense of consciously neglecting information from the data. The key observation is that this information might not be relevant for phase classification, assuming an informative choice of bases.

\section{Results} \label{sec:results}

\begin{figure*}[t]
    \centering
    \includegraphics{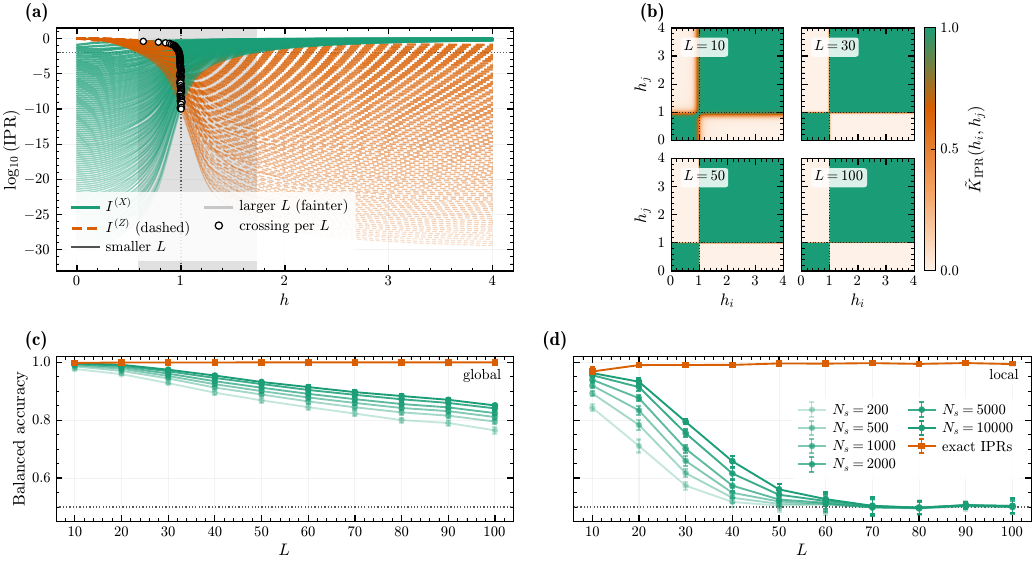}
    \caption{TFIM ground-state results for the normalized IPR kernel. (a) IPRs in the $X$ and $Z$ bases for $L=3,\dots,100$ on a $\log_{10}$ scale. Circles mark the crossing for each $L$, which approaches $h_c=1$ with increasing $L$. Shaded: range in which both IPRs at $L=100$ lie below $10^{-2}$ (dotted horizontal line). (b) Normalized Gram matrices for $L=10$, $30$, $50$, and $100$, with $h_c=1$ marked by dotted lines. (c) Global and (d) local ($\delta=0.25$) balanced accuracy versus $L$. Orange: exact IPRs, averaged over 100 train-test splits, standard deviation over splits. Green: finite-shot IPR estimates with $N_s$ shots, averaged over 10 splits and 10 noise realizations with standard deviation over realization means, more opaque curves correspond to larger $N_s$.}
    \label{fig:tfim_results}
\end{figure*}

\subsection{TFIM ground states}\label{sec:tfim}

The one-dimensional transverse-field Ising model (TFIM) is a well-studied quantum spin system that exhibits a quantum phase transition and provides an ideal first test of our IPR kernel. Its Hamiltonian is
\begin{equation}\label{eq:tfim_hamiltonian}
    H_{\mathrm{TFIM}}(h)=-J \sum_{i=1}^{L}Z_i Z_{i+1} - h \sum_{i=1}^{L}X_i,
\end{equation}
where $L$ denotes the number of spins in the chain. Here, the first term favors ferromagnetic order in the computational ($Z$) basis, while the transverse field $h$ favors polarization in the $X$ basis, competing with the $Z$-order and driving the system through a quantum phase transition at $h=J$. In our case, we set $J=1$ and limit $h\in[0,4]$.

We take ground states $\psi_{\mathrm{GS}}(h_i)$ of $H_{\mathrm{TFIM}}(h_i)$ and evaluate their inverse participation ratios in both the $Z$ and $X$ bases. Ground states are computed using matrix product states (MPS) via the density-matrix renormalization group (DMRG), which allows us to reach system sizes $L\in\{3,4,\dots,100\}$ \cite{white_dmrg_1992,white_dmrg_1993,schollwoeck_dmrg_2011}. The IPRs are then estimated by Monte Carlo sampling of bitstrings from the MPS Born distribution. A detailed description of the data generation procedure is given in \cref{app:data_generation}.

From smoothing-spline fits to the IPR curves (see \cref{app:data_generation}) we draw 4000 samples with varying transverse-field parameter $h$ and use those for the calculation of the full Gram matrix. The IPR curves are shown in \cref{fig:tfim_results}(a) and the Gram matrices for $L=10$, $30$, $50$, and $100$ in \cref{fig:tfim_results}(b); the Gram matrices we plot always include all data points.

We observe that the IPR curves cross close to $h=1$, which for $J=1$ is the quantum critical point $h_c$ of the one-dimensional TFIM in the thermodynamic limit. The observed behavior is a consequence of the Kramers-Wannier self-duality of the TFIM \cite{kramers_statistics_1941,stephan_renyi_entropy_2010}. The duality exchanges $Z_iZ_{i+1}$ and $X_i$ and maps $H_{\mathrm{TFIM}}(h)$ onto $h\,H_{\mathrm{TFIM}}(1/h)$. For ground states with periodic boundary conditions it implies $I^{(Z)}(h)=\frac{1}{2}I^{(X)}(1/h)$, where the factor $\frac{1}{2}$ stems from the twofold degeneracy of the $Z$-basis configurations under a global spin flip. The DMRG states at small $h$ are symmetry broken, so this factor is absent there. Up to this factor, the two curves are mirror images under $h\to1/h$. Since their slopes grow with $L$, the factor shifts the crossing below $h=1$ only for small $L$, and the crossing approaches $h_c=1$ with increasing $L$, see \cref{fig:tfim_results}(a). The same figure also shows that the IPRs decay exponentially with $L$. The decay is slow deep in the phase favoring the respective basis and fast in the opposite phase, where the IPRs approach the fully delocalized value $2^{-L}$.

The IPR kernel uses the opposite behavior of the two IPRs to predict the phase of new states, whose generating parameter $h_i$ may be unknown. To evaluate SVM performance, we use \texttt{scikit-learn}'s \texttt{SVC} following \cref{sec:ipr_kernel}, with $C=1$ and balanced class weights throughout. For our noiseless, densely interpolated data, larger $C$ can yield higher accuracies by weakening regularization, but such solutions are generally more sensitive to noise. To avoid numerical instabilities associated with the exponentially small raw IPR values, we use the normalized kernel,
\begin{equation} \label{eq:ipr_kernel_norm}
    \begin{aligned}
        \tilde{K}_{\mathrm{IPR}}&(\rho,\sigma)
         =
        \frac{
            \langle \mathbf{I}(\rho), \mathbf{I}(\sigma) \rangle
        }{
            \sqrt{\langle \mathbf{I}(\rho), \mathbf{I}(\rho) \rangle}
            \sqrt{\langle \mathbf{I}(\sigma), \mathbf{I}(\sigma) \rangle}
        }    \\
         & =
        \frac{
            \sum_{s=1}^{M}
            I^{(s)}(\rho)\,
            I^{(s)}(\sigma)
        }{
            \sqrt{\sum_{s=1}^{M} \left(I^{(s)}(\rho)\right)^2}
            \sqrt{\sum_{s=1}^{M} \left(I^{(s)}(\sigma)\right)^2}
        } .
    \end{aligned}
\end{equation}
This normalization removes the dependence on the overall magnitude of the IPR feature vectors. It therefore reduces the influence of raw IPR scale differences, which can make our Gram matrix ill-conditioned \cite{hsu_practical_guide_}, thus worsening the classification performance with increasing system size.

Furthermore, while always training on global data drawn from the full $h$ range, we show performance for two regimes: one where the test data is also drawn globally over the parameter set, i.e., all $\psi_{\mathrm{GS}}(h_i)$ where $h_i\in[0,4]$, and one where we limit that set to a region around the theoretical phase transition point $h_c=1$, i.e., $[h_c-\delta, h_c+\delta]$. In each case, we do a 90/10 train-test split of our 4000 data points and average over 100 different splits, making sure that class balance is taken into account. The corresponding performances are provided in terms of balanced accuracy (BA) where a BA of 0.5 means the classification results are essentially random. We provide their mean with standard deviation over the 100 splits for increasing $L$ in \cref{fig:tfim_results}(c) and (d) and in \cref{tab:svm_tfim} in \cref{app:supplementary}.

With exact IPRs, the normalized IPR kernel reaches near-perfect balanced accuracies for all $L>10$, also in the local region (\cref{tab:svm_tfim}), since the crossing lies close to $h_c=1$. However, due to the shot noise inherent in estimating the kernel values from a finite number of measurements, we cannot resolve IPR values below the shot-noise level $1/\sqrt{N_s}$. In our case, see \cref{fig:tfim_results}(a), IPRs close to the transition consistently fall below a realistic threshold. For $L=100$, both IPRs are smaller than $10^{-2}$ for $h\in[0.59,1.73]$ and smaller than $10^{-4}$ for $h\in[0.78,1.27]$. These ranges contain $h_c=1$ and widen with $L$, so the sampling budget sets how close to the transition the features can resolve states. In \cref{fig:tfim_results}(c) and (d) we therefore also show the performance in terms of balanced accuracy for increasing sampling budgets $N_s$. While the global performance holds up, the local performance around the phase transition quickly degrades with increasing system size and decreasing sampling budget. For $N_s=10^4$, corresponding to a resolution of about $10^{-2}$, the local balanced accuracy approaches chance level once the range in which both IPRs lie below $10^{-2}$ covers most of the local test region, as at $L=50$, where this range is $h\in[0.78,1.29]$.

In \cref{app:tfim} we provide further details and analysis of the sampling limit; we also discuss the scaling behavior of both the raw and normalized IPR kernels.

\begin{figure*}[t]
    \centering
    \includegraphics{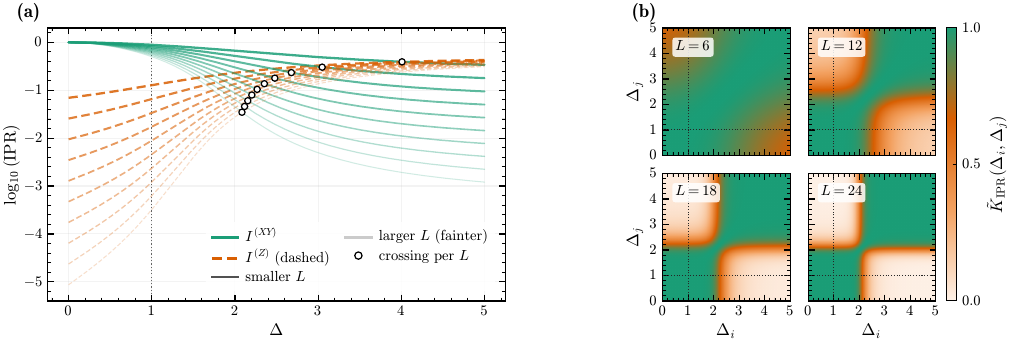}
    \caption{Heisenberg XXZ ground states with the IPR kernel built from the $XY$ and $Z$ bases. (a) IPRs for $L\in\{6,8,\dots,24\}$ on a $\log_{10}$ scale. Circles mark the crossing for each $L$, which moves towards $\Delta_c=1$ with increasing $L$ but remains well to its right. (b) Normalized Gram matrices for $L=6$, $12$, $18$, and $24$, with $\Delta_c=1$ marked by dotted lines.}
    \label{fig:xxz_results}
\end{figure*}

\subsection{XXZ ground states}\label{sec:xxz}

We now turn to a less trivial example, where the same principle is applied to a Hamiltonian decomposition beyond simple product bases. We consider the Heisenberg XXZ model without an external field \cite{yang_one_dimensional_chain_1966,misguich_inverse_participation_2016a,tsukerman_inverse_participation_2017},

\begin{equation}\label{eq:xxz_hamiltonian}
    H_{\text{XXZ}}=\sum_{i=1}^{L}\left(X_i X_{i+1}+Y_i Y_{i+1}+\Delta\, Z_i Z_{i+1}\right).
\end{equation}

The XXZ chain has two phase transitions, at $\Delta=-1$ to a ferromagnet and at $\Delta_c=1$ to the N\'eel antiferromagnet. For the zero-field chain, we restrict to $\Delta\geq0$ and use our IPR kernel to detect the latter. In order to properly capture the system's behavior, we follow the approach outlined in \cref{sec:basis_choice} and write the Hamiltonian as
\begin{equation}
    H=H_{XY}+\Delta H_{ZZ},
\end{equation}
with $H_{XY}=\sum_{i=1}^{L}\left(X_i X_{i+1}+Y_i Y_{i+1}\right)$ and $H_{ZZ}=\sum_{i=1}^{L}\left(Z_i Z_{i+1}\right)$. For $\Delta=0$, the model reduces to the XX chain described by $H_{XY}$. After a Jordan-Wigner transformation, $H_{XY}$ maps to a free-fermion Hamiltonian, which is diagonalized by a Fourier transform. The resulting eigenbasis is therefore a free-fermion occupation basis, which we refer to as the $XY$ basis. For large positive $\Delta$, the $ZZ$ interaction dominates and naturally motivates the computational $Z$ basis.

Ground states are computed via exact diagonalization (ED). For $\Delta\geq0$ they lie in the $S^z=0$ sector. ED limits us to sizes $L\in\{6,8,\dots,24\}$ but gives access to the full state vector, so the IPRs are evaluated exactly as $\sum_x|\langle x|\psi^{(s)}_{\mathrm{GS}}\rangle|^4$ without Monte Carlo sampling. Full details are given in \cref{app:data_generation}.

We draw 5000 points from the fitted data in a range of $\Delta\in[0,5]$. The IPR kernel for the $XY$ and $Z$ bases corresponds to the feature map 
\begin{equation}
    \mathbf{I}(\psi_{\mathrm{GS}}(\Delta)):=\big(I^{(XY)}(\psi_{\mathrm{GS}}(\Delta)),I^{(Z)}(\psi_{\mathrm{GS}}(\Delta))\big).
    \label{eq:xyz_features}
\end{equation}
In \cref{fig:xxz_results}(a) we plot the IPR curves for each numerically obtained system size. A notable observation is that the crossing of the two IPR curves shifts towards the left with increasing system size. However, since no duality relates the $XY$ and $Z$ bases, it is unclear if it will converge to the exact transition point $\Delta_c=1$ in the infinite-size limit.

\Cref{fig:xxz_results}(b) shows the corresponding Gram matrices. The phase blocks become sharper with increasing system size, but their boundary follows the shifted crossing to the right of $\Delta_c=1$. The SVM therefore performs well globally but fails locally. We provide the results and further analysis in \cref{app:xxz}.

Beyond the zero-field XXZ chain, we add a longitudinal field so that the total Hamiltonian reads
\begin{equation}
    H_{\mathrm{tot}} = H_{XY} + \Delta H_{ZZ} + h H_Z,
\end{equation}
with $H_Z = \sum_{i=1}^L Z_i$. The extra parameter promotes the transition points of the zero-field chain to a family of phase boundaries in the $(\Delta,h)$ plane, which we resolve from ED at $L=14$ using the same two-basis feature map as in \cref{eq:xyz_features}.

\begin{figure}[t]
    \centering
    \includegraphics[scale=1.14]{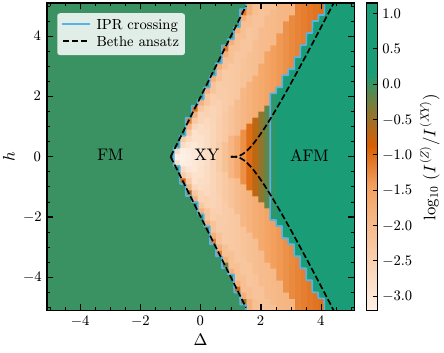}
    \caption{Phase diagram of the field-XXZ chain in the $(\Delta,h)$ plane, ED at $L=14$. Color: $\log_{10}(I^{(Z)}/I^{(XY)})$. The state is localized in the $Z$ basis in the N\'eel antiferromagnet (AFM, positive ratio), in the $XY$ basis in the critical phase (XY, negative ratio), and in both bases in the ferromagnet (FM, zero ratio). IPR boundary (light blue) vs.\ thermodynamic-limit Bethe-ansatz boundaries (black dashed): they coincide along the saturation line; along the AFM boundary, the finite-size spin gap shifts the IPR boundary to larger $|h|$, and near $h=0$ it follows the zero-field IPR crossing at $\Delta\approx2.3$.}
    \label{fig:xxz_field}
\end{figure}

\Cref{fig:xxz_field} shows the resulting diagram, with the color encoding the ratio $\log_{10}\!\big(I^{(Z)}/I^{(XY)}\big)$ of the two IPRs. This ratio acts as a natural indicator of which basis best resolves the ground state. It is positive (green) in the \emph{antiferromagnetic} (N\'eel) phase (AFM) at $\Delta\gtrsim1$, where the state is localized in the $Z$ basis, and negative (pale) throughout the central critical \emph{$XY$ phase} (XY), a Luttinger liquid whose free-fermion character makes the $XY$ basis the more localized one. In the \emph{ferromagnet} (FM), comprising the fully polarized phase and the line $h=0$, $\Delta<-1$, the ground state is localized in both bases with equal IPRs, and the ratio vanishes. The IPR ratio therefore separates the three phases on its own, and the predicted boundary (light blue), where the ratio turns negative, can be compared directly with the known thermodynamic-limit boundaries from the Bethe ansatz (black dashed) \cite{bethe_zur_1931,yang_one_dimensional_chain_1966,takahashi_thermodynamics_1999}.

The quality of the reconstruction is boundary-dependent. Since $H_Z$ commutes with $H_{XY}+\Delta H_{ZZ}$, the ground state within each magnetization sector does not depend on $h$, and the IPRs change only at level crossings between sectors. The phase boundaries at finite field are such level crossings, which the IPR ratio therefore locates directly. Along the ferromagnetic saturation line $|h|=2(1+\Delta)$ the fully polarized state is an exact product eigenstate, the level crossing is exact for any $L$, and the IPR boundary follows the analytical curve. Along the AFM boundary, the first level crossing at $L=14$ lies above the thermodynamic-limit value $h_{c1}$ of \cref{eq:hc1} because of the finite-size spin gap, e.g., at $h\approx3.0$ instead of $h_{c1}\approx2.45$ for $\Delta=3$. At $h=0$, in contrast, the transition between the XY and AFM phases at $\Delta_c=1$ is a Berezinskii-Kosterlitz-Thouless (BKT) transition \cite{berezinskii_destruction_1971,kosterlitz_thouless_ordering_1973,kosterlitz_critical_1974,giamarchi_quantum_2004} without a level crossing. There, the IPR boundary is the crossing of two smoothly varying IPRs, near $\Delta\approx2.3$ at $L=14$. As for the zero-field chain in \cref{fig:xxz_results}(a), this crossing drifts slowly towards $\Delta_c=1$ with increasing $L$, but no symmetry fixes it there.

\subsection{TFIM Gibbs states}\label{sec:thermal}

The previous sections showed that the IPR kernel separates ground states of different phases away from the transition. However, similar performance can also be obtained using the HS kernel $K_{\rm HS}(\psi_{\mathrm{GS}},\phi_{\mathrm{GS}})$ of \cref{eq:hs_kernel} \cite{sancho_lorente_quantum_kernels_2022}. We therefore turn to states away from the exact ground state, where we expect the two kernels to differ, since the HS kernel is sensitive to every difference between two states, while the IPR kernel compares only their localization in the chosen bases (\cref{sec:ipr_kernel}). In keeping with the previous example, we consider Gibbs states of the TFIM Hamiltonian \cite{alhambra_quantum_many_body_2023} defined as
\begin{equation}
    \rho(\beta,h)=\frac{e^{-\beta H_{\mathrm{TFIM}}(h)}}{Z(\beta,h)},
\end{equation}
where $\beta$ denotes the inverse temperature (in natural units taking $k_B = 1$) and $Z(\beta,h)=\Tr(e^{-\beta H_{\mathrm{TFIM}}(h)})$ is the partition function ensuring normalization, i.e., $\Tr(\rho)=1$. At fixed $\beta$, we evaluate $\tilde{K}_{\mathrm{IPR}}(\rho(h_i),\rho(h_j))$ using \cref{eq:mixed_ipr,eq:ipr_kernel_norm} in the $Z$ and $X$ bases, as in \cref{sec:tfim}, and varying transverse-field parameters $h\in[0,4]$. We then compare the results with the normalized HS kernel
\begin{equation} \label{eq:hs_kernel_norm}
\tilde{K}_{\mathrm{HS}}\big(\rho(h_i),\rho(h_j)\big)=\frac{\Tr(\rho(h_i)\rho(h_j))}{\sqrt{\Tr(\rho(h_i)^2)\Tr(\rho(h_j)^2)}},
\end{equation}
a slightly different measure from the Uhlmann fidelity widely used for mixed states in other approaches \cite{zanardi_mixed_state_fidelity_2007,bialonczyk_uhlmann_fidelity_2021}. We gather data for both kernels in a thermal range of $\beta \in \{0.1,0.2,\dots,2\}$. The Gibbs states are represented as matrix product operators (MPOs) obtained by imaginary-time evolution of the infinite-temperature state via time-evolving block decimation (TEBD). The IPR features are estimated by Monte Carlo sampling from the resulting MPO, and the HS kernel is evaluated by MPO contraction. Full details are given in \cref{app:data_generation}.

\begin{figure*}[t]
    \centering
    \includegraphics{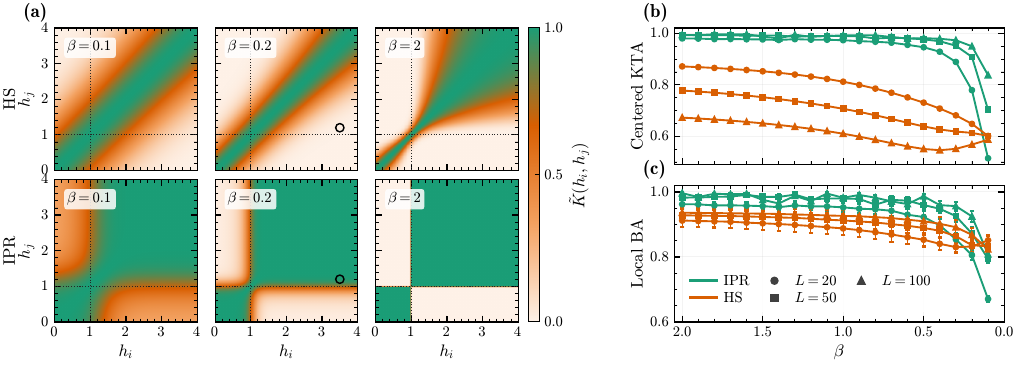}
    \caption{Phase classification of TFIM Gibbs states with the HS and IPR kernels. (a) Normalized Gram matrices at $L=100$ for $\beta=0.1$, $0.2$, and $2$, with $h_c=1$ marked by dotted lines. Circles mark the element $(h_i,h_j)=(3.5,1.2)$ discussed in the text. (b) Centered kernel-target alignment and (c) local balanced accuracy ($\delta=0.25$, averaged over 100 train-test splits, standard deviation over splits) versus $\beta$ for $L=20$, $50$, and $100$. Temperature increases to the right.}
    \label{fig:thermal_results}
\end{figure*}

First, we compare the two kernels in terms of their Gram matrices for $L=100$, shown in \cref{fig:thermal_results}(a). Close to the ground-state regime, i.e., at large $\beta$, we expect both Gram matrices to exhibit a phase-accurate block structure. As the temperature increases, the structure should gradually fade as the Gibbs states approach the maximally mixed state
\begin{equation}
    \rho(\beta,h) \xrightarrow[\beta\to 0]{} \frac{1}{2^L}\mathbb{1},
\end{equation}
and therefore become less distinguishable. A more detailed analysis of the temperature and system-size dependence of both kernels is given in \cref{app:thermal}.

With increasing temperature, the HS kernel Gram matrix loses its block structure faster than its IPR kernel counterpart. As an example of why this happens, consider $\beta=0.2$ and the matrix element $\tilde{K}_{\mathrm{HS/IPR}}(h_i=3.5,h_j=1.2)$, marked by circles in \cref{fig:thermal_results}(a). Both transverse-field inputs belong to the same phase, yet the HS kernel assigns the corresponding Gibbs states a similarity of only $0.007$, compared with $0.98$ for the IPR kernel. This reflects the fact that from the perspective of state overlap, the two density matrices can still be nearly orthogonal despite being in the same phase. However, their basis-dependent sparsity profiles remain aligned, which leads to the IPR kernel assigning them a higher similarity value.

\looseness=1 This difference between the expected classification performance of the two kernels can be quantified by the centered kernel-target alignment (KTA) \cite{cortes_algorithms_learning_2024}. It measures how well a given Gram matrix aligns with an ideal, label-derived target matrix under the Frobenius inner product. The target is $T_{ij}=y_i y_j$ with $y_i \in \{-1,+1\}$. For $N$ data points, centering uses $C=\mathbb{1}_N-\frac{1}{N}\mathbf{1}\mathbf{1}^{\mathsf T}$, where $\mathbf{1}$ is the all-ones vector, giving $K^{\mathrm{c}}=C\tilde K C$ and $T^{\mathrm{c}}=CTC$. We have
\begin{equation}
\mathrm{KTA}_{\mathrm{centered}}(\tilde K_{\mathrm{IPR/HS}},T)=
\frac{\langle K^{\mathrm{c}}_{\mathrm{IPR/HS}},T^{\mathrm{c}}\rangle_F}
{\|K^{\mathrm{c}}_{\mathrm{IPR/HS}}\|_F\,\|T^{\mathrm{c}}\|_F},
\end{equation}
which we plot in \cref{fig:thermal_results}(b) for various inverse temperatures. The two kernels show a clear difference in alignment with the phase labels.

To test whether this difference carries over to classification results, we train SVMs (using the same procedure as in \cref{sec:tfim}) for the same $\beta$ range, focusing in particular on the classification performance for test points drawn from the local region $\delta=0.25$ around the phase transition. The local balanced accuracies for $L=20$, $50$, and $100$ are shown in \cref{fig:thermal_results}(c), and the full results, including the global test region, are given in \cref{tab:svm_thermal}.

Both kernels achieve near-perfect global classification performance. However, this task is admittedly simple, since it amounts to separating points on a one-dimensional parameter line according to their position relative to the critical point. The local classification performance near the phase transition is therefore more revealing. Here, the HS kernel exhibits a smaller balanced accuracy than the IPR kernel for $\beta\geq0.3$. In \cref{app:thermal} we show that this behavior correlates with how sharply their corresponding decision functions change at their learned zero crossings. We also observe a different behavior: with increasing temperature, i.e., $\beta \rightarrow 0$, the IPR kernel shows a quicker decay than the HS kernel. We attribute this behavior to a faster loss of phase contrast in the IPR kernel and provide an analytical explanation in \cref{app:thermal}.

\section{Discussion}\label{sec:discussion}

\looseness=1 We introduced the IPR kernel, a basis-dependent kernel tailored to learning quantum phases of matter. Its novelty lies in comparing quantum states through their IPR profiles in selected physical bases, rather than through direct state overlap. For the TFIM, we observe that the IPR crossing approaches the critical point with increasing system size, as a consequence of the Kramers-Wannier duality. We also find that the exact IPR kernel retains phase-relevant structure for Gibbs states at finite temperature where the phase signal of the HS kernel is already degraded, suggesting robustness to thermal noise in the ground-state preparation.

\looseness=1 In the XXZ example, where no duality determines the crossing, the $XY$ and $Z$ bases produce a clear separation between the phases, but for the accessible system sizes the estimated transition remains displaced from the critical point $\Delta_c=1$. In a longitudinal field, in contrast, the phase boundaries are level crossings, which the IPRs locate directly at each system size. Without knowledge of the Hamiltonian one could start from an initial set of bases, e.g., $\mathcal{B}=\{X,Y,Z\}$, and check for the existence of opposing regimes using the corresponding IPRs. Developing a systematic algorithm for selecting suitable bases in Hamiltonian-agnostic settings would be a natural next step.

\looseness=1 An important distinction must also be made between the information contained in the exact kernel and the resources required to estimate it. The raw IPRs decrease exponentially with system size and more quickly when a state is delocalized in the chosen basis. Assuming a realistic fixed sampling budget, the kernel therefore only classifies states reliably if they are far from the transition. We also observe that the range of unresolved IPRs near the transition widens with system size. However, extrapolating the IPR curves from the deep phase regimes into this range could help estimate smaller IPR values, and the resulting crossing could precisely mark the transition point in cases where a symmetry or duality fixes it, as for the TFIM. We further note that, in the absence of explicit labeling, our approach can in principle be extended to an unsupervised learning task of identifying transition regions as the parameter ranges in which the IPRs in all physically relevant bases lie below the sampling resolution.

\looseness=1 More broadly, our results support a physics-informed view of quantum kernels: useful feature maps should be tailored to the structure of the problem at hand, rather than made maximally expressive by default. Future work could apply IPR-based features directly to quantum circuits and variational algorithms, in line with recent work on phase detection with variational states, localization-based trainability, and problem-informed ansätze \cite{cao_unveiling_quantum_2025,cao_exploiting_many_body_2025,srimahajariyapong_connecting_phases_2025,park_hamiltonian_variational_2024}. In this sense, the IPR kernel points towards problem-aware quantum machine learning and highlights the potential of quantum-computer-assisted learning for problems rooted in quantum many-body physics.

\section*{Data and code availability}

The numerical simulations were implemented in Python and Julia. Exact diagonalization was performed using the \texttt{NetKet} library in Python \cite{carleo_netket_2019,vicentini_netket3_2022}, and the matrix product state (MPS) calculations were carried out with the \texttt{ITensor} library in Julia \cite{fishman_itensor_2022}. The complete implementation and the raw data supporting the findings of this study are available at
\url{https://github.com/clemensLin/IPR-kernel}.
% TODO: confirm that the public repository contains Data/Raw before submission.

\section*{Acknowledgments}

The authors thank Juan Carrasquilla for his hospitality and support. C.L. acknowledges financial support from the Finnish Ministry of Education and Culture through the Quantum Doctoral Education Pilot Program (QDOC VN/3137/2024-OKM-4), and from the Research Council of Finland through the Finnish Quantum Flagship project (JYU 359240, University of Jyväskylä). 
A.P.-S. acknowledges the Swiss National Science Foundation for its support through the grant TMPFP2\_234085. 

\section*{Competing interests}

The authors declare that they have no competing interests relevant to the content of this article.
% --------------------------------------------
%\newpage
%\bibliographystyle{apsrev4-2} %added style, no need since titles in 4-2 added by default
\bibliography{refs}

@book{sachdev_quantum_phase_2011,
  title = {Quantum {{Phase Transitions}}},
  author = {Sachdev, Subir},
  year = 2011,
  edition = {2},
  publisher = {Cambridge University Press},
  doi = {10.1017/CBO9780511973765},
  url = {https://www.cambridge.org/core/books/quantum-phase-transitions/33C1C81500346005E54C1DE4223E5562},
  isbn = {978-0-521-51468-2}
}

@article{vojta_quantum_phase_2003,
  title = {Quantum Phase Transitions},
  author = {Vojta, Matthias},
  year = 2003,
  journal = {Reports on Progress in Physics},
  volume = {66},
  eprint = {cond-mat/0309604},
  pages = {2069--2110},
  issn = {0034-4885, 1361-6633},
  doi = {10.1088/0034-4885/66/12/R01},
  url = {http://arxiv.org/abs/cond-mat/0309604},
  archiveprefix = {arXiv}
}

@inproceedings{kempe_complexity_local_2005,
  title = {The {{Complexity}} of the {{Local Hamiltonian Problem}}},
  booktitle = {{{FSTTCS}} 2004: {{Foundations}} of {{Software Technology}} and {{Theoretical Computer Science}}},
  author = {Kempe, Julia and Kitaev, Alexei and Regev, Oded},
  editor = {Lodaya, Kamal and Mahajan, Meena},
  year = 2004,
  pages = {372--383},
  address = {Berlin, Heidelberg},
  doi = {10.1007/978-3-540-30538-5_31},
  isbn = {978-3-540-30538-5},
  langid = {english}
}

@article{hastings_area_law_2007,
  title = {An {{Area Law}} for {{One Dimensional Quantum Systems}}},
  author = {Hastings, M. B.},
  year = 2007,
  journal = {Journal of Statistical Mechanics: Theory and Experiment},
  volume = {2007},
  eprint = {0705.2024},
  primaryclass = {quant-ph},
  pages = {P08024-P08024},
  issn = {1742-5468},
  doi = {10.1088/1742-5468/2007/08/P08024},
  url = {http://arxiv.org/abs/0705.2024},
  archiveprefix = {arXiv}
}

@article{verstraete_matrix_product_2006,
  title = {Matrix Product States Represent Ground States Faithfully},
  author = {Verstraete, F. and Cirac, J. I.},
  year = 2006,
  journal = {Physical Review B},
  volume = {73},
  pages = {094423},
  doi = {10.1103/PhysRevB.73.094423},
  url = {https://link.aps.org/doi/10.1103/PhysRevB.73.094423}
}

@misc{landau_polynomial_time_algorithm_2013,
  title = {A Polynomial-Time Algorithm for the Ground State of {{1D}} Gapped Local {{Hamiltonians}}},
  author = {Landau, Zeph and Vazirani, Umesh and Vidick, Thomas},
  year = 2013,
  eprint = {1307.5143},
  primaryclass = {quant-ph},
  doi = {10.48550/arXiv.1307.5143},
  url = {http://arxiv.org/abs/1307.5143},
  archiveprefix = {arXiv}
}

@article{schuch_classifying_quantum_2011,
  title = {Classifying Quantum Phases Using Matrix Product States and Projected Entangled Pair States},
  author = {Schuch, Norbert and {P{\'e}rez-Garc{\'i}a}, David and Cirac, Ignacio},
  year = 2011,
  journal = {Physical Review B},
  volume = {84},
  pages = {165139},
  doi = {10.1103/PhysRevB.84.165139},
  url = {https://link.aps.org/doi/10.1103/PhysRevB.84.165139}
}

@article{caux_remarks_notion_2011,
  title = {Remarks on the Notion of Quantum Integrability},
  author = {Caux, J.-S. and Mossel, J.},
  year = 2011,
  journal = {Journal of Statistical Mechanics: Theory and Experiment},
  volume = {2011},
  eprint = {1012.3587},
  primaryclass = {cond-mat.str-el},
  pages = {P02023},
  issn = {1742-5468},
  doi = {10.1088/1742-5468/2011/02/P02023},
  url = {http://arxiv.org/abs/1012.3587},
  archiveprefix = {arXiv}
}

@incollection{brassard_quantum_amplitude_2002,
  author        = {Brassard, Gilles and H{\o}yer, Peter and Mosca, Michele and Tapp, Alain},
  title         = {Quantum amplitude amplification and estimation},
  booktitle     = {Quantum Computation and Information},
  editor        = {Lomonaco, Jr., Samuel J. and Brandt, Howard E.},
  series        = {Contemporary Mathematics},
  volume        = {305},
  pages         = {53-74},
  publisher     = {American Mathematical Society},
  address       = {Providence, RI},
  year          = {2002},
  doi           = {10.1090/conm/305/05215},
  eprint        = {quant-ph/0005055},
  archiveprefix = {arXiv}
}

@article{carrasquilla_machine_learning_2017,
  title = {Machine Learning Phases of Matter},
  author = {Carrasquilla, Juan and Melko, Roger G.},
  year = 2017,
  journal = {Nature Physics},
  volume = {13},
  pages = {431--434},
  issn = {1745-2481},
  doi = {10.1038/nphys4035},
  url = {https://www.nature.com/articles/nphys4035},
  copyright = {2017 Springer Nature Limited},
  langid = {english}
}

@article{carrasquilla_machine_learning_2020,
  title = {Machine Learning for Quantum Matter},
  author = {Carrasquilla, Juan},
  year = 2020,
  journal = {Advances in Physics: X},
  volume = {5},
  pages = {1797528},
  issn = {2374-6149},
  doi = {10.1080/23746149.2020.1797528},
  url = {https://www.tandfonline.com/doi/full/10.1080/23746149.2020.1797528},
  langid = {english}
}

@article{zhang_interpretable_machine_2019,
  title = {Interpretable {{Machine Learning Study}} of {{Many-Body Localization Transition}} in {{Disordered Quantum Ising Spin Chains}}},
  author = {Zhang, Wei and Wang, Lei and Wang, Ziqiang},
  year = 2019,
  journal = {Physical Review B},
  volume = {99},
  eprint = {1807.02954},
  primaryclass = {cond-mat},
  pages = {054208},
  issn = {2469-9950, 2469-9969},
  doi = {10.1103/PhysRevB.99.054208},
  url = {http://arxiv.org/abs/1807.02954},
  archiveprefix = {arXiv}
}

@article{huang_provably_efficient_2022,
  title = {Provably Efficient Machine Learning for Quantum Many-Body Problems},
  author = {Huang, Hsin-Yuan and Kueng, Richard and Torlai, Giacomo and Albert, Victor V. and Preskill, John},
  year = 2022,
  journal = {Science},
  volume = {377},
  pages = {eabk3333},
  issn = {0036-8075, 1095-9203},
  doi = {10.1126/science.abk3333},
  url = {https://www.science.org/doi/10.1126/science.abk3333},
  langid = {english}
}

@misc{scarato_hybrid_quantum_classical_2026,
  title = {Hybrid {{Quantum-Classical Neural Networks}} for {{Recognizing Quantum Phases}}},
  author = {Scarato, Colin and Kn{\"o}rzer, Johannes and Hoffmann, Markus K. and Sander, Leon C. and Hofele, Luca and Wang, Shengpu and Hanke, Kilian and Sathe, Ashay and Hagmann, Dominic and Flasby, Alexander and Hartmann, Michael J. and Zapletal, Petr and Wallraff, Andreas and Hellings, Christoph},
  year = 2026,
  eprint = {2606.28201},
  primaryclass = {quant-ph},
  doi = {10.48550/arXiv.2606.28201},
  url = {http://arxiv.org/abs/2606.28201},
  archiveprefix = {arXiv}
}

@misc{schoulepnikoff_discovering_quantum_2026,
  title = {Discovering Quantum Phenomena with {{Interpretable Machine Learning}}},
  author = {de Schoulepnikoff, Paulin and Nautrup, Hendrik Poulsen and Briegel, Hans J. and {Mu{\~n}oz-Gil}, Gorka},
  year = 2026,
  eprint = {2604.16015},
  primaryclass = {quant-ph},
  doi = {10.48550/arXiv.2604.16015},
  url = {http://arxiv.org/abs/2604.16015},
  archiveprefix = {arXiv}
}

@misc{bespalova_benchmarking_quantum_2026,
  title = {Benchmarking {{Quantum Convolutional Neural Networks}} and Their Classical Surrogates on Long-Range Strongly Correlated Models},
  author = {Bespalova, Tatiana A. and Schwengelbeck, R{\'e}ka and Mismetti, Laura and Schachenmayer, Johannes and Tavernelli, Ivano and Tacchino, Francesco},
  year = 2026,
  eprint = {2611.00255},
  primaryclass = {quant-ph},
  doi = {10.48550/arXiv.2611.00255},
  url = {http://arxiv.org/abs/2611.00255},
  archiveprefix = {arXiv}
}

@article{peruzzo_variational_eigenvalue_2014,
  title = {A Variational Eigenvalue Solver on a Photonic Quantum Processor},
  author = {Peruzzo, Alberto and McClean, Jarrod and Shadbolt, Peter and Yung, Man-Hong and Zhou, Xiao-Qi and Love, Peter J. and {Aspuru-Guzik}, Al{\'a}n and O'Brien, Jeremy L.},
  year = 2014,
  journal = {Nature Communications},
  volume = {5},
  eprint = {1304.3061},
  primaryclass = {quant-ph},
  pages = {4213},
  issn = {2041-1723},
  doi = {10.1038/ncomms5213},
  url = {http://arxiv.org/abs/1304.3061},
  archiveprefix = {arXiv}
}

@article{cerezo_variational_quantum_2021,
  title = {Variational {{Quantum Algorithms}}},
  author = {Cerezo, M. and Arrasmith, Andrew and Babbush, Ryan and Benjamin, Simon C. and Endo, Suguru and Fujii, Keisuke and McClean, Jarrod R. and Mitarai, Kosuke and Yuan, Xiao and Cincio, Lukasz and Coles, Patrick J.},
  year = 2021,
  journal = {Nature Reviews Physics},
  volume = {3},
  eprint = {2012.09265},
  primaryclass = {quant-ph},
  pages = {625--644},
  issn = {2522-5820},
  doi = {10.1038/s42254-021-00348-9},
  url = {http://arxiv.org/abs/2012.09265},
  archiveprefix = {arXiv}
}

@article{linke_measuring_renyi_2018,
  title = {Measuring the {{Renyi}} Entropy of a Two-Site {{Fermi-Hubbard}} Model on a Trapped Ion Quantum Computer},
  author = {Linke, Norbert M. and Johri, Sonika and Figgatt, Caroline and Landsman, Kevin A. and Matsuura, Anne Y. and Monroe, Christopher},
  year = 2018,
  journal = {Physical Review A},
  volume = {98},
  eprint = {1712.08581},
  primaryclass = {quant-ph},
  pages = {052334},
  issn = {2469-9926, 2469-9934},
  doi = {10.1103/PhysRevA.98.052334},
  url = {http://arxiv.org/abs/1712.08581},
  archiveprefix = {arXiv}
}

@misc{yang_probing_entanglement_2026,
  title         = {Probing Entanglement and Symmetries in Random States Using a Superconducting Quantum Processor},
  author        = {Yang, Jia-Nan and Joshi, Lata Kh and Ares, Filiberto and Han, Yihang and Zhang, Pengfei and Calabrese, Pasquale},
  year          = {2026},
  eprint        = {2601.22224},
  archiveprefix = {arXiv},
  primaryclass  = {quant-ph},
  doi           = {10.48550/arXiv.2601.22224},
  url           = {https://arxiv.org/abs/2601.22224}
}

@article{koh_measurement_induced_entanglement_2023,
  title = {Measurement-Induced Entanglement Phase Transition on a Superconducting Quantum Processor with Mid-Circuit Readout},
  author = {Koh, Jin Ming and Sun, Shi-Ning and Motta, Mario and Minnich, Austin J.},
  year = 2023,
  journal = {Nature Physics},
  volume = {19},
  pages = {1314--1319},
  issn = {1745-2481},
  doi = {10.1038/s41567-023-02076-6},
  url = {https://www.nature.com/articles/s41567-023-02076-6},
  copyright = {2023 The Author(s), under exclusive licence to Springer Nature Limited},
  langid = {english}
}

@article{tan_probing_quantum_2025,
  title = {Probing Quantum Phase Transitions via Short-Depth Quantum Circuits for Estimating Quantum Coherence and Discrete {{Berry}} Phases},
  author = {Tan, Kok Chuan and Liu, Deren and Bharti, Kishor and Bhowmick, Dhiman and Sengupta, Pinaki},
  year = 2025,
  journal = {Physical Review B},
  volume = {112},
  pages = {014108},
  doi = {10.1103/gydb-9jt2},
  url = {https://link.aps.org/doi/10.1103/gydb-9jt2}
}

@article{muser_provable_advantages_2024,
  title = {Provable Advantages of Kernel-Based Quantum Learners and Quantum Preprocessing Based on {{Grover}}'s Algorithm},
  author = {Muser, T. and Zapusek, E. and Belis, V. and Reiter, F.},
  year = 2024,
  journal = {Physical Review A},
  volume = {110},
  pages = {032434},
  doi = {10.1103/PhysRevA.110.032434},
  url = {https://link.aps.org/doi/10.1103/PhysRevA.110.032434}
}

@article{liu_rigorous_robust_2021,
  title = {A Rigorous and Robust Quantum Speed-up in Supervised Machine Learning},
  author = {Liu, Yunchao and Arunachalam, Srinivasan and Temme, Kristan},
  year = 2021,
  journal = {Nature Physics},
  volume = {17},
  pages = {1013--1017},
  issn = {1745-2481},
  doi = {10.1038/s41567-021-01287-z},
  url = {https://www.nature.com/articles/s41567-021-01287-z},
  copyright = {2021 The Author(s), under exclusive licence to Springer Nature Limited},
  langid = {english}
}

@article{wegner_inverse_participation_1980,
  title = {Inverse Participation Ratio in 2+{$\varepsilon$} Dimensions},
  author = {Wegner, F.},
  year = 1980,
  journal = {Zeitschrift f\"ur Physik B Condensed Matter},
  volume = {36},
  pages = {209--214},
  issn = {1431-584X},
  doi = {10.1007/BF01325284},
  url = {https://doi.org/10.1007/BF01325284},
  langid = {english}
}

@article{murphy_generalized_inverse_2011,
  title = {Generalized Inverse Participation Ratio as a Possible Measure of Localization for Interacting Systems},
  author = {Murphy, N. C. and Wortis, R. and Atkinson, W. A.},
  year = 2011,
  journal = {Physical Review B},
  volume = {83},
  pages = {184206},
  doi = {10.1103/PhysRevB.83.184206},
  url = {https://link.aps.org/doi/10.1103/PhysRevB.83.184206}
}

@article{liu_quantum_2025,
	title = {Quantum {Algorithms} for {Inverse} {Participation} {Ratio} {Estimation} in multi-qubit and multi-qudit systems},
	author = {Liu, Yingjian and Sierant, Piotr and Stornati, Paolo and Lewenstein, Maciej and Płodzień, Marcin},
	year = {2025},
	month = may,
	journal = {Physical Review A},
	volume = {111},
	number = {5},
	pages = {052614},
	doi = {10.1103/PhysRevA.111.052614},
	issn = {2469-9926, 2469-9934},
	url = {http://arxiv.org/abs/2405.03338},
	urldate = {2025-10-28},
	note = {arXiv:2405.03338 [quant-ph]}
}

@article{frey_probing_2024,
	title = {Probing {Hilbert} space fragmentation and the block inverse participation ratio},
	author = {Frey, Philipp and Mikhail, David and Rachel, Stephan and Hackl, Lucas},
	year = {2024},
	month = feb,
	journal = {Physical Review B},
	publisher = {American Physical Society},
	volume = {109},
	number = {6},
	pages = {064302},
	doi = {10.1103/PhysRevB.109.064302},
	url = {https://link.aps.org/doi/10.1103/PhysRevB.109.064302},
	urldate = {2026-01-19}
}

@article{stephan_renyi_entropy_2010,
  title = {Rényi Entropy of a Line in Two-Dimensional {{Ising}} Models},
  author = {St{\'e}phan, J.-M. and Misguich, G. and Pasquier, V.},
  year = 2010,
  journal = {Physical Review B},
  volume = {82},
  pages = {125455},
  doi = {10.1103/PhysRevB.82.125455},
  url = {https://link.aps.org/doi/10.1103/PhysRevB.82.125455}
}

@article{evers_anderson_transitions_2008,
  title = {Anderson Transitions},
  author = {Evers, Ferdinand and Mirlin, Alexander D.},
  year = 2008,
  journal = {Reviews of Modern Physics},
  volume = {80},
  pages = {1355--1417},
  doi = {10.1103/RevModPhys.80.1355},
  url = {https://link.aps.org/doi/10.1103/RevModPhys.80.1355}
}

@article{evers_fluctuations_inverse_2000,
  title = {Fluctuations of the {{Inverse Participation Ratio}} at the {{Anderson Transition}}},
  author = {Evers, F. and Mirlin, A. D.},
  year = 2000,
  journal = {Physical Review Letters},
  volume = {84},
  pages = {3690--3693},
  doi = {10.1103/PhysRevLett.84.3690},
  url = {https://link.aps.org/doi/10.1103/PhysRevLett.84.3690}
}

@article{rossignolo_localization_transitions_2019,
  title = {Localization Transitions and Mobility Edges in Coupled Aubry-André Chains},
  author = {Rossignolo, M. and Dell'Anna, L.},
  year = 2019,
  journal = {Physical Review B},
  volume = {99},
  pages = {054211},
  doi = {10.1103/PhysRevB.99.054211},
  url = {https://link.aps.org/doi/10.1103/PhysRevB.99.054211}
}

@article{pal_many_body_localization_2010,
  title = {Many-Body Localization Phase Transition},
  author = {Pal, Arijeet and Huse, David A.},
  year = 2010,
  journal = {Physical Review B},
  volume = {82},
  pages = {174411},
  doi = {10.1103/PhysRevB.82.174411},
  url = {https://link.aps.org/doi/10.1103/PhysRevB.82.174411}
}

@article{baena_entropies_ipr_2022,
  title = {Entropies and {{IPR}} as {{Markers}} for a {{Phase Transition}} in a {{Two-Level Model}} for {{Atom-Diatomic Molecule Coexistence}}},
  author = {Baena, Ignacio and {P{\'e}rez-Fern{\'a}ndez}, Pedro and {Rodr{\'i}guez-Gallardo}, Manuela and Arias, Jos{\'e} Miguel},
  year = 2022,
  journal = {Entropy},
  volume = {24},
  eprint = {2201.04571},
  primaryclass = {quant-ph},
  pages = {113},
  issn = {1099-4300},
  doi = {10.3390/e24010113},
  url = {http://arxiv.org/abs/2201.04571},
  archiveprefix = {arXiv}
}

@article{alet_many_body_localization_2018,
  title = {Many-Body Localization: An Introduction and Selected Topics},
  shorttitle = {Many-Body Localization},
  author = {Alet, Fabien and Laflorencie, Nicolas},
  year = 2018,
  journal = {Comptes Rendus. Physique},
  volume = {19},
  eprint = {1711.03145},
  primaryclass = {cond-mat},
  pages = {498--525},
  issn = {1631-0705, 1878-1535},
  doi = {10.1016/j.crhy.2018.03.003},
  url = {http://arxiv.org/abs/1711.03145},
  archiveprefix = {arXiv}
}

@article{bera_many_body_localization_2015,
  title = {Many-{{Body Localization Characterized}} from a {{One-Particle Perspective}}},
  author = {Bera, Soumya and Schomerus, Henning and {Heidrich-Meisner}, Fabian and Bardarson, Jens H.},
  year = 2015,
  journal = {Physical Review Letters},
  volume = {115},
  pages = {046603},
  issn = {0031-9007, 1079-7114},
  doi = {10.1103/PhysRevLett.115.046603},
  url = {https://link.aps.org/doi/10.1103/PhysRevLett.115.046603},
  copyright = {http://link.aps.org/licenses/aps-default-license},
  langid = {english}
}

@article{de_luca_ergodicity_2013,
	title = {Ergodicity breaking in a model showing many-body localization},
	author = {De Luca, A. and Scardicchio, A.},
	year = {2013},
	month = feb,
	journal = {EPL (Europhysics Letters)},
	volume = {101},
	number = {3},
	pages = {37003},
	doi = {10.1209/0295-5075/101/37003},
	issn = {0295-5075, 1286-4854},
	url = {https://iopscience.iop.org/article/10.1209/0295-5075/101/37003},
	urldate = {2025-11-24},
	copyright = {http://iopscience.iop.org/info/page/text-and-data-mining}
}

@article{brown_quantum_2008,
	title = {Quantum chaos, delocalization, and entanglement in disordered {Heisenberg} models},
	author = {Brown, Winton G. and Santos, Lea F. and Starling, David J. and Viola, Lorenza},
	year = {2008},
	month = feb,
	journal = {Physical Review E},
	volume = {77},
	number = {2},
	pages = {021106},
	doi = {10.1103/PhysRevE.77.021106},
	url = {https://link.aps.org/doi/10.1103/PhysRevE.77.021106},
	urldate = {2025-11-24},
	note = {Publisher: American Physical Society}
}

@misc{alam_onset_ergodicity_2026a,
  title = {Onset of {{Ergodicity Across Scales}} on a {{Digital Quantum Processor}}},
  author = {Alam, Faisal and Crichigno, Marcos and Crosson, Elizabeth and Flammia, Steven T. and Gambetta, Filippo Maria and Gordon, Max Hunter and Kreshchuk, Michael and Montanaro, Ashley and Nocera, Alberto and Santos, Raul A.},
  year = 2026,
  eprint = {2603.12236},
  primaryclass = {quant-ph},
  doi = {10.48550/arXiv.2603.12236},
  url = {http://arxiv.org/abs/2603.12236},
  archiveprefix = {arXiv}
}

@article{graf_quantum_2025,
  title = {Quantum {{Birthmarks}}: {{Ergodicity Breaking Beyond Scarring}}},
  shorttitle = {Quantum {{Birthmarks}}},
  author = {Graf, Anton M. and Atwood, Saul and Xiao, Mingxuan and Ketzmerick, Roland and Heller, Eric J. and {Keski-Rahkonen}, Joonas},
  year = 2026,
  journal = {Physical Review X},
  volume = {16},
  pages = {031063},
  issn = {2160-3308},
  doi = {10.1103/dhzb-28rb},
  url = {https://link.aps.org/doi/10.1103/dhzb-28rb},
  langid = {english}
}

@article{dalzell_random_quantum_2022,
  title = {Random {{Quantum Circuits Anticoncentrate}} in {{Log Depth}}},
  author = {Dalzell, Alexander M. and {Hunter-Jones}, Nicholas and Brand{\~a}o, Fernando G. S. L.},
  year = 2022,
  journal = {PRX Quantum},
  volume = {3},
  pages = {010333},
  issn = {2691-3399},
  doi = {10.1103/PRXQuantum.3.010333},
  url = {https://link.aps.org/doi/10.1103/PRXQuantum.3.010333},
  langid = {english}
}

@article{mari_counting_collisions_2024,
  title = {Counting Collisions in Random Circuit Sampling for Benchmarking Quantum Computers},
  author = {Mari, Andrea},
  year = 2024,
  journal = {APL Quantum},
  volume = {1},
  eprint = {2312.04222},
  primaryclass = {quant-ph},
  pages = {046113},
  issn = {2835-0103},
  doi = {10.1063/5.0219266},
  url = {http://arxiv.org/abs/2312.04222},
  archiveprefix = {arXiv}
}

@article{boixo_characterizing_quantum_2018,
  title = {Characterizing {{Quantum Supremacy}} in {{Near-Term Devices}}},
  author = {Boixo, Sergio and Isakov, Sergei V. and Smelyanskiy, Vadim N. and Babbush, Ryan and Ding, Nan and Jiang, Zhang and Bremner, Michael J. and Martinis, John M. and Neven, Hartmut},
  year = 2018,
  journal = {Nature Physics},
  volume = {14},
  eprint = {1608.00263},
  primaryclass = {quant-ph},
  pages = {595--600},
  issn = {1745-2473, 1745-2481},
  doi = {10.1038/s41567-018-0124-x},
  url = {http://arxiv.org/abs/1608.00263},
  archiveprefix = {arXiv}
}

@article{bremner_average_case_complexity_2016,
  title = {Average-Case Complexity versus Approximate Simulation of Commuting Quantum Computations},
  author = {Bremner, Michael J. and Montanaro, Ashley and Shepherd, Dan J.},
  year = 2016,
  journal = {Physical Review Letters},
  volume = {117},
  eprint = {1504.07999},
  primaryclass = {quant-ph},
  pages = {080501},
  issn = {0031-9007, 1079-7114},
  doi = {10.1103/PhysRevLett.117.080501},
  url = {http://arxiv.org/abs/1504.07999},
  archiveprefix = {arXiv}
}

@misc{aaronson_computational_complexity_2010,
  title = {The {{Computational Complexity}} of {{Linear Optics}}},
  author = {Aaronson, Scott and Arkhipov, Alex},
  year = 2010,
  eprint = {1011.3245},
  primaryclass = {quant-ph},
  doi = {10.48550/arXiv.1011.3245},
  url = {http://arxiv.org/abs/1011.3245},
  archiveprefix = {arXiv}
}

@article{lakshminarayan_entanglement_localization_2016,
  title = {Entanglement and Localization Transitions in Eigenstates of Interacting Chaotic Systems},
  author = {Lakshminarayan, Arul and Srivastava, Shashi C. L. and Ketzmerick, Roland and B{\"a}cker, Arnd and Tomsovic, Steven},
  year = 2016,
  journal = {Physical Review E},
  volume = {94},
  pages = {010205},
  doi = {10.1103/PhysRevE.94.010205},
  url = {https://link.aps.org/doi/10.1103/PhysRevE.94.010205}
}

@article{calixto_inverse_participation_2024,
  title = {Localization and Entanglement Characterization of Edge States in {{HgTe}} Quantum Wells in a Finite Strip Geometry},
  author = {Calixto, Manuel and Casta{\~n}os, Octavio},
  year = 2025,
  journal = {International Journal of Modern Physics B},
  volume = {39},
  pages = {2550263},
  issn = {0217-9792},
  doi = {10.1142/S0217979225502637},
  url = {https://www.worldscientific.com/doi/10.1142/S0217979225502637}
}

@article{freund_ability_entanglement_2026,
  title = {Ability of Entanglement and Purity to Help to Detect Systematic Experimental Errors},
  author = {Freund, Julia and Basso Basset, Francesco and Krieger, Tobias M. and Laneve, Alessandro and Beccaceci, Mattia and Rota, Michele B. and Buchinger, Quirin and da Silva, Saimon F. Covre and Stroj, Sandra and H{\"o}fling, Sven and {Huber-Loyola}, Tobias and Kueng, Richard and Rastelli, Armando and Trotta, Rinaldo and G{\"u}hne, Otfried},
  year = 2026,
  journal = {Physical Review A},
  volume = {113},
  eprint = {2503.09459},
  primaryclass = {quant-ph},
  pages = {022422},
  issn = {2469-9926, 2469-9934},
  doi = {10.1103/b74j-xxv2},
  url = {http://arxiv.org/abs/2503.09459},
  archiveprefix = {arXiv}
}

@article{brydges_probing_2019,
	title = {Probing {Rényi} entanglement entropy via randomized measurements},
	author = {Brydges, Tiff and Elben, Andreas and Jurcevic, Petar and Vermersch, Benoît and Maier, Christine and Lanyon, Ben P. and Zoller, Peter and Blatt, Rainer and Roos, Christian F.},
	year = {2019},
	month = apr,
	journal = {Science},
	volume = {364},
	number = {6437},
	pages = {260--263},
	doi = {10.1126/science.aau4963},
	url = {https://www.science.org/doi/10.1126/science.aau4963},
	urldate = {2025-10-30},
	note = {Publisher: American Association for the Advancement of Science}
}

@misc{perez_salinas_multiple_basis_representation_2025,
	title = {Multiple-Basis Representation of Quantum States},
	author = {{P{\'e}rez-Salinas}, Adri{\'a}n and Emonts, Patrick and Tura, Jordi and Dunjko, Vedran},
	year = 2025,
	doi = {10.48550/arXiv.2411.03110},
	url = {http://arxiv.org/abs/2411.03110},
	eprint = {2411.03110},
	primaryclass = {quant-ph},
	archiveprefix = {arXiv}
}

@article{durt_mutually_unbiased_2010,
  title = {On Mutually Unbiased Bases},
  author = {Durt, Thomas and Englert, Berthold-Georg and Bengtsson, Ingemar and {\.Z}yczkowski, Karol},
  year = 2010,
  journal = {International Journal of Quantum Information},
  volume = {08},
  pages = {535--640},
  issn = {0219-7499},
  doi = {10.1142/S0219749910006502},
  url = {https://www.worldscientific.com/doi/abs/10.1142/S0219749910006502}
}

@article{sancho_lorente_quantum_kernels_2022,
  title = {Quantum Kernels to Learn the Phases of Quantum Matter},
  author = {{Sancho-Lorente}, Teresa and {Rom{\'a}n-Roche}, Juan and Zueco, David},
  year = 2022,
  journal = {Physical Review A},
  volume = {105},
  pages = {042432},
  issn = {2469-9926, 2469-9934},
  doi = {10.1103/PhysRevA.105.042432},
  url = {https://link.aps.org/doi/10.1103/PhysRevA.105.042432},
  langid = {english}
}

@article{wu_quantum_phase_2023,
  title = {Quantum {{Phase Recognition}} via {{Quantum Kernel Methods}}},
  author = {Wu, Yusen and Wu, Bujiao and Wang, Jingbo and Yuan, Xiao},
  year = 2023,
  journal = {Quantum},
  volume = {7},
  eprint = {2111.07553},
  primaryclass = {quant-ph},
  pages = {981},
  issn = {2521-327X},
  doi = {10.22331/q-2023-04-17-981},
  url = {http://arxiv.org/abs/2111.07553},
  archiveprefix = {arXiv},
  langid = {english}
}

@misc{ali_finite_size_resource_2026,
  title = {Finite-Size Resource Scaling for Learning Quantum Phase Transitions with Fidelity-Based Support Vector Machines},
  author = {Ali, Aaqib and Scala, Giovanni and Lupo, Cosmo and Mandarino, Antonio},
  year = 2026,
  eprint = {2603.18211},
  primaryclass = {quant-ph},
  doi = {10.48550/arXiv.2603.18211},
  url = {http://arxiv.org/abs/2603.18211},
  archiveprefix = {arXiv}
}

@misc{khosrojerdi_learning_classify_2024,
  title = {Learning to {{Classify Quantum Phases}} of {{Matter}} with a {{Few Measurements}}},
  author = {Khosrojerdi, Mehran and Pereira, Jason L. and Cuccoli, Alessandro and Banchi, Leonardo},
  year = 2024,
  eprint = {2409.05188},
  primaryclass = {quant-ph},
  doi = {10.48550/arXiv.2409.05188},
  url = {http://arxiv.org/abs/2409.05188},
  archiveprefix = {arXiv}
}

@article{zanardi_ground_state_2006,
  title = {Ground State Overlap and Quantum Phase Transitions},
  author = {Zanardi, Paolo and Paunkovi{\'c}, Nikola},
  year = 2006,
  journal = {Physical Review E},
  volume = {74},
  pages = {031123},
  doi = {10.1103/PhysRevE.74.031123},
  url = {https://link.aps.org/doi/10.1103/PhysRevE.74.031123}
}

@inproceedings{damski_fidelity_approach_2015,
  title = {Fidelity Approach to Quantum Phase Transitions in Quantum {{Ising}} Model},
  booktitle = {Quantum {{Criticality}} in {{Condensed Matter}}},
  author = {Damski, Bogdan},
  year = 2015,
  eprint = {1509.03051},
  primaryclass = {quant-ph},
  pages = {159--182},
  doi = {10.1142/9789814704090_0006},
  url = {http://arxiv.org/abs/1509.03051},
  archiveprefix = {arXiv}
}

@article{gu_fidelity_approach_2010,
  title = {Fidelity Approach to Quantum Phase Transitions},
  author = {Gu, Shi-Jian},
  year = 2010,
  journal = {International Journal of Modern Physics B},
  volume = {24},
  eprint = {0811.3127},
  primaryclass = {quant-ph},
  pages = {4371--4458},
  issn = {0217-9792, 1793-6578},
  doi = {10.1142/S0217979210056335},
  url = {http://arxiv.org/abs/0811.3127},
  archiveprefix = {arXiv}
}

@article{rossini_ground_state_fidelity_2018,
  title = {Ground-State Fidelity at First-Order Quantum Transitions},
  author = {Rossini, Davide and Vicari, Ettore},
  year = 2018,
  journal = {Physical Review E},
  volume = {98},
  pages = {062137},
  issn = {2470-0045, 2470-0053},
  doi = {10.1103/PhysRevE.98.062137},
  url = {https://link.aps.org/doi/10.1103/PhysRevE.98.062137},
  langid = {english}
}

@book{scholkopf_learning_2001,
	title = {Learning with {Kernels}: {Support} {Vector} {Machines}, {Regularization}, {Optimization}, and {Beyond}},
	shorttitle = {Learning with {Kernels}},
	author = {Schölkopf, Bernhard and Smola, Alexander J.},
	year = {2001},
	month = dec,
	publisher = {The MIT Press},
	doi = {10.7551/mitpress/4175.001.0001},
	isbn = {978-0-262-25693-3},
	url = {https://direct.mit.edu/books/monograph/1821/Learning-with-KernelsSupport-Vector-Machines},
	urldate = {2025-03-31}
}

@misc{ghojogh_reproducing_2021,
	title = {Reproducing {Kernel} {Hilbert} {Space}, {Mercer}'s {Theorem}, {Eigenfunctions}, {Nyström} {Method}, and {Use} of {Kernels} in {Machine} {Learning}: {Tutorial} and {Survey}},
	shorttitle = {Reproducing {Kernel} {Hilbert} {Space}, {Mercer}'s {Theorem}, {Eigenfunctions}, {Nyström} {Method}, and {Use} of {Kernels} in {Machine} {Learning}},
	author = {Ghojogh, Benyamin and Ghodsi, Ali and Karray, Fakhri and Crowley, Mark},
	year = {2021},
	month = jun,
	publisher = {arXiv},
	doi = {10.48550/arXiv.2106.08443},
	url = {http://arxiv.org/abs/2106.08443},
	urldate = {2025-06-09},
	note = {arXiv:2106.08443 [stat]}
}

@inproceedings{boser_training_1992,
	title = {A training algorithm for optimal margin classifiers},
	author = {Boser, Bernhard E. and Guyon, Isabelle M. and Vapnik, Vladimir N.},
	year = {1992},
	month = jul,
	booktitle = {Proceedings of the fifth annual workshop on {Computational} learning theory},
	publisher = {Association for Computing Machinery},
	address = {New York, NY, USA},
	series = {{COLT} '92},
	pages = {144--152},
	doi = {10.1145/130385.130401},
	isbn = {978-0-89791-497-0},
	url = {https://dl.acm.org/doi/10.1145/130385.130401},
	urldate = {2026-02-23}
}

@article{cortes_algorithms_learning_2024,
  title = {Algorithms for {{Learning Kernels Based}} on {{Centered Alignment}}},
  author = {Cortes, Corinna and Mohri, Mehryar and Rostamizadeh, Afshin},
  year = 2012,
  journal = {Journal of Machine Learning Research},
  volume = {13},
  pages = {795--828},
  issn = {1533-7928},
  url = {http://jmlr.org/papers/v13/cortes12a.html}
}

@article{hofmann_kernel_methods_2008,
  title = {Kernel Methods in Machine Learning},
  author = {Hofmann, Thomas and Sch{\"o}lkopf, Bernhard and Smola, Alexander J.},
  year = 2008,
  journal = {The Annals of Statistics},
  volume = {36},
  pages = {1171-1220},
  eprint = {math/0701907},
  issn = {0090-5364},
  doi = {10.1214/009053607000000677},
  url = {http://arxiv.org/abs/math/0701907},
  archiveprefix = {arXiv}
}

@article{cervantes_comprehensive_survey_2020,
  title = {A Comprehensive Survey on Support Vector Machine Classification: {{Applications}}, Challenges and Trends},
  shorttitle = {A Comprehensive Survey on Support Vector Machine Classification},
  author = {Cervantes, Jair and {Garcia-Lamont}, Farid and {Rodr{\'i}guez-Mazahua}, Lisbeth and Lopez, Asdrubal},
  year = 2020,
  journal = {Neurocomputing},
  volume = {408},
  pages = {189--215},
  issn = {0925-2312},
  doi = {10.1016/j.neucom.2019.10.118},
  url = {https://www.sciencedirect.com/science/article/pii/S0925231220307153}
}

@techreport{hsu_practical_guide_,
  title = {A {{Practical Guide}} to {{Support Vector Classification}}},
  author = {Hsu, Chih-Wei and Chang, Chih-Chung and Lin, Chih-Jen},
  institution = {Department of Computer Science and Information Engineering,
                 National Taiwan University},
  url = {https://www.csie.ntu.edu.tw/~cjlin/papers/guide/guide.pdf},
  year = {2003}
}

@article{havlicek_supervised_2019,
	title = {Supervised learning with quantum-enhanced feature spaces},
	author = {Havlíček, Vojtěch and Córcoles, Antonio D. and Temme, Kristan and Harrow, Aram W. and Kandala, Abhinav and Chow, Jerry M. and Gambetta, Jay M.},
	year = {2019},
	month = mar,
	journal = {Nature},
	volume = {567},
	number = {7747},
	pages = {209--212},
	doi = {10.1038/s41586-019-0980-2},
	issn = {1476-4687},
	url = {https://www.nature.com/articles/s41586-019-0980-2},
	urldate = {2025-03-19},
	copyright = {2019 The Author(s), under exclusive licence to Springer Nature Limited},
	note = {Publisher: Nature Publishing Group}
}

@misc{schuld_supervised_2021,
	title = {Supervised quantum machine learning models are kernel methods},
	author = {Schuld, Maria},
	year = {2021},
	month = apr,
	publisher = {arXiv},
	doi = {10.48550/arXiv.2101.11020},
	url = {http://arxiv.org/abs/2101.11020},
	urldate = {2025-03-11},
	note = {arXiv:2101.11020 [quant-ph]}
}

@article{thanasilp_exponential_2024,
	title = {Exponential concentration in quantum kernel methods},
	author = {Thanasilp, Supanut and Wang, Samson and Cerezo, M. and Holmes, Zoë},
	year = {2024},
	month = jun,
	journal = {Nature Communications},
	publisher = {Nature Publishing Group},
	volume = {15},
	number = {1},
	pages = {5200},
	doi = {10.1038/s41467-024-49287-w},
	issn = {2041-1723},
	url = {https://www.nature.com/articles/s41467-024-49287-w},
	urldate = {2025-02-21},
	copyright = {2024 The Author(s)}
}

@article{buhrman_quantum_fingerprinting_2001,
  title = {Quantum Fingerprinting},
  author = {Buhrman, Harry and Cleve, Richard and Watrous, John and de Wolf, Ronald},
  year = 2001,
  journal = {Physical Review Letters},
  volume = {87},
  eprint = {quant-ph/0102001},
  pages = {167902},
  issn = {0031-9007, 1079-7114},
  doi = {10.1103/PhysRevLett.87.167902},
  url = {http://arxiv.org/abs/quant-ph/0102001},
  archiveprefix = {arXiv}
}

@article{ekert_direct_estimations_2002,
  title = {Direct Estimations of Linear and Non-Linear Functionals of a Quantum State},
  author = {Ekert, Artur K. and Alves, Carolina Moura and Oi, Daniel K. L. and Horodecki, Michal and Horodecki, Pawel and Kwek, L. C.},
  year = 2002,
  journal = {Physical Review Letters},
  volume = {88},
  eprint = {quant-ph/0203016},
  pages = {217901},
  issn = {0031-9007, 1079-7114},
  doi = {10.1103/PhysRevLett.88.217901},
  url = {http://arxiv.org/abs/quant-ph/0203016},
  archiveprefix = {arXiv}
}

@article{pfeuty_one_dimensional_ising_1970,
  title = {The One-Dimensional {{Ising}} Model with a Transverse Field},
  author = {Pfeuty, Pierre},
  year = 1970,
  journal = {Annals of Physics},
  volume = {57},
  pages = {79--90},
  issn = {0003-4916},
  doi = {10.1016/0003-4916(70)90270-8},
  url = {https://www.sciencedirect.com/science/article/pii/0003491670902708}
}

@article{bethe_zur_1931,
  title = {Zur {{Theorie}} der {{Metalle}}. {{I}}. {{Eigenwerte}} und {{Eigenfunktionen}} der linearen {{Atomkette}}},
  author = {Bethe, Hans},
  year = 1931,
  journal = {Zeitschrift f{\"u}r Physik},
  volume = {71},
  pages = {205--226},
  doi = {10.1007/BF01341708}
}

@article{yang_one_dimensional_chain_1966,
  title = {One-{{Dimensional Chain}} of {{Anisotropic Spin-Spin Interactions}}. {{I}}. {{Proof}} of {{Bethe}}'s {{Hypothesis}} for {{Ground State}} in a {{Finite System}}},
  author = {Yang, C. N. and Yang, C. P.},
  year = 1966,
  journal = {Physical Review},
  volume = {150},
  pages = {321--327},
  doi = {10.1103/PhysRev.150.321},
  url = {https://link.aps.org/doi/10.1103/PhysRev.150.321}
}

@book{takahashi_thermodynamics_1999,
  title = {Thermodynamics of One-Dimensional Solvable Models},
  author = {Takahashi, Minoru},
  year = 1999,
  publisher = {Cambridge University Press},
  doi = {10.1017/CBO9780511524332}
}

@article{berezinskii_destruction_1971,
  title = {Destruction of Long-Range Order in One-Dimensional and Two-Dimensional Systems Possessing a Continuous Symmetry Group. {{II}}. {{Quantum Systems}}},
  author = {Berezinskii, V. L.},
  year = 1972,
  journal = {Soviet Physics JETP},
  volume = {34},
  number = {3},
  pages = {610--616}
}

@article{kosterlitz_thouless_ordering_1973,
  title = {Ordering, Metastability and Phase Transitions in Two-Dimensional Systems},
  author = {Kosterlitz, J. M. and Thouless, D. J.},
  year = 1973,
  journal = {Journal of Physics C: Solid State Physics},
  volume = {6},
  number = {7},
  pages = {1181--1203},
  doi = {10.1088/0022-3719/6/7/010}
}

@article{kosterlitz_critical_1974,
  title = {The Critical Properties of the Two-Dimensional {XY} Model},
  author = {Kosterlitz, J. M.},
  year = 1974,
  journal = {Journal of Physics C: Solid State Physics},
  volume = {7},
  number = {6},
  pages = {1046--1060},
  doi = {10.1088/0022-3719/7/6/005}
}

@book{giamarchi_quantum_2004,
  title = {Quantum Physics in One Dimension},
  author = {Giamarchi, Thierry},
  year = 2004,
  publisher = {Oxford University Press},
  doi = {10.1093/acprof:oso/9780198525004.001.0001}
}

@article{misguich_inverse_participation_2016a,
  title = {Inverse Participation Ratios in the {{XXZ}} Spin Chain},
  author = {Misguich, Gr{\'e}goire and Pasquier, Vincent and Luck, Jean-Marc},
  year = 2016,
  journal = {Physical Review B},
  volume = {94},
  pages = {155110},
  issn = {2469-9950, 2469-9969},
  doi = {10.1103/PhysRevB.94.155110},
  url = {https://link.aps.org/doi/10.1103/PhysRevB.94.155110},
  copyright = {http://link.aps.org/licenses/aps-default-license},
  langid = {english}
}

@article{tsukerman_inverse_participation_2017,
  title = {Inverse {{Participation Ratios}} in the {{XX}} Spin Chain},
  author = {Tsukerman, Emmanuel},
  year = 2017,
  journal = {Physical Review B},
  volume = {95},
  eprint = {1611.01806},
  primaryclass = {cond-mat},
  pages = {115121},
  issn = {2469-9950, 2469-9969},
  doi = {10.1103/PhysRevB.95.115121},
  url = {http://arxiv.org/abs/1611.01806},
  archiveprefix = {arXiv}
}

@article{white_dmrg_1992,
  title = {Density Matrix Formulation for Quantum Renormalization Groups},
  author = {White, Steven R.},
  year = 1992,
  journal = {Physical Review Letters},
  volume = {69},
  number = {19},
  pages = {2863--2866},
  doi = {10.1103/PhysRevLett.69.2863}
}

@article{white_dmrg_1993,
  title = {Density-Matrix Algorithms for Quantum Renormalization Groups},
  author = {White, Steven R.},
  year = 1993,
  journal = {Physical Review B},
  volume = {48},
  number = {14},
  pages = {10345--10356},
  doi = {10.1103/PhysRevB.48.10345}
}

@article{schollwoeck_dmrg_2011,
  title = {The Density-Matrix Renormalization Group in the Age of Matrix Product States},
  author = {Schollwöck, Ulrich},
  year = 2011,
  journal = {Annals of Physics},
  volume = {326},
  number = {1},
  pages = {96--192},
  doi = {10.1016/j.aop.2010.09.012}
}

@article{ferris_perfect_sampling_2012,
  title = {Perfect Sampling with Unitary Tensor Networks},
  author = {Ferris, Andrew J. and Vidal, Guifre},
  journal = {Phys. Rev. B},
  volume = {85},
  pages = {165146},
  year = {2012},
  doi = {10.1103/PhysRevB.85.165146}
}

@article{carleo_netket_2019,
  title = {{{NetKet}}: {{A}} Machine Learning Toolkit for Many-Body Quantum Systems},
  shorttitle = {{{NetKet}}},
  author = {Carleo, Giuseppe and Choo, Kenny and Hofmann, Damian and Smith, James E. T. and Westerhout, Tom and Alet, Fabien and Davis, Emily J. and Efthymiou, Stavros and Glasser, Ivan and Lin, Sheng-Hsuan and Mauri, Marta and Mazzola, Guglielmo and Mendl, Christian B. and {van Nieuwenburg}, Evert and O'Reilly, Ossian and Th{\'e}veniaut, Hugo and Torlai, Giacomo and Vicentini, Filippo and Wietek, Alexander},
  year = 2019,
  journal = {SoftwareX},
  volume = {10},
  pages = {100311},
  issn = {2352-7110},
  doi = {10.1016/j.softx.2019.100311},
  url = {https://www.sciencedirect.com/science/article/pii/S2352711019300974}
}

@article{vicentini_netket3_2022,
  title = {{{NetKet}} 3: {{Machine Learning Toolbox}} for {{Many-Body Quantum Systems}}},
  shorttitle = {{{NetKet}} 3},
  author = {Vicentini, Filippo and Hofmann, Damian and Szab{\'o}, Attila and Wu, Dian and Roth, Christopher and Giuliani, Clemens and Pescia, Gabriel and Nys, Jannes and {Vargas-Calderon}, Vladimir and Astrakhantsev, Nikita and Carleo, Giuseppe},
  year = 2022,
  journal = {SciPost Physics Codebases},
  eprint = {2112.10526},
  primaryclass = {quant-ph},
  pages = {7},
  issn = {2949-804X},
  doi = {10.21468/SciPostPhysCodeb.7},
  url = {http://arxiv.org/abs/2112.10526},
  archiveprefix = {arXiv}
}

@article{fishman_itensor_2022,
  title = {The {{ITensor Software Library}} for {{Tensor Network Calculations}}},
  author = {Fishman, Matthew and White, Steven and Stoudenmire, E. Miles},
  year = 2022,
  journal = {SciPost Physics Codebases},
  pages = {4},
  issn = {2949-804X},
  doi = {10.21468/SciPostPhysCodeb.4},
  url = {https://scipost.org/10.21468/SciPostPhysCodeb.4},
  langid = {english}
}

@article{alhambra_quantum_many_body_2023,
  title = {Quantum Many-Body Systems in Thermal Equilibrium},
  author = {Alhambra, {\'A}lvaro M.},
  year = 2023,
  journal = {PRX Quantum},
  volume = {4},
  eprint = {2204.08349},
  primaryclass = {quant-ph},
  pages = {040201},
  issn = {2691-3399},
  doi = {10.1103/PRXQuantum.4.040201},
  url = {http://arxiv.org/abs/2204.08349},
  archiveprefix = {arXiv}
}

@article{sachdev_universal_1996,
	title = {Universal, finite temperature, crossover functions of the quantum transition in the {Ising} chain in a transverse field},
	author = {Sachdev, Subir},
	year = {1996},
	month = apr,
	journal = {Nuclear Physics B},
	volume = {464},
	number = {3},
	pages = {576--595},
	doi = {10.1016/0550-3213(95)00657-5},
	issn = {05503213},
	url = {http://arxiv.org/abs/cond-mat/9509147},
	urldate = {2026-05-18},
	note = {arXiv:cond-mat/9509147}
}

@article{zanardi_mixed_state_fidelity_2007,
  title = {Mixed-State Fidelity and Quantum Criticality at Finite Temperature},
  author = {Zanardi, Paolo and Quan, H. T. and Wang, Xiaoguang and Sun, C. P.},
  year = 2007,
  journal = {Physical Review A},
  volume = {75},
  pages = {032109},
  issn = {1050-2947, 1094-1622},
  doi = {10.1103/PhysRevA.75.032109},
  url = {https://link.aps.org/doi/10.1103/PhysRevA.75.032109},
  copyright = {http://link.aps.org/licenses/aps-default-license},
  langid = {english}
}

@article{bialonczyk_uhlmann_fidelity_2021,
  title = {Uhlmann {{Fidelity}} and {{Fidelity Susceptibility}} for {{Integrable Spin Chains}} at {{Finite Temperature}}: {{Exact Results}}},
  shorttitle = {Uhlmann {{Fidelity}} and {{Fidelity Susceptibility}} for {{Integrable Spin Chains}} at {{Finite Temperature}}},
  author = {Bia{\l}o{\'n}czyk, Micha{\l} and {G{\'o}mez-Ruiz}, Fernando Javier and del Campo, Adolfo},
  year = 2021,
  journal = {New Journal of Physics},
  volume = {23},
  eprint = {2105.05055},
  primaryclass = {quant-ph},
  pages = {093033},
  issn = {1367-2630},
  doi = {10.1088/1367-2630/ac23f0},
  url = {http://arxiv.org/abs/2105.05055},
  archiveprefix = {arXiv}
}

@article{frerot_reconstructing_2019,
	title = {Reconstructing the quantum critical fan of strongly correlated systems using quantum correlations},
	author = {Frérot, Irénée and Roscilde, Tommaso},
	year = {2019},
	month = feb,
	journal = {Nature Communications},
	volume = {10},
	number = {1},
	pages = {577},
	doi = {10.1038/s41467-019-08324-9},
	issn = {2041-1723},
	url = {http://arxiv.org/abs/1805.03140},
	urldate = {2026-06-10},
	note = {arXiv:1805.03140 [cond-mat.str-el]}
}

@misc{zhang_critical_2019,
	title = {Critical behaviors of the entanglement and participation entropy near the many-body localization transition in a disordered quantum spin chain},
	author = {Zhang, Wei and Wang, Ziqiang},
	year = {2019},
	month = aug,
	publisher = {arXiv},
	doi = {10.48550/arXiv.1908.04039},
	url = {http://arxiv.org/abs/1908.04039},
	urldate = {2026-06-10},
	note = {arXiv:1908.04039 [cond-mat.str-el]}
}

@article{wu_crossovers_critical_2018,
  title = {Crossovers and Critical Scaling in the One-Dimensional Transverse-Field {{Ising}} Model},
  author = {Wu, Jianda and Zhu, Lijun and Si, Qimiao},
  year = 2018,
  journal = {Physical Review B},
  volume = {97},
  pages = {245127},
  doi = {10.1103/PhysRevB.97.245127},
  url = {https://link.aps.org/doi/10.1103/PhysRevB.97.245127}
}

@misc{srimahajariyapong_connecting_phases_2025,
  title = {Connecting Phases of Matter to the Flatness of the Loss Landscape in Analog Variational Quantum Algorithms},
  author = {Srimahajariyapong, Kasidit and Thanasilp, Supanut and Chotibut, Thiparat},
  year = 2025,
  eprint = {2506.13865},
  primaryclass = {quant-ph},
  doi = {10.48550/arXiv.2506.13865},
  url = {http://arxiv.org/abs/2506.13865},
  archiveprefix = {arXiv}
}

@article{cao_exploiting_many_body_2025,
  title = {Exploiting Many-Body Localization for Scalable Variational Quantum Simulation},
  author = {Cao, Chenfeng and Zhou, Yeqing and Tannu, Swamit and Shannon, Nic and Joynt, Robert},
  year = 2025,
  journal = {Quantum},
  volume = {9},
  eprint = {2404.17560},
  primaryclass = {quant-ph},
  pages = {1942},
  issn = {2521-327X},
  doi = {10.22331/q-2025-12-12-1942},
  url = {http://arxiv.org/abs/2404.17560},
  archiveprefix = {arXiv},
  langid = {english}
}

@article{cao_unveiling_quantum_2025,
  title = {Unveiling Quantum Phase Transitions from Traps in Variational Quantum Algorithms},
  author = {Cao, Chenfeng and Gambetta, Filippo Maria and Montanaro, Ashley and Santos, Raul A.},
  year = 2025,
  journal = {npj Quantum Information},
  volume = {11},
  pages = {93},
  issn = {2056-6387},
  doi = {10.1038/s41534-025-01038-5},
  url = {https://www.nature.com/articles/s41534-025-01038-5},
  copyright = {2025 The Author(s)},
  langid = {english}
}

@article{park_hamiltonian_variational_2024,
  title = {Hamiltonian Variational Ansatz without Barren Plateaus},
  author = {Park, Chae-Yeun and Killoran, Nathan},
  year = 2024,
  journal = {Quantum},
  volume = {8},
  pages = {1239},
  doi = {10.22331/q-2024-02-01-1239},
  url = {https://quantum-journal.org/papers/q-2024-02-01-1239/},
  langid = {british}
}

@article{kramers_statistics_1941,
  title = {Statistics of the {{Two-Dimensional Ferromagnet}}. {{Part I}}},
  author = {Kramers, H. A. and Wannier, G. H.},
  year = 1941,
  journal = {Physical Review},
  volume = {60},
  pages = {252--262},
  doi = {10.1103/PhysRev.60.252}
}
% --------------------------------------------

% --- End of Main Text ---
\appendix
\onecolumngrid

\crefalias{section}{appendix}
\crefalias{subsection}{appendix}
\crefalias{subsubsection}{appendix}

\section{Data generation}
\label{app:data_generation}

We generate four datasets: ground states of the TFIM, ground states of the XXZ chain without and with a longitudinal field, and Gibbs states of the TFIM. For each we compute multi-basis IPR features and, for the Gibbs states, also the HS kernel. \Cref{tab:datasets} summarizes the system sizes, parameter grids, and numerical methods. All Hamiltonians are those of \cref{eq:tfim_hamiltonian,eq:xxz_hamiltonian} with $J=1$ and periodic boundary conditions, $Z_{L+1}\equiv Z_1$, unless stated otherwise.

\begin{table}[htb]
    \centering
    \footnotesize
    \caption{Overview of the generated datasets. All datasets use periodic boundary conditions. The last column lists the subset that enters the SVM experiments after the post-processing of \cref{app:postprocessing}. Grids include both end points. MC denotes Monte Carlo sampling.}
    \label{tab:datasets}
    \begin{tabular}{@{}llllll@{}}
        \toprule
        Dataset & Method & $L$ & Grid (step) & IPRs & Classification subset \\
        \midrule
        TFIM ground states & DMRG & $3,4,\dots,100$ & $h\in[0,4]$ (0.2) & MC, $N_{\mathrm{MC}}=10^4$ & $L\geq10$ in steps of 10; 4000 pts \\
        XXZ ground states & ED & $6,8,\dots,24$ & $\Delta\in[0,5]$ (0.1) & exact & all $L$; 5000 pts \\
        XXZ in a field & ED & $14$ & $\Delta,h\in[-5,5]$ (0.2) & exact & not used \\
        TFIM Gibbs, IPR & TEBD & $10,20,\dots,100$ & $h\in[0,4]$, $\beta\in[0,6]$ (0.1) & MC, $N_{\mathrm{MC}}=2000$ & $L=20,50,100$; $\beta\in[0.1,2]$; 4000 pts \\
        TFIM Gibbs, HS & TEBD & $10,20,\dots,100$ & $h\in[0,4]$, $\beta\in[0,2]$ (0.1) & -- & $L=20,50,100$; $\beta\in[0.1,2]$; 4000 pts \\
        \bottomrule
    \end{tabular}
\end{table}

\subsection{TFIM ground states}\label{app:data_tfim}

We compute ground states $\ket{\psi_{\mathrm{GS}}(h)}$ of \cref{eq:tfim_hamiltonian} as MPS with two-site DMRG in \texttt{ITensors.jl}~\cite{fishman_itensor_2022}. The periodic bond between sites $L$ and $1$ is included as a long-range term in the MPO. Each run starts from a random MPS and uses 10 sweeps with maximal bond dimension $\chi_{\max}=400$, truncation cutoff $10^{-10}$, and noise $10^{-8}$.

Enumerating all $2^L$ amplitudes is infeasible at these sizes, so we estimate the IPR (\cref{eq:ipr}) by Monte Carlo sampling. Writing $p^{(s)}(x)=|\langle x|U_s^\dagger|\psi\rangle|^2$, we have
\begin{equation}
    I^{(s)}(\psi) = \sum_{x} p^{(s)}(x)^2 = \mathbb{E}_{x\sim p^{(s)}}\big[p^{(s)}(x)\big] \approx \frac{1}{N_{\mathrm{MC}}}\sum_{n=1}^{N_{\mathrm{MC}}} p^{(s)}(x_n),
    \label{eq:ipr_mc}
\end{equation}
with $x_n$ drawn by perfect sampling from the MPS~\cite{ferris_perfect_sampling_2012}. We use the $Z$ basis, $U_Z=\mathbb{1}$, and the $X$ basis, $U_X=\mathrm{H}^{\otimes L}$, which we apply to the MPS before sampling.

\subsection{XXZ ground states}\label{app:data_xxz}

For the XXZ chain we use Lanczos exact diagonalization (ED) in \texttt{NetKet}~\cite{vicentini_netket3_2022}, which gives the full state vector for $L\leq24$. The IPRs are then evaluated exactly, without sampling. We use three bases.

\emph{$Z$ and $X$ bases.} We diagonalize \cref{eq:xxz_hamiltonian} in the full $2^L$-dimensional Hilbert space. For $\Delta\geq0$ and even $L$ the ground state is unique and lies in the $S^z=0$ sector. The $X$-basis IPR follows from applying $\mathrm{H}^{\otimes L}$ to this state.

\emph{$XY$ basis.} With the Jordan-Wigner transformation $c_i=\bigl(\prod_{j<i}Z_j\bigr)\tfrac{1}{2}(X_i-\mathrm{i}Y_i)$, for which $n_i=c_i^\dagger c_i=\tfrac{1}{2}(1+Z_i)$, and the Fourier transform $c_k=L^{-1/2}\sum_j e^{\mathrm{i}kj}c_j$, the XXZ chain in a longitudinal field $h\sum_iZ_i$ becomes
\begin{equation}
    H_{\mathrm{XXZ}}^{(\mathrm{F})}
    =\sum_k\left(-4\cos k-4\Delta+2h\right)n_k
    +\frac{4\Delta}{L}\sum_{p,k,q}e^{-\mathrm{i}q}\,c_{p+q}^\dagger c_p\,c_k^\dagger c_{k+q}
    +(\Delta-h)L .
    \label{eq:xxz_fermions}
\end{equation}
The first term is the free-fermion dispersion of $H_{XY}$, so the momentum occupation states $\ket{\{n_k\}}$ diagonalize $H_{\mathrm{XXZ}}$ at $\Delta=h=0$. They define the $XY$ basis. The $ZZ$ coupling adds the momentum-exchange term. On a ring, the Jordan-Wigner string makes the fermions antiperiodic for an even fermion number $N_F$ and periodic for odd $N_F$, so $k\in\frac{2\pi}{L}(\mathbb{Z}+\frac12)$ or $k\in\frac{2\pi}{L}\mathbb{Z}$, respectively. We diagonalize \cref{eq:xxz_fermions} in the sector of fixed $N_F$ that contains the $Z$-basis ground state. At $h=0$ this is half filling, $N_F=L/2$, which corresponds to $S^z=0$. The resulting ground-state vector directly gives the amplitudes in the $XY$ basis. We confirm that both diagonalizations give the same ground-state energy.

\emph{Longitudinal field.} For the phase diagram in \cref{fig:xxz_field} we repeat both diagonalizations at $L=14$ on a $51\times51$ grid in $(\Delta,h)$. The fermion number is read off from the magnetization of the $Z$-basis ground state. In the fully polarized sectors, $N_F\in\{0,L\}$, and on the degenerate ferromagnetic line $h=0$, $\Delta<-1$, the two bases coincide, and we reuse the $Z$-basis state.

\emph{Analytical phase boundaries.} We compare with the Bethe ansatz phase boundaries of the XXZ chain in a field~\cite{bethe_zur_1931,yang_one_dimensional_chain_1966}. In our Pauli-operator convention the boundary between the critical $XY$ phase and the fully polarized phase is
\begin{equation}
    h_{c2} = 2(1 + \Delta),
    \label{eq:hc2}
\end{equation}
and, for $\Delta>1$, the boundary between the antiferromagnetic and $XY$ phases is
\begin{equation}
    h_{c1} = 2\sinh\eta \sum_{k=-\infty}^{\infty} \frac{(-1)^k}{\cosh(k\eta)}, \qquad \eta = \operatorname{arccosh}\Delta.
    \label{eq:hc1}
\end{equation}

\subsection{TFIM Gibbs states}\label{app:data_thermal}

We represent the Gibbs states $\rho_\beta(h)=e^{-\beta H_{\mathrm{TFIM}}(h)}/\Tr e^{-\beta H_{\mathrm{TFIM}}(h)}$ as MPOs. Starting from the identity at $\beta=0$, we apply second-order Trotterized imaginary-time evolution to both sides, $\rho\mapsto e^{-\delta\tau H/2}\rho\, e^{-\delta\tau H/2}$, with step $\delta\tau=0.01$, and normalize the result. Each step increases $\beta$ by $\delta\tau$, so a single run per $h$ passes through all $\beta$ values on the grid.

\emph{IPR features.} For the IPR dataset we use $\chi_{\max}=400$ and truncation cutoff $10^{-8}$. The IPR of a mixed state in basis $s$ follows by replacing $p^{(s)}(x)$ in \cref{eq:ipr_mc} with the diagonal $p^{(s)}(x)=\bra{x}U_s^\dagger\rho_\beta(h)U_s\ket{x}$. We sample this diagonal exactly, site by site from its conditional marginals, which we obtain by contracting the MPO with partial traces. For the $X$ basis we first apply $\mathrm{H}^{\otimes L}$ to both sides of the MPO.

\emph{HS kernel.} For the HS dataset we use $\chi_{\max}=200$ and truncation cutoff $10^{-12}$ during the evolution. On the grid $\beta\leq2$ the bond dimension reaches at most 143, so only the cutoff truncates. We then evaluate
\begin{equation}
    \tilde{K}_{\mathrm{HS}}\big(\rho_\beta(h_i),\rho_\beta(h_j)\big) =
    \frac{\Tr\big(\rho_\beta(h_i)\,\rho_{\beta}(h_j)\big)}
    {\sqrt{\Tr\big(\rho_\beta(h_i)^2\big)\,\Tr\big(\rho_{\beta}(h_j)^2\big)}}
\end{equation}
as the exact Hilbert-Schmidt inner product of the two MPOs, without further truncation. The normalization of the evolved MPOs drops out of this ratio. At $\beta=0$ all states coincide and both kernels are trivial, so the analysis starts at $\beta=0.1$.

\subsection{Post-processing for classification}\label{app:postprocessing}

The SVM experiments use densely resampled data, obtained in three steps.
\begin{enumerate}[(i)]
    \item \emph{IPR curves.} For each $L$ and basis we fit a cubic smoothing spline (\texttt{make\_smoothing\_spline} in \texttt{scipy.interpolate}, smoothing strength chosen by generalized cross-validation) to $\log_{10}I^{(s)}$ as a function of the control parameter. The fit is done on the logarithm because the IPRs span many orders of magnitude.
    \item \emph{Resampling.} We evaluate the spline on a uniform grid and exponentiate back, so the features are raw IPRs. The grids contain 4000 points in $h\in[0,4]$ for the TFIM, 5000 points in $\Delta\in[0,5]$ for the XXZ chain, and 4000 points in $h\in[0,4]$ for each $\beta\in\{0.1,\dots,2.0\}$ of the thermal IPR dataset with $L\in\{20,50,100\}$.
    \item \emph{Gram matrices.} From the resampled features we build the normalized IPR kernel of \cref{eq:ipr_kernel_norm}. For the thermal HS kernel we interpolate the raw $41\times41$ Gram matrix bilinearly (\texttt{RegularGridInterpolator}) onto the same $4000\times4000$ grid and then normalize and symmetrize it again.
\end{enumerate}
The Gram matrices in \cref{fig:tfim_results,fig:xxz_results,fig:thermal_results,fig:tfim_grams,fig:xxz_xz_kernel} use the resampled data, those in \cref{fig:thermal_l_sweep} the raw data. Because training and test points are drawn from the same interpolated curves, the balanced accuracies measure how well the kernel separates the two phases along these curves rather than generalization to independently generated states. The finite-shot noise model of \cref{app:noise} is applied on top of the resampled features.

\section{Supplementary results}\label{app:supplementary}

In this appendix we provide further results for the examples in \cref{sec:results}.

\subsection{Support vector classification}\label{app:svm}

The kernel-based learning algorithm we use is support vector classification (SVC). Given a training set $\{(x_i,y_i)\}_{i=1}^{N}$ with binary labels $y_i\in\{-1,+1\}$, SVC learns a maximum-margin classifier using only kernel evaluations between training points. These evaluations define the Gram matrix
$[K]_{ij}=K(x_i,x_j)$ with which the following dual soft-margin optimization problem is solved:
\begin{equation}
    \underset{\boldsymbol{\alpha}\in\mathbb{R}^N}{\mathrm{maximize}}
    \quad
    W(\boldsymbol{\alpha})
    =
    \sum_{i=1}^{N}\alpha_i
    -
    \frac{1}{2}
    \sum_{i,j=1}^{N}
    \alpha_i\alpha_j y_i y_j K_{ij},
\end{equation}
subject to
\begin{equation}
    0 \leq \alpha_i \leq C,
    \quad \text{and} \quad
    \sum_{i=1}^{N}\alpha_i y_i = 0 .
\end{equation}
Here, the coefficients $\alpha_i$ are the dual variables, and $C$ is the
regularization parameter controlling the trade-off between margin size and
training-set violations. Training points with nonzero $\alpha_i$ are called support
vectors and determine the learned classifier.

For a new test point $x$, the corresponding decision function is
\begin{equation} 
    \label{eq:decision_function}
    f(x)
    =
    \sum_{i\in \mathrm{SV}}
    \alpha_i y_i K(x_i,x)
    + b ,
\end{equation}
where $b$ is the learned offset and the sum runs over the support vectors. The
predicted class is then determined by the sign of the decision function,
\begin{equation}
    \hat{y}=\mathrm{sign}\bigl(f(x)\bigr).
\end{equation}
Thus, points with $f(x)>0$ are assigned to one class, while points with $f(x)<0$
are assigned to the other. In our case, the learned decision crossing $f(x)=0$ will hence
be interpreted as the boundary between the two phases learned by the SVC.

\begin{figure}[t]
    \centering
    \includegraphics{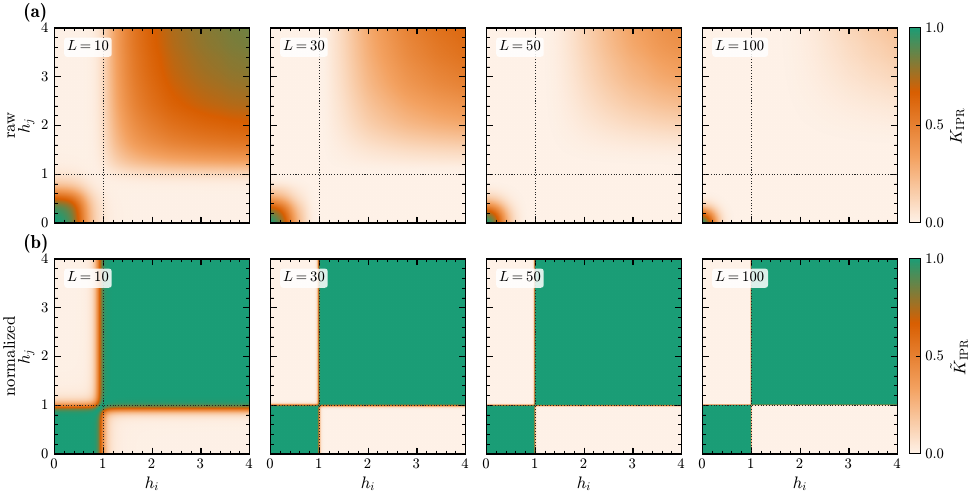}
    \caption{TFIM Gram matrices of the (a) raw and (b) normalized IPR kernel for $L=10$, $30$, $50$, and $100$, with $h_c=1$ marked by dotted lines.}
    \label{fig:tfim_grams}
\end{figure}

\subsection{TFIM ground states}\label{app:tfim}

The SVM performance for the normalized IPR kernel can be found in \cref{fig:tfim_results}(c) and (d) with precise values given in \cref{tab:svm_tfim}. In general the data is slightly noisy due to the Monte Carlo approach we used to obtain it, thus the choice of interpolation function is also an important one which can lead to varying crossings of the IPR curves yielding small differences in classification performances. 

\subsubsection{Raw versus normalized kernel}\label{app:raw_kernel}

In \cref{fig:tfim_results}(a) we provided the results for the IPR curves of system sizes $L=3,\dots,100$. The exponential decay in the IPRs of the opposite basis is clearly visible but it is important to note that resolving the crossing of the IPR curves requires values down to approximately $10^{-10}$ for $L=100$. Consider our raw kernel in the TFIM case:
\begin{equation}
    K_{\mathrm{IPR}}(h_i,h_j)
    =
    I^{(Z)}(h_i) I^{(Z)}(h_j)
    +
    I^{(X)}(h_i) I^{(X)}(h_j).
\end{equation}
The smallest values in the Gram matrix will be those between states of opposite phase regimes, e.g., in the ideal limiting case, we have
\begin{equation}
\begin{aligned}
    K_{\mathrm{IPR}}(0,h \to \infty)
    =
    I^{(Z)}(0) I^{(Z)}(h \to \infty)
    +
    I^{(X)}(0) I^{(X)}(h \to \infty)
    =
    1\cdot 2^{-L}
    +
    2^{-L}\cdot 1
    =
    2^{-(L-1)}.
\end{aligned}
\end{equation}
Assuming the IPR at the crossing is $I_{\mathrm{c}}$, the diagonal raw kernel value becomes
\begin{equation}
\begin{aligned}
    K_{\mathrm{IPR}}(h_c,h_c)
    =
    \left(I^{(Z)}(h_c)\right)^2
    +
    \left(I^{(X)}(h_c)\right)^2 
    =
    I_{\mathrm{c}}^2 + I_{\mathrm{c}}^2
    =
    2 I_{\mathrm{c}}^2 .
\end{aligned}
\end{equation}
Thus, while same-phase blocks correspond to the highest and opposite-phase blocks to the lowest values, the raw diagonal entries near the transition can also become small. This distinction disappears in the normalized IPR kernel, whose diagonal entries satisfy
\begin{equation}
    \tilde{K}_{\mathrm{IPR}}(h,h)=1
\end{equation}
by construction. Therefore, the normalized kernel emphasizes the relative direction of the IPR feature vectors, whereas the raw kernel also retains information about their absolute magnitude; however, its accurate experimental estimation still requires resolving the small raw kernel values entering the normalization.

To illustrate this fact we compare the Gram matrices of the raw and normalized IPR kernels in \cref{fig:tfim_grams}. The phase transition becomes sharper with increasing system size $L$; however, when looking at the raw IPR values, the points close to the phase transition become too small to resolve with realistic finite precision.

\begin{table}[t]
    \centering
    \begin{tabular}{|r|l|l|}
        \hline
        $L$ & Global BA & Local BA \\
        \hline
        \hline
        10  & $0.9978 \pm 0.0024$ & $0.9681 \pm 0.0141$ \\
        \hline
        20  & $0.9991 \pm 0.0016$ & $0.9902 \pm 0.0062$ \\
        \hline
        30  & $0.9991 \pm 0.0016$ & $0.9907 \pm 0.0059$ \\
        \hline
        40  & $0.9992 \pm 0.0016$ & $0.9908 \pm 0.0055$ \\
        \hline
        50  & $0.9996 \pm 0.0012$ & $0.9959 \pm 0.0032$ \\
        \hline
        60  & $0.9995 \pm 0.0013$ & $0.9952 \pm 0.0034$ \\
        \hline
        70  & $0.9996 \pm 0.0010$ & $0.9969 \pm 0.0027$ \\
        \hline
        80  & $0.9995 \pm 0.0012$ & $0.9943 \pm 0.0040$ \\
        \hline
        90  & $0.9996 \pm 0.0009$ & $0.9972 \pm 0.0026$ \\
        \hline
        100 & $0.9994 \pm 0.0013$ & $0.9938 \pm 0.0034$ \\
        \hline
    \end{tabular}
    \caption{Balanced accuracy scores including standard deviations over 100 splits for the normalized TFIM IPR kernel with local region $\delta=0.25$.}
    \label{tab:svm_tfim}
\end{table}

\subsubsection{Finite-shot noise model}\label{app:noise}

In order to test SVM performance under finite measurement shot samples $N_s$, we perturb the exact IPR features using additive Gaussian noise with standard deviation $1/\sqrt{N_s}$. For each transverse-field value $h_j$ and basis $s \in \{X,Z\}$, we define
\begin{equation}
    \hat{I}^{(s)}(h_j)
    =
    I^{(s)}(h_j)
    +
    \varepsilon^{(s)}_{j},
    \qquad
    \varepsilon^{(s)}_{j}
    \sim
    \mathcal{N}\!\left(0, \frac{1}{N_s}\right).
\end{equation}

To ensure numerical stability and preserve the physical range of the IPR, we clip the noisy estimates to the interval $[1/N_s, 1]$. The lower bound $1/N_s$ corresponds to the smallest nonzero scale resolvable with $N_s$ samples.

The noisy feature vectors
\begin{equation}
    \hat{\mathbf{I}}(h_j)
    =
    \big(
    \hat{I}^{(X)}(h_j),
    \hat{I}^{(Z)}(h_j)
    \big)
\end{equation}
are then used to construct the Gram matrix
\begin{equation}
    \hat{K}_{ij}
    =
    \hat{\mathbf{I}}(h_i)
    \cdot
    \hat{\mathbf{I}}(h_j),
\end{equation}
with normalization performed using the same noisy features. In the following we will work with the normalized kernel.

To assess the impact of finite-shot noise, we repeat the noisy dataset generation for multiple independent noise realizations. For a fixed shot number $N_s$, we draw $R$ independent noise seeds and construct $R$ noisy datasets according to the Gaussian model described above. For each realization, we evaluate the SVM using the same procedure as before (reduced to 10 train-test splits), yielding a split-averaged balanced accuracy $\mu_r(L)$ for each system size $L$.

The reported performance is obtained by averaging over noise realizations,
\begin{equation}
    \mu(L)
    =
    \frac{1}{R}
    \sum_{r=1}^{R}
    \mu_r(L),
\end{equation}
with uncertainty estimated as the sample standard deviation across noise realizations,
\begin{equation}
    \sigma(L)
    =
    \sqrt{
        \frac{1}{R-1}
        \sum_{r=1}^{R}
        \big(
        \mu_r(L) - \mu(L)
        \big)^2
    }.
\end{equation}

We set $R=10$ and show the results in \cref{fig:tfim_results}(c) and (d) with precise values found in \cref{tab:svm_tfim_noisy}. Each data point represents balanced accuracy averaged over 10 train-test splits and 10 independent finite-shot noise realizations. As expected, both global and local performance degrade as the shot budget $N_s$ decreases. Notably, the local classification performance approaches the random baseline more rapidly, reflecting its higher sensitivity to measurement noise.

\begin{table*}[t]
    \centering
    \small
    \setlength{\tabcolsep}{4pt}
    \renewcommand{\arraystretch}{1.15}

    \begin{tabular}{|c||c||c|}
        \hline
        \multicolumn{1}{|c||}{$N_s=200$} &
        \multicolumn{1}{c||}{$N_s=500$} &
        \multicolumn{1}{c|}{$N_s=1000$} \\
        \hline\hline

        \begin{tabular}{r|l|l}
            $L$ & Global BA & Local BA \\
            \hline\hline
            10  & $0.9766 \pm 0.0039$ & $0.8430 \pm 0.0122$ \\
            20  & $0.9589 \pm 0.0040$ & $0.7107 \pm 0.0225$ \\
            30  & $0.9279 \pm 0.0024$ & $0.5737 \pm 0.0141$ \\
            40  & $0.8949 \pm 0.0068$ & $0.5174 \pm 0.0188$ \\
            50  & $0.8684 \pm 0.0080$ & $0.5048 \pm 0.0207$ \\
            60  & $0.8449 \pm 0.0107$ & $0.5091 \pm 0.0361$ \\
            70  & $0.8226 \pm 0.0086$ & $0.4984 \pm 0.0250$ \\
            80  & $0.8001 \pm 0.0088$ & $0.4994 \pm 0.0241$ \\
            90  & $0.7902 \pm 0.0098$ & $0.5043 \pm 0.0167$ \\
            100 & $0.7648 \pm 0.0116$ & $0.5012 \pm 0.0224$
        \end{tabular}
        &
        \begin{tabular}{r|l|l}
            $L$ & Global BA & Local BA \\
            \hline\hline
            10  & $0.9858 \pm 0.0020$ & $0.8924 \pm 0.0091$ \\
            20  & $0.9702 \pm 0.0039$ & $0.7834 \pm 0.0185$ \\
            30  & $0.9441 \pm 0.0038$ & $0.6172 \pm 0.0147$ \\
            40  & $0.9126 \pm 0.0052$ & $0.5319 \pm 0.0193$ \\
            50  & $0.8886 \pm 0.0057$ & $0.5108 \pm 0.0225$ \\
            60  & $0.8647 \pm 0.0101$ & $0.5091 \pm 0.0305$ \\
            70  & $0.8444 \pm 0.0077$ & $0.4988 \pm 0.0287$ \\
            80  & $0.8269 \pm 0.0089$ & $0.4949 \pm 0.0230$ \\
            90  & $0.8155 \pm 0.0068$ & $0.5018 \pm 0.0103$ \\
            100 & $0.7951 \pm 0.0069$ & $0.5000 \pm 0.0203$
        \end{tabular}
        &
        \begin{tabular}{r|l|l}
            $L$ & Global BA & Local BA \\
            \hline\hline
            10  & $0.9903 \pm 0.0018$ & $0.9197 \pm 0.0076$ \\
            20  & $0.9775 \pm 0.0031$ & $0.8334 \pm 0.0141$ \\
            30  & $0.9525 \pm 0.0046$ & $0.6594 \pm 0.0163$ \\
            40  & $0.9241 \pm 0.0052$ & $0.5491 \pm 0.0175$ \\
            50  & $0.9004 \pm 0.0065$ & $0.5189 \pm 0.0222$ \\
            60  & $0.8784 \pm 0.0078$ & $0.5114 \pm 0.0298$ \\
            70  & $0.8595 \pm 0.0081$ & $0.4987 \pm 0.0304$ \\
            80  & $0.8420 \pm 0.0067$ & $0.4958 \pm 0.0215$ \\
            90  & $0.8311 \pm 0.0061$ & $0.5047 \pm 0.0086$ \\
            100 & $0.8108 \pm 0.0065$ & $0.5001 \pm 0.0216$
        \end{tabular}
        \\
        \hline\hline

        \multicolumn{1}{|c||}{$N_s=2000$} &
        \multicolumn{1}{c||}{$N_s=5000$} &
        \multicolumn{1}{c|}{$N_s=10000$} \\
        \hline\hline

        \begin{tabular}{r|l|l}
            $L$ & Global BA & Local BA \\
            \hline\hline
            10  & $0.9935 \pm 0.0014$ & $0.9395 \pm 0.0085$ \\
            20  & $0.9836 \pm 0.0023$ & $0.8760 \pm 0.0113$ \\
            30  & $0.9607 \pm 0.0039$ & $0.7011 \pm 0.0137$ \\
            40  & $0.9345 \pm 0.0059$ & $0.5738 \pm 0.0201$ \\
            50  & $0.9121 \pm 0.0046$ & $0.5258 \pm 0.0211$ \\
            60  & $0.8909 \pm 0.0071$ & $0.5147 \pm 0.0296$ \\
            70  & $0.8724 \pm 0.0086$ & $0.4993 \pm 0.0299$ \\
            80  & $0.8557 \pm 0.0062$ & $0.4972 \pm 0.0217$ \\
            90  & $0.8441 \pm 0.0070$ & $0.5063 \pm 0.0106$ \\
            100 & $0.8243 \pm 0.0064$ & $0.5016 \pm 0.0253$
        \end{tabular}
        &
        \begin{tabular}{r|l|l}
            $L$ & Global BA & Local BA \\
            \hline\hline
            10  & $0.9959 \pm 0.0010$ & $0.9552 \pm 0.0061$ \\
            20  & $0.9885 \pm 0.0021$ & $0.9142 \pm 0.0128$ \\
            30  & $0.9689 \pm 0.0031$ & $0.7548 \pm 0.0130$ \\
            40  & $0.9458 \pm 0.0049$ & $0.6158 \pm 0.0207$ \\
            50  & $0.9247 \pm 0.0027$ & $0.5422 \pm 0.0204$ \\
            60  & $0.9043 \pm 0.0072$ & $0.5208 \pm 0.0270$ \\
            70  & $0.8875 \pm 0.0076$ & $0.5036 \pm 0.0306$ \\
            80  & $0.8721 \pm 0.0077$ & $0.4970 \pm 0.0223$ \\
            90  & $0.8594 \pm 0.0072$ & $0.5067 \pm 0.0110$ \\
            100 & $0.8408 \pm 0.0046$ & $0.5033 \pm 0.0255$
        \end{tabular}
        &
        \begin{tabular}{r|l|l}
            $L$ & Global BA & Local BA \\
            \hline\hline
            10  & $0.9969 \pm 0.0007$ & $0.9627 \pm 0.0053$ \\
            20  & $0.9913 \pm 0.0019$ & $0.9327 \pm 0.0112$ \\
            30  & $0.9744 \pm 0.0033$ & $0.7944 \pm 0.0091$ \\
            40  & $0.9541 \pm 0.0045$ & $0.6582 \pm 0.0189$ \\
            50  & $0.9318 \pm 0.0032$ & $0.5608 \pm 0.0195$ \\
            60  & $0.9143 \pm 0.0067$ & $0.5272 \pm 0.0252$ \\
            70  & $0.8968 \pm 0.0082$ & $0.5041 \pm 0.0290$ \\
            80  & $0.8834 \pm 0.0077$ & $0.4976 \pm 0.0219$ \\
            90  & $0.8703 \pm 0.0065$ & $0.5059 \pm 0.0110$ \\
            100 & $0.8509 \pm 0.0046$ & $0.5038 \pm 0.0260$
        \end{tabular}
        \\
        \hline
    \end{tabular}

    \caption{Balanced accuracy scores averaged over 10 train-test splits, with standard deviations over $R=10$ noise seeds, for the noisy normalized TFIM IPR kernel with local region $\delta=0.25$.}
    \label{tab:svm_tfim_noisy}
\end{table*}

\subsection{XXZ ground states}\label{app:xxz}

\begin{figure}[t]
    \centering
    \includegraphics{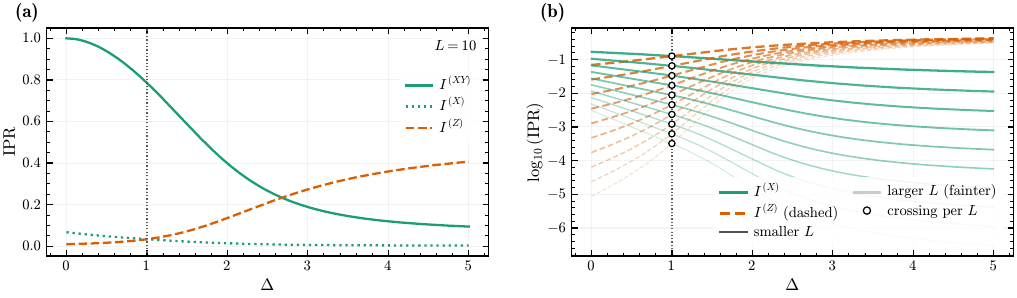}
    \caption{IPRs of Heisenberg XXZ ground states. (a) IPRs in the $XY$, $X$ (dotted), and $Z$ (dashed) bases for $L=10$. (b) IPRs in the $X$ and $Z$ bases for $L\in\{6,8,\dots,24\}$ on a $\log_{10}$ scale. For every $L$ the crossing (circles) lies at $\Delta_c=1$, as enforced by the SU(2) symmetry of the isotropic point.}
    \label{fig:xxz_ipr_xz}
\end{figure}

Here we give more details on the Heisenberg model example. First, we show data for the naive approach of taking the $X$ and $Z$ bases, i.e., building our kernel with the feature map 
\begin{equation}
     \mathbf{I}(\psi_{\mathrm{GS}}(\Delta)):=\big(I^{(X)}(\psi_{\mathrm{GS}}(\Delta)),I^{(Z)}(\psi_{\mathrm{GS}}(\Delta))\big).
\end{equation}
In \cref{fig:xxz_ipr_xz}(a) we show the IPRs for all considered bases at $L=10$. Although the $X$-basis IPRs are much smaller than the $XY$-basis IPRs, they cross the $Z$-basis IPRs at $\Delta_c=1$. This crossing is fixed by symmetry. At $\Delta=1$ the Hamiltonian is the isotropic Heisenberg chain, which is invariant under global SU(2) rotations. For even $L$ its ground state is a unique singlet and is therefore invariant under the global rotation that maps $Z$ to $X$, so that $I^{(X)}=I^{(Z)}$ holds exactly at $\Delta=1$ for every $L$. \Cref{fig:xxz_ipr_xz}(b) confirms this: for all $L\in\{6,8,\dots,24\}$ the crossing stays at $\Delta=1$, in contrast to the TFIM, where the crossing approaches $h_c=1$ only with increasing system size.

\begin{figure}[t]
    \centering
    \includegraphics{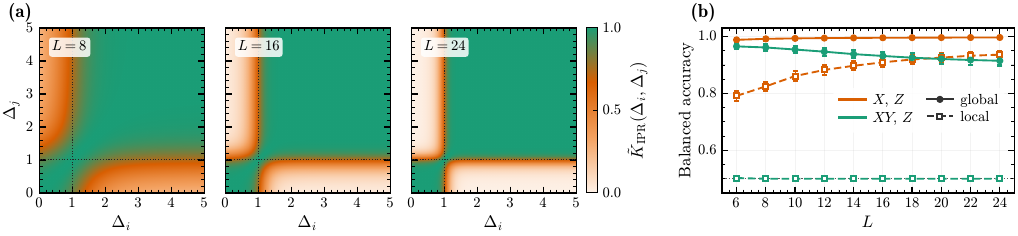}
    \caption{IPR kernel built from the $X$ and $Z$ bases for Heisenberg XXZ ground states. (a) Normalized Gram matrices for $L=8$, $16$, and $24$, with $\Delta_c=1$ marked by dotted lines. (b) Global (filled circles, solid) and local ($\delta=0.25$, open squares, dashed) balanced accuracy versus $L$ for the $X$/$Z$ kernel and the $XY$/$Z$ kernel of \cref{sec:xxz}, averaged over 100 train-test splits with standard deviation over splits.}
    \label{fig:xxz_xz_kernel}
\end{figure}

Consequently, the Gram matrices of the $X$ and $Z$ kernel in \cref{fig:xxz_xz_kernel}(a) develop a sharp block structure at $\Delta=1$, and the kernel reaches a high local balanced accuracy. This performance reflects the symmetry-enforced crossing and not a signature of the transition. An IPR crossing alone is therefore not sufficient evidence for a phase transition. In addition, the small $X$-basis IPRs would require a larger sampling budget.

The SVM results for both basis set choices are provided in \cref{fig:xxz_xz_kernel}(b) and \cref{tab:svm_xxz}. We have lowered the test fraction to 0.05 since there exist 5000 instead of 4000 points in the dataset now and a test fraction of 0.1 would make the selection of local test points deterministic.

As expected, the results for the $XY$ and $Z$ approach show a random guess in the local region since those IPR curves cross outside the local region for the examined system sizes. Furthermore, while we normally expect balanced accuracy to increase with system size, the global performance for the $XY$ and $Z$ case does decay. This is seemingly due to the learned transition sharpening but doing so away from the critical point $\Delta_c=1$, meaning the normalized kernel increasingly forces the decision boundary towards the wrong crossing and we lose the ``benefit'' of misplaced classification boundaries. 

\begin{table}[t]
    \centering
    \begin{tabular}{|r|l|l||r|l|l|}
        \hline
        \multicolumn{3}{|c||}{Heisenberg, $X$ and $Z$ basis}
        &
        \multicolumn{3}{c|}{Heisenberg, $XY$ and $Z$ basis} \\
        \hline
        $L$ & Global BA & Local BA
        &
        $L$ & Global BA & Local BA \\
        \hline
        \hline
        6  & $0.9883 \pm 0.0071$ & $0.7913 \pm 0.0168$
        &
        6  & $0.9653 \pm 0.0108$ & $0.5024 \pm 0.0042$ \\
        \hline
        8  & $0.9918 \pm 0.0059$ & $0.8248 \pm 0.0158$
        &
        8  & $0.9614 \pm 0.0117$ & $0.5000 \pm 0.0000$ \\
        \hline
        10 & $0.9934 \pm 0.0049$ & $0.8610 \pm 0.0171$
        &
        10 & $0.9542 \pm 0.0133$ & $0.5000 \pm 0.0000$ \\
        \hline
        12 & $0.9944 \pm 0.0045$ & $0.8824 \pm 0.0171$
        &
        12 & $0.9466 \pm 0.0138$ & $0.5000 \pm 0.0000$ \\
        \hline
        14 & $0.9952 \pm 0.0043$ & $0.8971 \pm 0.0162$
        &
        14 & $0.9381 \pm 0.0146$ & $0.5000 \pm 0.0000$ \\
        \hline
        16 & $0.9955 \pm 0.0043$ & $0.9092 \pm 0.0150$
        &
        16 & $0.9316 \pm 0.0152$ & $0.5000 \pm 0.0000$ \\
        \hline
        18 & $0.9959 \pm 0.0042$ & $0.9193 \pm 0.0151$
        &
        18 & $0.9255 \pm 0.0160$ & $0.5000 \pm 0.0000$ \\
        \hline
        20 & $0.9962 \pm 0.0038$ & $0.9258 \pm 0.0146$
        &
        20 & $0.9209 \pm 0.0162$ & $0.5000 \pm 0.0000$ \\
        \hline
        22 & $0.9964 \pm 0.0036$ & $0.9324 \pm 0.0143$
        &
        22 & $0.9174 \pm 0.0162$ & $0.5000 \pm 0.0000$ \\
        \hline
        24 & $0.9965 \pm 0.0035$ & $0.9362 \pm 0.0138$
        &
        24 & $0.9147 \pm 0.0165$ & $0.5000 \pm 0.0000$ \\
        \hline
    \end{tabular}
    \caption{Averaged balanced accuracy scores including standard deviations over 100 splits for the Heisenberg IPR kernel with local region $\delta=0.25$. Left: $X$ and $Z$ basis. Right: $XY$ and $Z$ basis.}
    \label{tab:svm_xxz}
\end{table}

\subsection{TFIM Gibbs states}\label{app:thermal}

\subsubsection{IPR and purity of mixed states}\label{app:mixed_ipr}

We now give further details on the behavior of the IPR kernel for mixed states. Starting from \cref{eq:mixed_ipr}, we can relate $I^{(s)}(\rho)$ to the purity of a density matrix $\Tr(\rho^2)$. For convenience, we denote by $(\cdot)_D$ the diagonal part of a matrix, obtained by keeping only its diagonal entries. Then
\begin{equation}
I^{(s)}(\rho)=\sum_x \left(\rho_{xx}^{(s)}\right)^2=\Tr[\Big(\big(\rho^{(s)}\big)_{D}\Big)^{2}].
\end{equation}
Thus, $I^{(s)}(\rho)$ is the purity contained in the diagonal part of $\rho^{(s)}$. Since the full purity is invariant under basis changes, we obtain
\begin{equation}
    \begin{aligned}
        \Tr(\rho^2)
         & =
        \Tr(
            \left(
            \rho^{(s)}
            \right)^2
        )         \\
         & =
        \sum_{x}
        \bra{x}
        U_s^\dagger \rho U_s
        U_s^\dagger \rho U_s
        \ket{x}         \\
         & =
        \sum_{x,y}
        \bra{x}
        U_s^\dagger \rho U_s
        \ket{y}
        \bra{y}
        U_s^\dagger \rho U_s
        \ket{x}         \\
         & =
        \sum_{x,y}
        \rho^{(s)}_{xy}
        \rho^{(s)}_{yx} \\
         & =
        \sum_{x}
        (
        \rho^{(s)}_{xx}
        )^2
        +
        \sum_{x\neq y}
        \rho^{(s)}_{xy}
        \rho^{(s)}_{yx} \\
         & =
        I^{(s)}(\rho)
        +
        \sum_{x\neq y}
        \abs{\rho^{(s)}_{xy}}^2.
    \end{aligned}
\end{equation}
Since $\rho$ is Hermitian, $\Tr(\rho^2) \geq I^{(s)}(\rho)$ and $\Tr(\rho^2)\Tr(\sigma^2)\geq I^{(s)}(\rho)I^{(s)}(\sigma)$ with equality being achieved only if $\rho$ and $\sigma$ are diagonal in the basis specified by $U_s$.

Summing over the $M$ bases used in the IPR kernel yields
\begin{equation}
\frac{1}{M}
K_{\rm IPR}(\rho,\sigma)
\leq
\Tr(\rho^2)
\Tr(\sigma^2).
\end{equation}
On the other hand, the HS kernel satisfies the Cauchy-Schwarz bound
\begin{equation}
K_{\rm HS}(\rho,\sigma)^2
=
\Tr(\rho\sigma)^2
\leq
\Tr(\rho^2)
\Tr(\sigma^2).
\end{equation}
Thus, both the IPR and HS kernels are controlled by the same purity-based upper bound. However, because $K_{\rm IPR}$ depends on the chosen bases, there is no general ordering between the two kernels.

The same statement can also be phrased through the Schur-Horn theorem. Let $V$ be the unitary diagonalizing $\rho$, such that 
\begin{equation}
V^\dagger \rho V = D ,
\end{equation}
where $D$ is diagonal and contains the eigenvalues of $\rho$. Then, for any basis transformation $U$, the Schur-Horn theorem implies
\begin{equation}
\left(
U^\dagger \rho U
\right)_D
\prec
\left(
V^\dagger \rho V
\right)_D ,
\end{equation}
where both sides are interpreted as vectors of diagonal entries. The purity contained in the diagonal is thus maximized in an eigenbasis of $\rho$. If $\rho$ is the reduced density matrix of a larger pure state, e.g., $\rho_A=\Tr_B(\rho_{AB})$, the purity is related to the bipartite entanglement between the subsystem and its complement. In such a case, the chosen basis $U$ determines how much purity-based information about entanglement remains visible after restricting to diagonal measurement statistics.

This relation can equivalently be expressed through the participation
entropy of Rényi order $q=2$,
\begin{equation}
    S_{P,2}^{(s)}(\rho)
    =
    -\log I^{(s)}(\rho)
    =
    -\log \Tr[
        \Big(\big(\rho^{(s)}\big)_{D}\Big)^{2}
    ].
\end{equation}
Connections between localization, participation entropies, purities, and
entanglement have been explored widely in the literature \cite{lakshminarayan_entanglement_localization_2016, calixto_inverse_participation_2024, tan_probing_quantum_2025, freund_ability_entanglement_2026, brydges_probing_2019, koh_measurement_induced_entanglement_2023, yang_probing_entanglement_2026}.

\subsubsection{Gram matrices and classification performance}\label{app:thermal_svm}
\begin{figure}[t]
    \centering
    \includegraphics{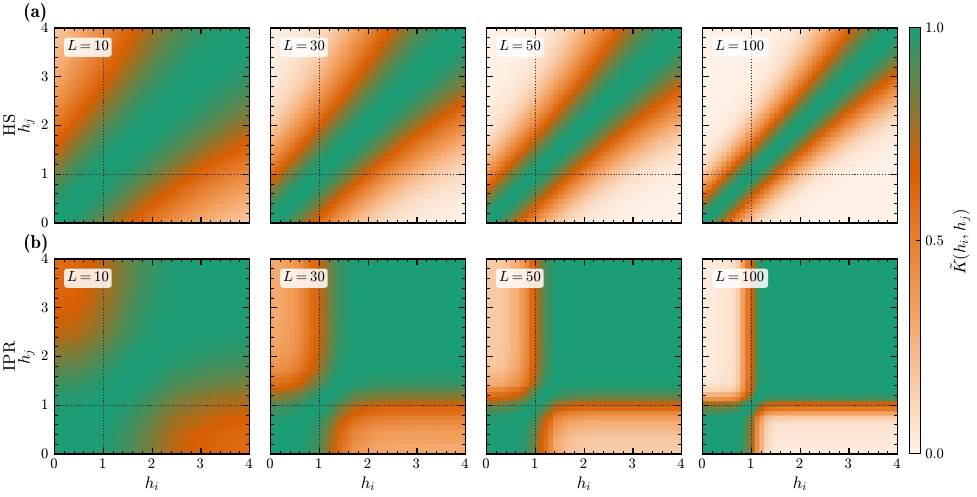}
    \caption{Normalized (a) HS-kernel and (b) IPR-kernel Gram matrices for TFIM Gibbs states at $\beta=0.2$ and increasing system size $L$.}
    \label{fig:thermal_l_sweep}
\end{figure}

Here we go into more detail for both the HS and IPR kernels used in \cref{sec:thermal}. In \cref{fig:thermal_l_sweep} we compare the Gram matrices of both kernels at $\beta=0.2$ for increasing system sizes $L$. Regarding the IPR kernel we again see its structure become clearer with increasing system size, an effect already observed in \cref{fig:tfim_grams}. The HS kernel follows the effects of exponential concentration, as discussed in \cite{thanasilp_exponential_2024}. In short, due to the increasing size of the Hilbert space, states become increasingly orthogonal, leading to the off-diagonal values in the HS-kernel Gram matrices converging to zero. While this ``dimensional curse'' is also prevalent in our case, where IPRs become exponentially small in system size, it mainly concerns the region around the phase transition.

We run HS- and IPR-kernel SVMs for system sizes $L\in\{20,50,100\}$ and $\beta\in\{0.1,0.2,\dots,2.0\}$ using a 90/10 train-test split ratio. The exact balanced-accuracy results for the thermal phase-classification SVMs corresponding to \cref{fig:thermal_results}(c) are reported in \cref{tab:svm_thermal}. For each inverse temperature $\beta$, the SVM is trained on $N_{\mathrm{splits}}=100$ random train-test splits. For a given split $n$, the SVM decision function follows \cref{eq:decision_function}. In the thermal setting, the input states are TFIM Gibbs states parametrized by the transverse field $h$, i.e., we have
\begin{equation}
    f_n(\rho_\beta(h))
    =
    \sum_{j\in \mathrm{SV}_n}
    \alpha^{(n)}_j y^{(n)}_j
    \tilde{K}_{\mathrm{IPR}}\!\left(
        \rho_\beta(h^{(n)}_j),
        \rho_\beta(h)
    \right)
    +
    b_n .
\end{equation}
Analogously, for the HS kernel we replace $\tilde{K}_{\mathrm{IPR}}$ by $\tilde{K}_{\mathrm{HS}}$ from \cref{eq:hs_kernel_norm}.

\begin{figure}[t]
    \centering
    \includegraphics{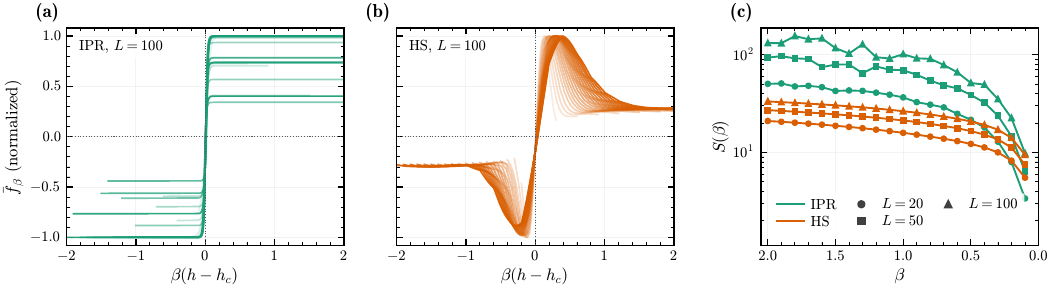}
    \caption{(a) IPR-kernel and (b) HS-kernel global decision functions at $L=100$, normalized by their maximum absolute value and plotted against $\beta(h-h_c)$. Curves for smaller $\beta$ are fainter. (c) Steepness $S(\beta)$ of the decision function at its zero crossing for $L=20$, $50$, and $100$.}
    \label{fig:decision_functions}
\end{figure}

After training globally, the decision function is evaluated on the full sampled $h$ grid and averaged pointwise over all splits:
\begin{equation}
    \bar f_\beta(h_i)
    =
    \frac{1}{N_{\mathrm{splits}}}
    \sum_{n=1}^{N_{\mathrm{splits}}}
    f_n(\rho_\beta(h_i)).
\end{equation}
For visual comparison, each averaged curve is normalized by its maximum absolute value,
\begin{equation}
    \tilde f_\beta(h_i)
    =
    \frac{\bar f_\beta(h_i)}
    {\max_j |\bar f_\beta(h_j)|}.
\end{equation}
This normalization preserves the sign and zero crossing of the decision function, leaving the learned phase boundary unchanged. Finally, the normalized curves are plotted as a function of the rescaled variable
\begin{equation}
    \beta(h-h_c),
    \quad \text{where} \quad h_c=1
\end{equation}
and shown in \cref{fig:decision_functions}. The motivation for this rescaling comes from finite-temperature quantum-critical scaling \cite{sachdev_universal_1996,sachdev_quantum_phase_2011,wu_crossovers_critical_2018}. Close to a quantum critical point, the correlation length diverges as
\begin{equation}
\xi \sim |h-h_c|^{-\nu},
\end{equation}
where $\nu$ is the correlation-length critical exponent. The corresponding energy gap scales as
\begin{equation}
\Delta_{\mathrm{g}} \sim \xi^{-z} \sim |h-h_c|^{z\nu},
\end{equation}
where $z$ is the dynamical critical exponent. At finite temperature, the thermal energy scale $T=1/\beta$ cuts off the quantum-critical behavior once it becomes comparable to $\Delta_{\mathrm{g}}$. The crossover region is therefore expected to satisfy
\begin{equation}
T \sim |h-h_c|^{z\nu}.
\end{equation}
For the one-dimensional TFIM we have $z=1$ and $\nu=1$ and get
\begin{equation}
\frac{1}{\beta}\sim|h-h_c|.
\end{equation}
If therefore the decision functions for different $\beta$ approximately align under this rescaling, it would indicate that the corresponding kernel captures the finite-temperature critical crossover more faithfully.

We can see that the decision function is noticeably steeper in the IPR case. In \cref{fig:decision_functions}(c) we also plot the steepness $S(\beta)$ of the split-averaged decision function $\bar{f}_\beta(h)$ as the gradient where it crosses its learned decision boundary, i.e.
\begin{equation}
S(\beta)=\left|\frac{\partial \bar f_\beta(h)}{\partial h}\right|_{h=h^*(\beta)},
\end{equation}
with $\bar{f}_\beta(h^*)=0$. A steeper decision function separates the phases more sharply at the learned boundary, consistent with the higher local balanced accuracy of the IPR kernel.

\subsubsection{Quantum critical fan from IPRs}\label{app:critical_fan}

At finite temperature, the zero-temperature quantum critical point is no longer observed as a sharp transition, but broadens into a quantum critical fan \cite{frerot_reconstructing_2019}. In the $(h,\beta)$ plane this corresponds to a thermal crossover cone whose width is governed by the finite-temperature scaling variable discussed above, namely $\beta \sim {|h - h_c|}^{-1}$. We now ask whether this cone can be recovered directly from the IPR features themselves.

\begin{figure}[t]
    \centering
    \includegraphics{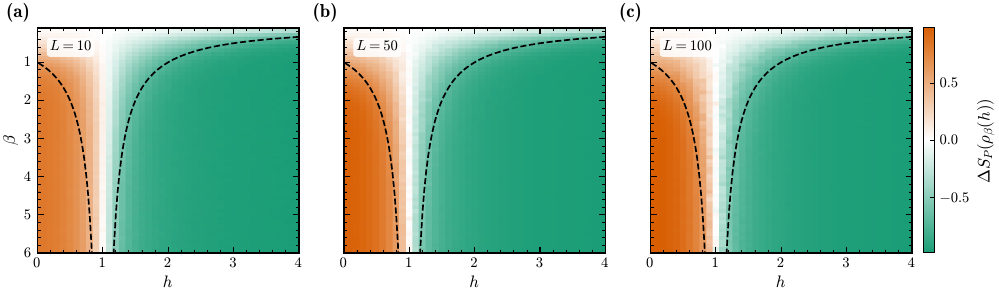}
    \caption{Thermal phase diagram $\Delta S_P(\rho_\beta(h))$ extracted from the IPRs for (a) $L=10$, (b) $L=50$, and (c) $L=100$. Dashed black lines correspond to $\beta = {|h - h_c|}^{-1}$.}
    \label{fig:thermal_phase_diagram}
\end{figure}

Following along the lines of \cite{zhang_critical_2019}, we define the basis-dependent participation entropy of Rényi order $q$ as
\begin{equation}
S^{(s)}_{P,q}(\rho)
=
\frac{1}{1-q}
\log\left(I^{(s)}_q(\rho)\right)
\overset{q=2}{=}
-\log\left(I^{(s)}(\rho)\right),
\end{equation}
where $I^{(s)}_q(\rho)$ denotes the corresponding basis-dependent IPR, and $I^{(s)}(\rho)\equiv I^{(s)}_2(\rho)$ is the $q=2$ IPR used throughout this work. Next, in order to obtain a thermal cone in terms of our IPRs we define the normalized participation-entropy difference as
\begin{equation}
\begin{aligned}
\Delta S_P(\rho_\beta(h))
&=
\frac{
S^{(X)}_{P,2}(\rho_\beta(h))
-
S^{(Z)}_{P,2}(\rho_\beta(h))
}{
S^{(X)}_{P,2}(\rho_\beta(h))
+
S^{(Z)}_{P,2}(\rho_\beta(h))
}
=
\frac{
\log I^{(Z)}(\rho_\beta(h))
-
\log I^{(X)}(\rho_\beta(h))
}{
\left|\log I^{(Z)}(\rho_\beta(h))\right|
+
\left|\log I^{(X)}(\rho_\beta(h))\right|
}.
\end{aligned}
\end{equation}
Thus $\Delta S_P<0$ corresponds to stronger $X$-basis localization, while $\Delta S_P>0$ corresponds to stronger $Z$-basis localization. Following the finite-temperature scaling argument of the previous section, the crossover region is expected to be bounded by $\beta = {|h - h_c|}^{-1}$. We therefore plot a thermal phase diagram of $\Delta S_P(\rho_\beta(h))$ for $L=10$, $50$, and $100$ in \cref{fig:thermal_phase_diagram}, where the dashed lines indicate the expected bounds.

\begin{table*}[t]
    \centering
    \scriptsize
    \setlength{\tabcolsep}{2pt}
    \renewcommand{\arraystretch}{0.95}

    % ---------- top row: L=20 and L=50 ----------
    \begin{minipage}{0.49\textwidth}
        \centering
        \textbf{$L=20$}

        \vspace{0.25em}

        \resizebox{\linewidth}{!}{%
        \begin{tabular}{|r|c|c|c|c|}
            \hline
            $\beta$ & IPR-G & IPR-L & HS-G & HS-L \\
            \hline\hline
            0.1 & $0.9638\pm0.0100$ & $0.6706\pm0.0080$ & $0.9851\pm0.0068$ & $0.8512\pm0.0148$ \\
            0.2 & $0.9882\pm0.0060$ & $0.8055\pm0.0151$ & $0.9899\pm0.0057$ & $0.8321\pm0.0163$ \\
            0.3 & $0.9929\pm0.0046$ & $0.8545\pm0.0205$ & $0.9916\pm0.0051$ & $0.8304\pm0.0165$ \\
            0.4 & $0.9950\pm0.0039$ & $0.8972\pm0.0193$ & $0.9924\pm0.0047$ & $0.8404\pm0.0196$ \\
            0.5 & $0.9960\pm0.0034$ & $0.9230\pm0.0188$ & $0.9929\pm0.0045$ & $0.8530\pm0.0209$ \\
            0.6 & $0.9963\pm0.0032$ & $0.9300\pm0.0170$ & $0.9932\pm0.0046$ & $0.8611\pm0.0216$ \\
            0.7 & $0.9969\pm0.0028$ & $0.9421\pm0.0158$ & $0.9937\pm0.0045$ & $0.8693\pm0.0224$ \\
            0.8 & $0.9972\pm0.0026$ & $0.9468\pm0.0149$ & $0.9940\pm0.0043$ & $0.8764\pm0.0220$ \\
            0.9 & $0.9972\pm0.0026$ & $0.9485\pm0.0139$ & $0.9944\pm0.0041$ & $0.8826\pm0.0216$ \\
            1.0 & $0.9976\pm0.0024$ & $0.9526\pm0.0135$ & $0.9946\pm0.0042$ & $0.8878\pm0.0211$ \\
            1.1 & $0.9977\pm0.0024$ & $0.9555\pm0.0127$ & $0.9947\pm0.0040$ & $0.8917\pm0.0209$ \\
            1.2 & $0.9976\pm0.0024$ & $0.9555\pm0.0123$ & $0.9949\pm0.0040$ & $0.8952\pm0.0212$ \\
            1.3 & $0.9978\pm0.0024$ & $0.9586\pm0.0122$ & $0.9950\pm0.0039$ & $0.8986\pm0.0212$ \\
            1.4 & $0.9976\pm0.0024$ & $0.9531\pm0.0123$ & $0.9952\pm0.0037$ & $0.9013\pm0.0203$ \\
            1.5 & $0.9977\pm0.0024$ & $0.9565\pm0.0127$ & $0.9953\pm0.0037$ & $0.9036\pm0.0198$ \\
            1.6 & $0.9977\pm0.0024$ & $0.9579\pm0.0119$ & $0.9954\pm0.0037$ & $0.9057\pm0.0193$ \\
            1.7 & $0.9979\pm0.0024$ & $0.9590\pm0.0117$ & $0.9955\pm0.0037$ & $0.9085\pm0.0194$ \\
            1.8 & $0.9978\pm0.0024$ & $0.9585\pm0.0118$ & $0.9956\pm0.0037$ & $0.9096\pm0.0190$ \\
            1.9 & $0.9981\pm0.0022$ & $0.9619\pm0.0116$ & $0.9956\pm0.0037$ & $0.9108\pm0.0194$ \\
            2.0 & $0.9980\pm0.0022$ & $0.9622\pm0.0120$ & $0.9956\pm0.0037$ & $0.9121\pm0.0188$ \\
            \hline
        \end{tabular}%
        }
    \end{minipage}
    \hfill
    \begin{minipage}{0.49\textwidth}
        \centering
        \textbf{$L=50$}

        \vspace{0.25em}

        \resizebox{\linewidth}{!}{%
        \begin{tabular}{|r|c|c|c|c|}
            \hline
            $\beta$ & IPR-G & IPR-L & HS-G & HS-L \\
            \hline\hline
            0.1 & $0.9849\pm0.0068$ & $0.7945\pm0.0129$ & $0.9887\pm0.0060$ & $0.8331\pm0.0167$ \\
            0.2 & $0.9937\pm0.0045$ & $0.8719\pm0.0202$ & $0.9923\pm0.0047$ & $0.8361\pm0.0201$ \\
            0.3 & $0.9962\pm0.0034$ & $0.9258\pm0.0168$ & $0.9934\pm0.0046$ & $0.8641\pm0.0219$ \\
            0.4 & $0.9972\pm0.0027$ & $0.9483\pm0.0143$ & $0.9943\pm0.0042$ & $0.8806\pm0.0216$ \\
            0.5 & $0.9978\pm0.0024$ & $0.9598\pm0.0128$ & $0.9947\pm0.0040$ & $0.8893\pm0.0217$ \\
            0.6 & $0.9984\pm0.0020$ & $0.9731\pm0.0117$ & $0.9949\pm0.0039$ & $0.8960\pm0.0213$ \\
            0.7 & $0.9985\pm0.0020$ & $0.9758\pm0.0109$ & $0.9953\pm0.0037$ & $0.9010\pm0.0204$ \\
            0.8 & $0.9980\pm0.0022$ & $0.9609\pm0.0107$ & $0.9954\pm0.0037$ & $0.9061\pm0.0199$ \\
            0.9 & $0.9987\pm0.0018$ & $0.9786\pm0.0096$ & $0.9956\pm0.0037$ & $0.9093\pm0.0201$ \\
            1.0 & $0.9989\pm0.0017$ & $0.9800\pm0.0087$ & $0.9958\pm0.0036$ & $0.9123\pm0.0198$ \\
            1.1 & $0.9988\pm0.0017$ & $0.9800\pm0.0087$ & $0.9958\pm0.0036$ & $0.9147\pm0.0192$ \\
            1.2 & $0.9987\pm0.0018$ & $0.9772\pm0.0082$ & $0.9958\pm0.0036$ & $0.9177\pm0.0181$ \\
            1.3 & $0.9992\pm0.0016$ & $0.9919\pm0.0055$ & $0.9959\pm0.0035$ & $0.9202\pm0.0181$ \\
            1.4 & $0.9990\pm0.0017$ & $0.9847\pm0.0078$ & $0.9960\pm0.0034$ & $0.9216\pm0.0178$ \\
            1.5 & $0.9990\pm0.0016$ & $0.9852\pm0.0076$ & $0.9961\pm0.0034$ & $0.9231\pm0.0179$ \\
            1.6 & $0.9991\pm0.0016$ & $0.9909\pm0.0061$ & $0.9962\pm0.0034$ & $0.9248\pm0.0183$ \\
            1.7 & $0.9990\pm0.0016$ & $0.9858\pm0.0069$ & $0.9962\pm0.0033$ & $0.9259\pm0.0181$ \\
            1.8 & $0.9990\pm0.0016$ & $0.9849\pm0.0072$ & $0.9963\pm0.0032$ & $0.9265\pm0.0184$ \\
            1.9 & $0.9990\pm0.0016$ & $0.9828\pm0.0065$ & $0.9964\pm0.0032$ & $0.9276\pm0.0182$ \\
            2.0 & $0.9990\pm0.0016$ & $0.9828\pm0.0070$ & $0.9964\pm0.0032$ & $0.9281\pm0.0181$ \\
            \hline
        \end{tabular}%
        }
    \end{minipage}

    \vspace{0.75em}

    % ---------- bottom row: L=100 centered ----------
    \begin{minipage}{0.52\textwidth}
        \centering
        \textbf{$L=100$}

        \vspace{0.25em}

        \resizebox{\linewidth}{!}{%
        \begin{tabular}{|r|c|c|c|c|}
            \hline
            $\beta$ & IPR-G & IPR-L & HS-G & HS-L \\
            \hline\hline
            0.1 & $0.9906\pm0.0054$ & $0.8079\pm0.0142$ & $0.9910\pm0.0052$ & $0.8237\pm0.0158$ \\
            0.2 & $0.9960\pm0.0034$ & $0.9247\pm0.0178$ & $0.9937\pm0.0045$ & $0.8681\pm0.0217$ \\
            0.3 & $0.9977\pm0.0024$ & $0.9565\pm0.0136$ & $0.9948\pm0.0040$ & $0.8916\pm0.0214$ \\
            0.4 & $0.9979\pm0.0024$ & $0.9594\pm0.0117$ & $0.9953\pm0.0037$ & $0.9021\pm0.0199$ \\
            0.5 & $0.9989\pm0.0017$ & $0.9869\pm0.0073$ & $0.9956\pm0.0036$ & $0.9099\pm0.0198$ \\
            0.6 & $0.9980\pm0.0022$ & $0.9627\pm0.0084$ & $0.9958\pm0.0036$ & $0.9145\pm0.0191$ \\
            0.7 & $0.9984\pm0.0020$ & $0.9709\pm0.0075$ & $0.9959\pm0.0035$ & $0.9197\pm0.0181$ \\
            0.8 & $0.9989\pm0.0017$ & $0.9824\pm0.0072$ & $0.9960\pm0.0034$ & $0.9224\pm0.0178$ \\
            0.9 & $0.9992\pm0.0016$ & $0.9908\pm0.0058$ & $0.9962\pm0.0033$ & $0.9252\pm0.0183$ \\
            1.0 & $0.9989\pm0.0016$ & $0.9822\pm0.0060$ & $0.9963\pm0.0032$ & $0.9264\pm0.0181$ \\
            1.1 & $0.9995\pm0.0010$ & $0.9951\pm0.0050$ & $0.9964\pm0.0031$ & $0.9278\pm0.0182$ \\
            1.2 & $0.9984\pm0.0020$ & $0.9721\pm0.0060$ & $0.9964\pm0.0030$ & $0.9295\pm0.0179$ \\
            1.3 & $0.9991\pm0.0016$ & $0.9881\pm0.0048$ & $0.9965\pm0.0030$ & $0.9308\pm0.0176$ \\
            1.4 & $0.9977\pm0.0023$ & $0.9646\pm0.0056$ & $0.9965\pm0.0030$ & $0.9323\pm0.0176$ \\
            1.5 & $0.9986\pm0.0018$ & $0.9757\pm0.0048$ & $0.9965\pm0.0030$ & $0.9333\pm0.0174$ \\
            1.6 & $0.9995\pm0.0013$ & $0.9941\pm0.0038$ & $0.9966\pm0.0029$ & $0.9338\pm0.0174$ \\
            1.7 & $0.9993\pm0.0013$ & $0.9931\pm0.0038$ & $0.9967\pm0.0030$ & $0.9344\pm0.0171$ \\
            1.8 & $0.9994\pm0.0013$ & $0.9946\pm0.0035$ & $0.9966\pm0.0030$ & $0.9351\pm0.0170$ \\
            1.9 & $0.9990\pm0.0016$ & $0.9844\pm0.0047$ & $0.9967\pm0.0028$ & $0.9356\pm0.0166$ \\
            2.0 & $0.9996\pm0.0010$ & $0.9963\pm0.0037$ & $0.9967\pm0.0029$ & $0.9361\pm0.0166$ \\
            \hline
        \end{tabular}%
        }
    \end{minipage}

    \caption{Balanced-accuracy (BA) scores for the thermal TFIM phase-classification task using the normalized IPR and HS kernels. Results are shown for $L=20,50,100$ as a function of inverse temperature $\beta$, with global (G) and local (L) test regions. Values are averaged over 100 train-test splits and reported with standard deviations, using a 90/10 train-test fraction.}
    \label{tab:svm_thermal}
\end{table*}

\subsubsection{High-temperature decay of the kernels}\label{app:kernel_decay}

In this section we analytically compare the high-temperature decay of the normalized HS and IPR kernels. Recall that for TFIM Gibbs states the two normalized kernels are given by
\begin{equation}
\label{eq:normalized_hs_kernel_comparison}
\tilde{K}_{\mathrm{HS}}\bigl(\rho_\beta(h_i),\rho_\beta(h_j)\bigr)
=
\frac{
\Tr(\rho_\beta(h_i)\rho_\beta(h_j))
}{
\sqrt{
\Tr(\rho_\beta(h_i)^2)
\Tr(\rho_\beta(h_j)^2)
}
}
\end{equation}
and
\begin{equation}
\label{eq:normalized_ipr_kernel_comparison}
\tilde{K}_{\mathrm{IPR}}\bigl(\rho_\beta(h_i),\rho_\beta(h_j)\bigr)
=
\frac{
I^{(X)}(\rho_\beta(h_i)) I^{(X)}(\rho_\beta(h_j))
+
I^{(Z)}(\rho_\beta(h_i)) I^{(Z)}(\rho_\beta(h_j))
}{
\sqrt{
\left(I^{(X)}(\rho_\beta(h_i))\right)^2
+
\left(I^{(Z)}(\rho_\beta(h_i))\right)^2
}
\sqrt{
\left(I^{(X)}(\rho_\beta(h_j))\right)^2
+
\left(I^{(Z)}(\rho_\beta(h_j))\right)^2
}
}.
\end{equation}
Our goal here is to understand how quickly these two similarity measures become insensitive to phase information as the temperature increases. To isolate this effect, we evaluate both kernels between two limiting TFIM regimes, namely the Ising-dominated point $h=0$ and the transverse-field-dominated limit $h\to\infty$. Hence, we expect the kernel values to be zero for large $\beta$ (close to the ground state) and one for $\beta=0$ (both limits are maximally mixed states).

\paragraph{Limit $h \to \infty$.} The spectrum of the Hamiltonian dominated by the transverse field is computable by the following procedure. The eigenbasis is the Hadamard basis, and the energy is the difference between the numbers of spins in the $\ket -$ and $\ket +$ states, as $E(x) = -L + 2 w(x)$, with $w(x)$ being the Hamming weight, that is, the number of $(-)$ after transformation into the Hadamard basis. Since the Hamming weight follows a binomial distribution, the energies $E_k$ and degeneracies $d_k$ are
\begin{equation}
    E_k = -L + 2k \qquad\text{and}\qquad  d_k = \binom{L}{k},  \qquad\text{where}\qquad  k \in \{0, 1, 2, \ldots, L \}.
\end{equation}
Starting from $H_{\mathrm{TFIM}}(h)$ and rescaling the inverse temperature as $\beta/h$, we obtain the nontrivial limiting Gibbs state through
\begin{equation}
    \lim_{h\to\infty} \rho_{\beta/h}(h)
    =
    \frac{e^{-\beta H_X}}{\Tr(e^{-\beta H_X})},
    \qquad
    H_X=-\sum_i X_i.
\end{equation}
Since $\exp(-\beta E_k)=\exp(\beta L)\exp(-2\beta k)$ and the factor $\exp(\beta L)$ cancels against the same factor in the partition function, we can equivalently use the shifted Boltzmann weights $\exp(-2\beta k)$. Thus,
\begin{align}
    \rho(\beta / h, h \to \infty) & = \frac{1}{Z^{(X)}_\beta}\sum_{k = 0}^L\exp\left( -2 \beta k \right) \sum_{\substack{x\in\{0,1\}^L\\ w(x)=k}}  \mathrm{H}^{\otimes L}\ket x \bra x  \mathrm{H}^{\otimes L} \\
    Z_\beta^{(X)}                     & = \sum_{k = 0}^L\exp\left( -2 \beta k \right) \binom{L}{k} = \left( 1 + \exp(-2\beta)\right)^L,
\end{align}
where the last step is obtained by identifying the sum as in the binomial theorem.

\begin{figure}[t]
    \centering
    \includegraphics{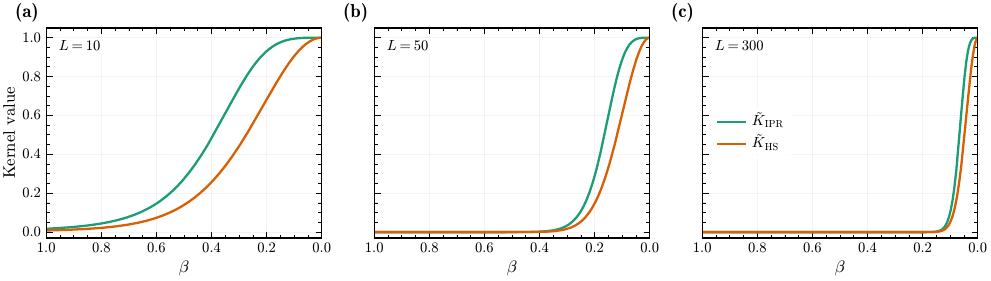}
    \caption{Analytical normalized kernels between the Gibbs states at $h=0$ and $h \to \infty$ as a function of $\beta$ for (a) $L=10$, (b) $L=50$, and (c) $L=300$, where the correct value would be zero. Temperature increases to the right.}
    \label{fig:analytical_kernels}
\end{figure}

\paragraph{Limit $h = 0$.} To compute the spectrum of the Hamiltonian dominated by the Ising term we can follow the same procedure as before with a small change. We transform the bitstrings under the mapping
\begin{equation}
    \mathcal D: (x_1, x_2, \ldots, x_L) \mapsto \mathcal D(x)=([x_1 \neq x_2], [x_2 \neq x_3], \ldots, [x_L \neq x_1]),
\end{equation}
and we operate over the Hamming weight of the transformed bitstrings. The Hamming weights in this case can only be even, and the multiplicity is doubled. Thus, we obtain
\begin{equation}
    E_k = -L + 2k \qquad\text{and}\qquad  d_k = 2 \binom{L}{k}, \qquad\text{where}\qquad k \in \{0, 2, 4, \ldots, L\}.
\end{equation}
Therefore, the Gibbs states of $H_{\mathrm{Ising}}=-\sum_i Z_iZ_{i+1}$ are
\begin{align}
    \rho(\beta, h = 0) & = \frac{1}{Z_\beta^{(Z)}}\sum_{\substack{k = 0 \\ 2 | k}}^L\exp\left( -2 \beta k \right) \sum_{x, w(\mathcal D(x))=k} \ket{x} \bra{x} \\
    Z_\beta^{(Z)}          & = 2 \sum_{\substack{k = 0                    \\ 2 | k}}^L\exp\left( -2 \beta k \right) \binom{L}{k} = \left( 1 +\exp(-2\beta)\right)^L + \left( 1 - \exp(-2\beta)\right)^L.
\end{align}
The kernels are now easy to compute.
\begin{align}
    I^{(Z)}(\rho(\beta, h = 0))            & = \frac{Z^{(Z)}_{2\beta}}{(Z^{(Z)}_\beta)^2} \equiv I^{(Z)}_0 \\
    I^{(X)}(\rho(\beta, h = 0))            & = 2^{-L}                                        \\
    I^{(Z)}(\rho(\beta / h, h \to \infty)) & = 2^{-L}                                        \\
    I^{(X)}(\rho(\beta / h, h \to \infty)) & =\frac{Z^{(X)}_{2\beta}}{(Z^{(X)}_\beta)^2} \equiv I^{(X)}_\infty,
\end{align}
and thus
\begin{align}
    \tilde{K}_{\rm IPR} & = \frac{1}{2^L} \frac{I^{(Z)}_0 + I^{(X)}_\infty}{\sqrt{4^{-L} + \big(I^{(Z)}_0\big)^2}\sqrt{4^{-L} + \big(I^{(X)}_\infty\big)^2}}        \\
    \tilde{K}_{\rm HS}  & = \frac{1}{2^L} \frac{Z^{(Z)}_\beta Z^{(X)}_\beta}{\sqrt{Z^{(Z)}_{2\beta}}\sqrt{Z^{(X)}_{2\beta}}}.
\end{align}
In both cases, the kernels can be approximated by Gaussians in the limit of small $\beta$ and large $L$. Notice that, for $\beta \to 0$ 
\begin{align}
    Z_\beta^{(Z)} \approx Z_\beta^{(X)} \approx 2^L\left( 1 - \beta + \beta^2 + \mathcal O(\beta^3)\right)^L.
\end{align}
Furthermore, we can approximate this function in exponential form as
\begin{equation}
    Z_\beta \approx 2^L \exp(L\ln(1 - \beta + \beta^2)) \approx 2^L \exp(-L \beta) \exp(\beta^2 L / 2),
\end{equation}
which gives us the final kernel approximations
\begin{align}
    \tilde{K}_{\rm IPR} & \approx \frac{2 \exp(\beta^2 L)}{1 + \exp(2\beta^2 L)} \geq \exp(-\frac{\beta^4 L^2}{2}) \\
    \tilde{K}_{\rm HS}  & \approx \exp(-\beta^2 L),
\end{align}
meaning near $\beta=0$, the IPR kernel stays closer to one than the HS kernel. In \cref{fig:analytical_kernels} we also depict the full plots of $ \tilde{K}_{\rm IPR}$ and $\tilde{K}_{\rm HS}$. We note again that their ground-state values should be close to zero; the slower decay of $\tilde{K}_{\rm IPR}$ with increasing $\beta$ is therefore the reason for the worse performance of the IPR kernel close to $\beta=0$ in the thermal SVM results of \cref{fig:thermal_results}(c), an effect that is more pronounced for smaller system sizes.

\section{Quantum circuits for IPR kernel evaluation}\label{app:circuits}

In this section we provide quantum circuits that calculate the IPR kernel and encode it in the probability of measuring $\ket{0}^{\otimes m}_{A}$ on an $m$-qubit ancilla register. The circuits are based on work done in \cite{liu_quantum_2025}, where the authors provide quantum circuits for inverse participation ratios for a single quantum state. We extend their work by allowing the circuits to process two states simultaneously, yielding the IPR kernel as a result.

\subsection{Known basis unitaries}\label{app:circuits_known}

\begin{figure}[htb]
    \centering
    \begin{quantikz}[row sep=0.5cm, column sep=0.7cm]
        \lstick{$\ket{0}_{A}$} &
        \gate{\mathrm{H}} &
        \ctrl{1} &
        \gate{\mathrm{H}} &
        \meter[wires=1]{}\\
        \lstick{$\ket{\psi}$}  &
        \qw &
        \gate[wires=2]{\text{SWAP}} &
        \qw &
        \qw\\
        \lstick{$\ket{\psi}$} &
        \ctrl{1} &
        \qw &
        \qw &
        \qw\\
        \lstick{$\ket{0}^{\otimes L}$} &
        \targ{} &
        \qw &
        \qw &
        \qw\\
    \end{quantikz}
    \caption{Circuit to obtain the IPR of $\ket{\psi}$ in the computational basis. The IPR is given by $2P_{0,A}-1$, where $P_{0,A}$ is the probability of measuring the ancilla $A$ in the zero state.}
    \label{fig:circuit_single_ipr}
\end{figure}
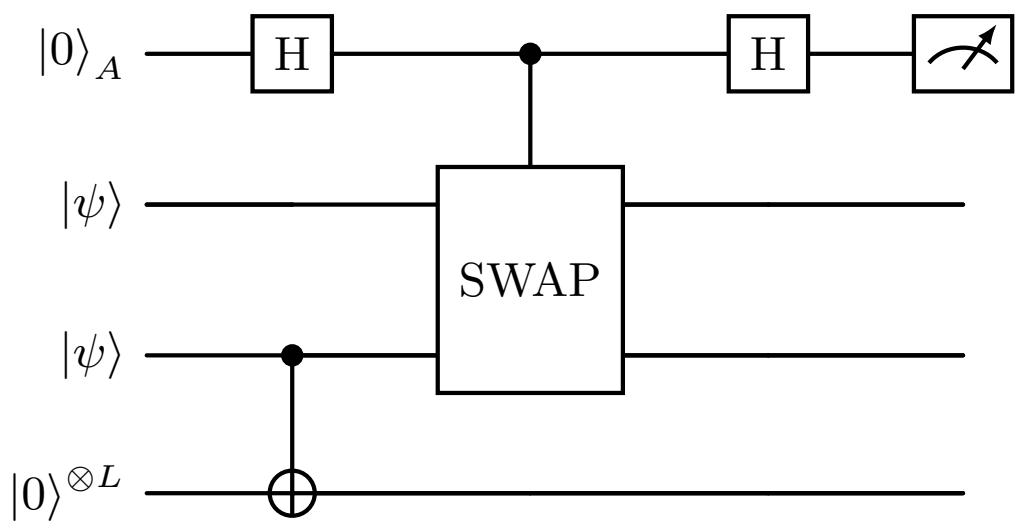

In general, a distinction is made between settings where the relevant basis is known explicitly and those where it is defined as the eigenbasis of a Hamiltonian. In the former case, the computational basis IPR of an $L$-qubit quantum state $\ket{\psi}$ can be obtained by using two copies of $\ket{\psi}$ and executing the circuit depicted in \cref{fig:circuit_single_ipr}. 

For the computational basis we have
\begin{equation}
    P_{0,A}=\frac{1}{2}+\frac{\sum_i |c_i|^{4}}{2}=\frac{1}{2}+\frac{1}{2}I^{(Z)}(\psi),
\end{equation}
where $c_i=\braket{i}{\psi}$ are the amplitudes of $\ket{\psi}$ in the computational basis, and the derivation follows from the general $q$-IPR circuit analysis in \cite{liu_quantum_2025}, specialized here to $q=2$. The previous circuit can easily be extended to arbitrary bases by acting with a basis transformation unitary $U_{s}^\dagger$ on both copies of $\ket{\psi}$ before the CNOT acts. The result of the circuit will yield the IPR of $\ket{\psi}$ in the new basis.

\begin{figure}[t]
    \centering
    \begin{quantikz}[row sep=0.3cm, column sep=0.3cm]
        \lstick{$\ket{0}_{A}$} &
        \gate{\mathrm{H}} &
        \qw &
        \qw &
        \qw &
        \qw &
        \ctrl{4} &
        \gate{\mathrm{H}} &
        \meter[wires=1]{}\\
        \lstick{$\ket{\psi}$}  &
        \qw &
        \gate{U_a^\dagger} &
        \gate{U_b^\dagger} &
        \qw &
        \qw &
        \gate[wires=2]{\text{SWAP}} &
        \qw &
        \qw\\
        \lstick{$\ket{\psi}$}  &
        \qw &
        \gate{U_a^\dagger} &
        \gate{U_b^\dagger} &
        \ctrl{4} &
        \qw &
        \qw &
        \qw &
        \qw\\
        \lstick{$\ket{\phi}$}  &
        \qw &
        \gate{U_a^\dagger} &
        \gate{U_b^\dagger} &
        \qw &
        \qw &
        \gate[wires=2]{\text{SWAP}} &
        \qw &
        \qw\\
        \lstick{$\ket{\phi}$}  &
        \qw &
        \gate{U_a^\dagger} &
        \gate{U_b^\dagger} &
        \qw &
        \ctrl{3} &
        \qw &
        \qw &
        \qw\\
        \lstick{$\ket{0}_\mathcal{B}$}  &
        \gate{\mathrm{H}} &
        \octrl{-4} &
        \ctrl{-4} &
        \qw &
        \qw &
        \qw &
        \qw &
        \qw\\
        \lstick{$\ket{0}^{\otimes L}$}  &
        \qw&
        \qw &
        \qw &
        \targ &
        \qw &
        \qw &
        \qw &
        \qw &
        \qw\\
        \lstick{$\ket{0}^{\otimes L}$}  &
        \qw&
        \qw &
        \qw &
        \qw &
        \targ &
        \qw &
        \qw &
        \qw &
        \qw\\
    \end{quantikz}
    \caption{IPR-kernel circuit for estimating $K(\psi,\phi)$.}
    \label{fig:circuit_known_basis}
\end{figure}
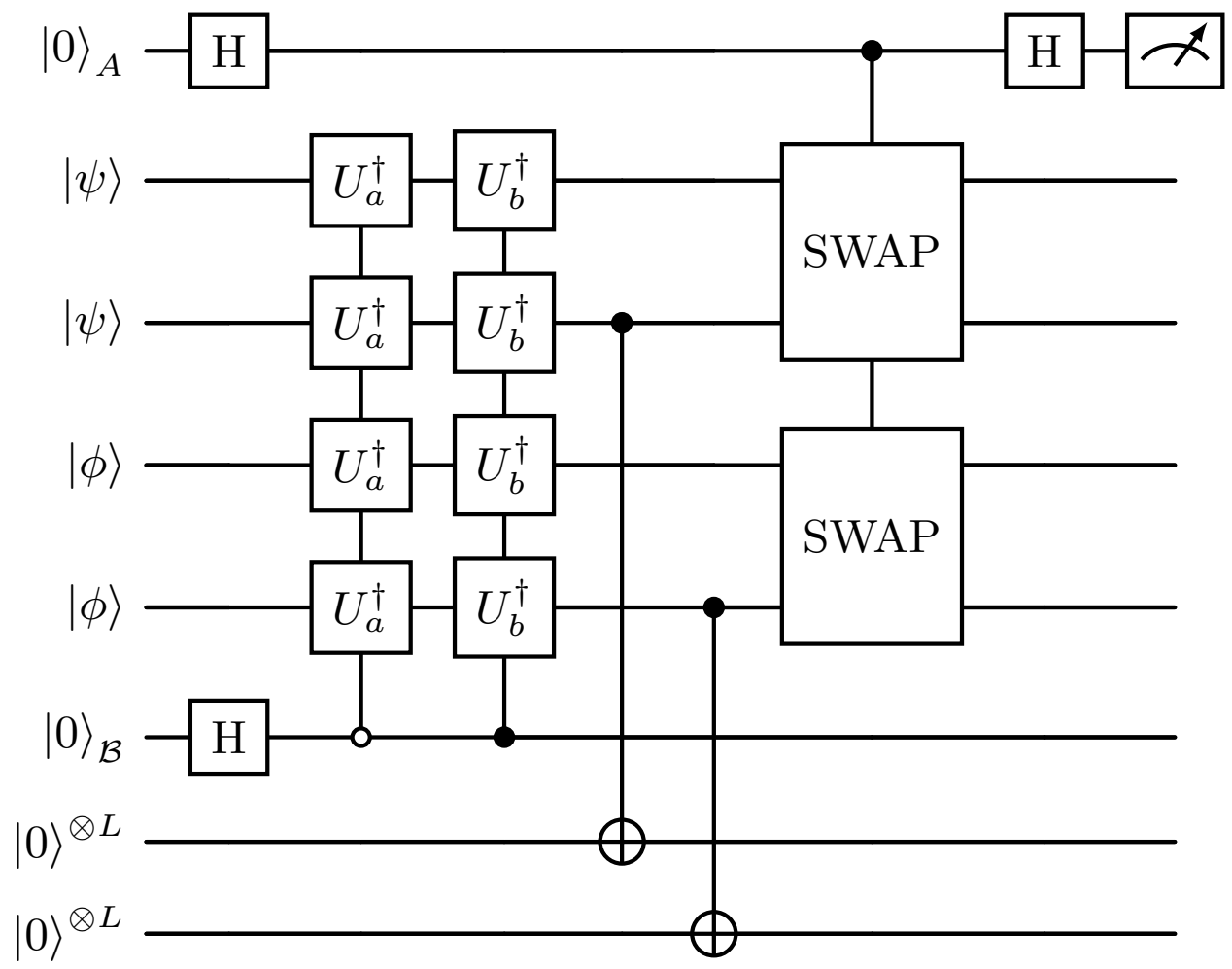

Using the previous results, the IPR kernel for two bases $a$ and $b$ can thus be obtained by executing the circuit in \cref{fig:circuit_known_basis} and again measuring the probability of the ancilla being in the $\ket{0}_{A}$ state. The two-basis approach necessitates a ``basis-selector'' ancilla $\ket{0}_{\mathcal{B}}$ which coherently implements either $U_{a}^\dagger$ or $U_{b}^\dagger$ through a control operation. Denoting our states in the computational basis as $\ket{\psi}=\sum_i c_i\ket{i}$ and $\ket{\phi}=\sum_k d_k\ket{k}$, or equivalently in basis $a$ as $\ket{\psi}=\sum_i c_i^{(a)}\ket{a_i} =\sum_i c_i^{(a)}U_a\ket{i}$ and analogously for $\ket{\phi}$, we obtain
\begin{equation}
    \begin{aligned}
        \ket{\Psi_0}
                    & =\ket{+}_A\ket{+}_{\mathcal B}\Big(\sum_{ijkl} c_i c_j d_k d_l \ket{i}\ket{j}\ket{k}\ket{l}\Big)\ket{0^L}\ket{0^L}
        \nonumber                                                                                                                                                     \\
        & \rightarrow \ket{+}_A\frac{1}{\sqrt2}\Big[\ket{0}_{\mathcal B}\sum_{i j k l} c_i^{(a)} c_j^{(a)} d_k^{(a)} d_l^{(a)}\ket{i}\ket{j}\ket{k}\ket{l}
        +\ket{1}_{\mathcal B}\sum_{i j k l} c_i^{(b)} c_j^{(b)} d_k^{(b)} d_l^{(b)}\ket{i}\ket{j}\ket{k}\ket{l}\Big]\ket{0^L}\ket{0^L}                        \\
        & \rightarrow
        \frac{1}{2}\Big[
        \ket{0}_A\Big(
        \ket{0}_{\mathcal B}\sum_{i j k l} c_i^{(a)} c_j^{(a)} d_k^{(a)} d_l^{(a)}
        \ket{i}\ket{j}\ket{k}\ket{l}\ket{j}\ket{l}
        +
        \ket{1}_{\mathcal B}\sum_{i j k l} c_i^{(b)} c_j^{(b)} d_k^{(b)} d_l^{(b)}
        \ket{i}\ket{j}\ket{k}\ket{l}\ket{j}\ket{l}
        \Big)\nonumber                                                                                                                                                \\
                    & \quad+\ket{1}_A\Big(\ket{0}_{\mathcal B}\sum_{i j k l} c_i^{(a)} c_j^{(a)} d_k^{(a)} d_l^{(a)}\ket{j}\ket{i}\ket{l}\ket{k}\ket{j}\ket{l}
        +\ket{1}_{\mathcal B}\sum_{i j k l} c_i^{(b)} c_j^{(b)} d_k^{(b)} d_l^{(b)}\ket{j}\ket{i}\ket{l}\ket{k}\ket{j}\ket{l}\Big)\Big]                   \\
        & \rightarrow \frac{1}{2}(\ket{+}_A\ket{\Omega_+}+\ket{-}_A\ket{\Omega_-})=:\ket{\Psi_{\text{out}}}.
    \end{aligned}
\end{equation}
\begin{figure*}[t]
    \centering
    \begin{quantikz}[row sep=0.35cm, column sep=0.45cm]
        \lstick{$\ket{0}^{\otimes m}_{\mathcal A_{\psi}}$}
        & \gate{\mathrm{H}^{\otimes m}}
        & \ctrl{2}
        & \ctrl{2}
        &
        &
        & \gate{\mathrm{QFT}_{m}}
        & \meter{}
        \\
        \lstick{$\ket{0}^{\otimes m}_{\mathcal A_{\phi}}$}
        & \gate{\mathrm{H}^{\otimes m}}
        &
        &
        & \ctrl{3}
        & \ctrl{3}
        & \gate{\mathrm{QFT}_{m}}
        & \meter{}
        \\
        \lstick{$\ket{\psi}$}
        &
        & \gate[2]{U_a^{x}\otimes (U_a^\dagger)^{x}}
        & \gate[2]{U_b^{x}\otimes (U_b^\dagger)^{x}}
        &
        &
        &
        &
        \\
        \lstick{$\ket{\psi}$}
        &
        &
        &
        &
        &
        &
        &
        \\
        \lstick{$\ket{\phi}$}
        &
        &
        &
        & \gate[2]{U_a^{y}\otimes (U_a^\dagger)^{y}}
        & \gate[2]{U_b^{y}\otimes (U_b^\dagger)^{y}}
        &
        &
        \\
        \lstick{$\ket{\phi}$}
        &
        &
        &
        &
        &
        &
        &
        \\
        \lstick{$\ket{0}_{\mathcal B}$}
        & \gate{\mathrm{H}}
        & \octrl{-4}
        & \ctrl{-4}
        & \octrl{-2}
        & \ctrl{-2}
        &
        &
    \end{quantikz}

    \caption{IPR-kernel circuit with two independent $m$-qubit phase-estimation registers. The register $\mathcal A_{\psi}$ controls the forward and adjoint time evolution $U_s=\exp(-\mathrm{i}tH_s)$ of the two copies of $\ket{\psi}$, while $\mathcal A_{\phi}$ independently controls the two copies of $\ket{\phi}$. The basis-selector ancilla $\mathcal B$ coherently selects between $H_a$ and $H_b$, and separate quantum Fourier transforms (QFTs) are applied to the two phase-estimation registers.}
    \label{fig:circuit_hamiltonian}
\end{figure*}
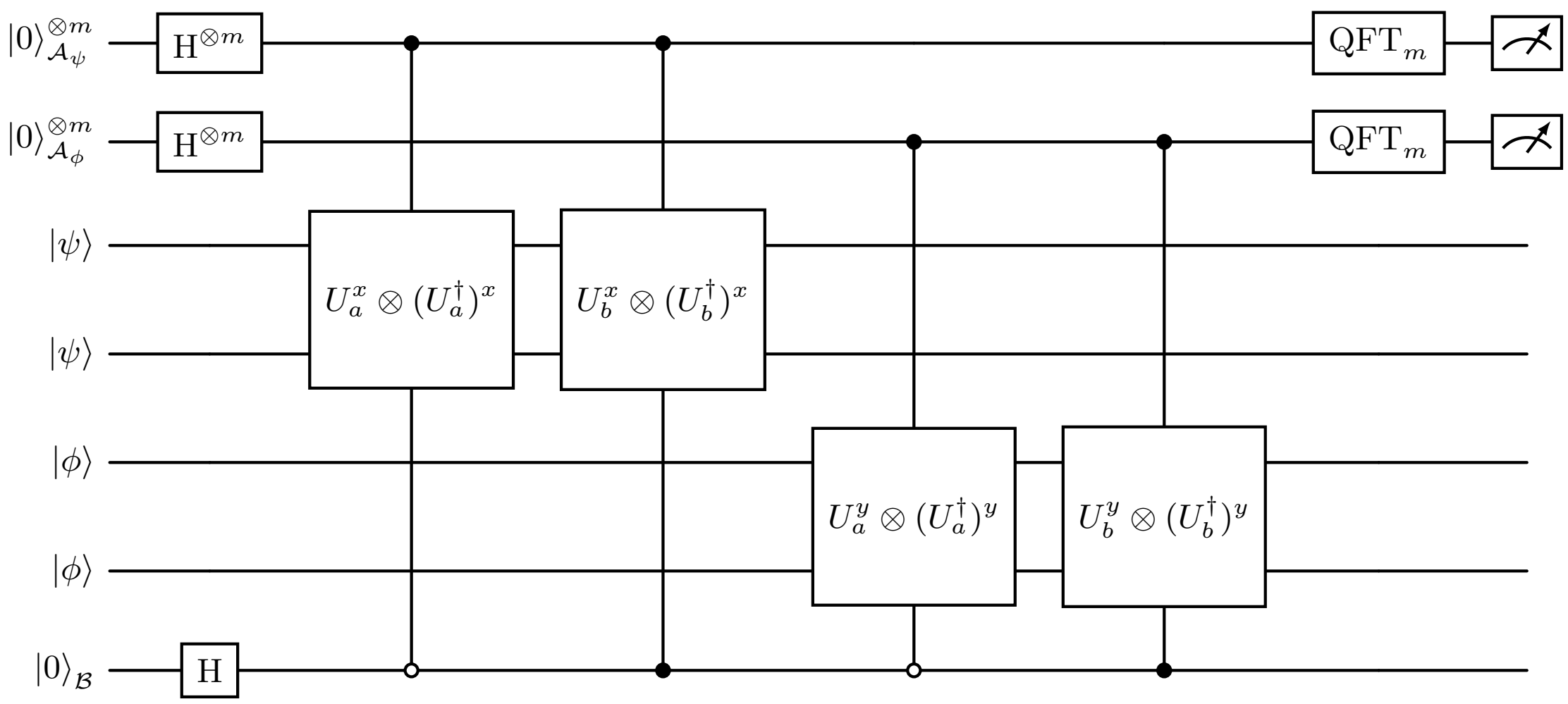
Here, $\ket{\Omega_+}$ and $\ket{\Omega_-}$ denote the states in parentheses multiplying $\ket{0}_A$ and $\ket{1}_A$ in the third line, respectively, and the last step applies the final Hadamard gate on $A$. Noting $\braket{\Omega_+}{\Omega_+} = 2 = \braket{\Omega_-}{\Omega_-}$ and
\begin{equation}
    \ket{\Psi_{\text{out}}}=\frac{1}{2\sqrt{2}}[\ket{0}_A(\ket{\Omega_+}+\ket{\Omega_-})+\ket{1}_A(\ket{\Omega_+}-\ket{\Omega_-})]
\end{equation}
with
\begin{equation}
    \begin{aligned}
        \braket{\Omega_-}{\Omega_+}
         & =
        \sum_{i j k l}\sum_{i' j' k' l'}
        \big(c_{i'}^{(a)} c_{j'}^{(a)} d_{k'}^{(a)} d_{l'}^{(a)}\big)^{*}
        \big(c_i^{(a)} c_j^{(a)} d_k^{(a)} d_l^{(a)}\big)
        \braket{j'}{i}\,
        \braket{i'}{j}\,
        \braket{l'}{k}\,
        \braket{k'}{l}\,
        \braket{j'}{j}\,
        \braket{l'}{l}
        \\
         & \quad+
        \sum_{i j k l}\sum_{i' j' k' l'}
        \big(c_{i'}^{(b)} c_{j'}^{(b)} d_{k'}^{(b)} d_{l'}^{(b)}\big)^{*}
        \big(c_i^{(b)} c_j^{(b)} d_k^{(b)} d_l^{(b)}\big)
        \braket{j'}{i}\,
        \braket{i'}{j}\,
        \braket{l'}{k}\,
        \braket{k'}{l}\,
        \braket{j'}{j}\,
        \braket{l'}{l}
        \\
         & =
        \sum_{i,k}
        \big(c_i^{(a)} c_i^{(a)} d_k^{(a)} d_k^{(a)}\big)^{*}
        \big(c_i^{(a)} c_i^{(a)} d_k^{(a)} d_k^{(a)}\big)
        +
        \sum_{i,k}
        \big(c_i^{(b)} c_i^{(b)} d_k^{(b)} d_k^{(b)}\big)^{*}
        \big(c_i^{(b)} c_i^{(b)} d_k^{(b)} d_k^{(b)}\big)
        \\
         & =
        \sum_{i} \abs{c_i^{(a)}}^{4}
        \sum_{k} \abs{d_k^{(a)}}^{4}
        +
        \sum_{i} \abs{c_i^{(b)}}^{4}
        \sum_{k} \abs{d_k^{(b)}}^{4}
        \\
         & =
        I^{(a)}(\psi) I^{(a)}(\phi)
        +
        I^{(b)}(\psi) I^{(b)}(\phi)
        =
        \sum_{s\in\{a,b\}} I^{(s)}(\psi) I^{(s)}(\phi),
    \end{aligned}
\end{equation}
we get
\begin{equation}
    \begin{aligned}
        P_{0,A}
         & = \bra{\Psi_{\mathrm{out}}}
        \big( \ketbra{0}{0} \otimes \mathbb{1} \big)
        \ket{\Psi_{\mathrm{out}}}                                                     \\
         & =\frac{1}{8}(\bra{\Omega_+}+\bra{\Omega_-})(\ket{\Omega_+}+\ket{\Omega_-}) \\
         & =\frac{1}{8}(4+\braket{\Omega_+}{\Omega_-}+\braket{\Omega_-}{\Omega_+})    \\
         & =\frac{1}{2}+\frac{1}{4} K_{\mathrm{IPR}}(\psi,\phi),
    \end{aligned}
\end{equation}
or $K_{\mathrm{IPR}}(\psi,\phi)=4P_{0,A}-2$. This circuit extends to a multi-basis IPR kernel by enlarging the basis-selector register: a register of $\lceil \log_2 M \rceil$ qubits coherently selects among $M$ basis unitaries.

Furthermore, when computing the normalized version of the IPR kernel we have to run the circuit three times (assuming sufficient samples for each), once for $K_{\mathrm{IPR}}(\psi,\phi)$, $K_{\mathrm{IPR}}(\psi,\psi)$ and $K_{\mathrm{IPR}}(\phi,\phi)$. Combining their values we get the normalized IPR kernel through
\begin{equation}
    \tilde{K}_{\mathrm{IPR}}(\psi,\phi)
    =
    \frac{
        K_{\mathrm{IPR}}(\psi,\phi)
    }{
        \sqrt{
            K_{\mathrm{IPR}}(\psi,\psi)\,
            K_{\mathrm{IPR}}(\phi,\phi)
        }
    }.
\end{equation}

\subsection{Hamiltonian eigenbasis}\label{app:circuits_hamiltonian}

The quantum circuit for the IPR kernel in a Hamiltonian's eigenbasis is depicted in \cref{fig:circuit_hamiltonian}.
In the non-degenerate case, phase estimation resolves the individual energy eigenstates and the circuit estimates the standard IPR in that eigenbasis. If the Hamiltonian has degeneracies, its eigenbasis is not uniquely defined: any unitary rotation within a degenerate eigenspace gives an equally valid eigenbasis. A phase-estimation-based circuit therefore resolves the energy eigenspaces themselves. The corresponding quantity is a block version of the IPR, defined using projectors onto degenerate eigenspaces. Given a degenerate Hamiltonian with spectral decomposition \cite{liu_quantum_2025},
\begin{equation}
H=\sum_{j}\varepsilon_j\sum_{\alpha=1}^{d_j}\ket{\varepsilon_j^\alpha}\bra{\varepsilon_j^\alpha}
=\sum_{j}\varepsilon_j P_j,
\end{equation}
where $d_j$ is the dimension of the degenerate eigenspace associated with $\varepsilon_j$ and $P_j$ is the corresponding projector, our circuit measures the generalized IPR as
\begin{equation} \label{eq:block_ipr}
I^{(H)}(\psi) =\sum_{j}\left(\langle\psi|P_j|\psi\rangle\right)^{2}.
\end{equation}
Writing $\ket{\psi}$ in the eigenbasis of the corresponding Hamiltonian as $\ket{\psi}=\sum_j \sum_{\alpha=1}^{d_j} c_{j,\alpha}\ket{\varepsilon_j^\alpha}$, this quantity measures how strongly $\ket{\psi}$ is concentrated over the degenerate eigenspace blocks of $H$. In the non-degenerate case, where $d_j=1$ for all $j$, it reduces to the standard IPR. A related block-IPR construction can be found in \cite{frey_probing_2024}.

In the following we limit ourselves to the non-degenerate case. Here, $U_{a,b}=\exp(-\mathrm{i} t H_{a,b})$ denote the time-evolution unitaries corresponding to $H_a$ and $H_b$, respectively, rather than the basis changes of \cref{app:circuits_known}. The value of the IPR kernel is obtained from the probability of finding both $m$-qubit ancilla registers in the state $\ket{0}^{\otimes m}$. With $N_m=2^m$ and writing $\ket{\psi}=\sum_i c_i^{(s)}\ket{\varepsilon_i^{(s)}}$ and $\ket{\phi}=\sum_k d_k^{(s)}\ket{\varepsilon_k^{(s)}}$ for $s\in\{a,b\}$, where $\ket{\varepsilon_i^{(s)}}$ is the eigenstate of $H_s$ with eigenvalue $\varepsilon_i^{(s)}$, we have

\begin{equation}
    \begin{aligned}
        \ket{\Psi_0}
         & =
        \frac{1}{N_m}
        \sum_{x,y=0}^{N_m-1}
        \ket{x}
        \otimes
        \ket{y}
        \otimes
        \ket{+}_{\mathcal B}
        \otimes
        \ket{\psi}
        \otimes
        \ket{\psi}
        \otimes
        \ket{\phi}
        \otimes
        \ket{\phi}
        \\[0.5ex]
         & \rightarrow
        \frac{1}{N_m\sqrt{2}}
        \sum_{x,y=0}^{N_m-1}
        \ket{x}
        \otimes
        \ket{y}
        \otimes
        \Big[
            \ket{0}_{\mathcal B}
            \big(U_a^{x}\ket{\psi}\big)
            \otimes
            \big((U_a^\dagger)^{x}\ket{\psi}\big)
            \otimes
            \big(U_a^{y}\ket{\phi}\big)
            \otimes
        \big((U_a^\dagger)^{y}\ket{\phi}\big)            \\
         & \qquad\qquad\qquad\qquad\quad+
            \ket{1}_{\mathcal B}
            \big(U_b^{x}\ket{\psi}\big)
            \otimes
            \big((U_b^\dagger)^{x}\ket{\psi}\big)
            \otimes
            \big(U_b^{y}\ket{\phi}\big)
            \otimes
            \big((U_b^\dagger)^{y}\ket{\phi}\big)
            \Big]
        \\[0.5ex]
         & =
        \frac{1}{N_m\sqrt{2}}
        \sum_{x,y=0}^{N_m-1}
        \ket{x}
        \otimes
        \ket{y}
        \otimes
        \Big[
        \ket{0}_{\mathcal B}
        \sum_{i j k l}
        c_i^{(a)} c_j^{(a)} d_k^{(a)} d_l^{(a)}
        e^{-\mathrm{i}t\big(
                x(\varepsilon_i^{(a)}-\varepsilon_j^{(a)})
                +y(\varepsilon_k^{(a)}-\varepsilon_l^{(a)})
                \big)}
        \ket*{\varepsilon_i^{(a)}}
        \ket*{\varepsilon_j^{(a)}}
        \ket*{\varepsilon_k^{(a)}}
        \ket*{\varepsilon_l^{(a)}}
        \\
         & \qquad\qquad\qquad\qquad\quad+
        \ket{1}_{\mathcal B}
        \sum_{i j k l}
        c_i^{(b)} c_j^{(b)} d_k^{(b)} d_l^{(b)}
        e^{-\mathrm{i}t\big(
                x(\varepsilon_i^{(b)}-\varepsilon_j^{(b)})
                +y(\varepsilon_k^{(b)}-\varepsilon_l^{(b)})
                \big)}
        \ket*{\varepsilon_i^{(b)}}
        \ket*{\varepsilon_j^{(b)}}
        \ket*{\varepsilon_k^{(b)}}
        \ket*{\varepsilon_l^{(b)}}
        \Big]
        \\[0.5ex]
         & \rightarrow
        \frac{1}{N_m^2\sqrt{2}}
        \sum_{q,r=0}^{N_m-1}
        \sum_{x,y=0}^{N_m-1}
        e^{\,\mathrm{i}\frac{2\pi}{N_m}(qx+ry)}
        \ket{q}
        \otimes
        \ket{r}
        \otimes
        \Big[
        \ket{0}_{\mathcal B}
        \sum_{i j k l}
        c_i^{(a)} c_j^{(a)} d_k^{(a)} d_l^{(a)}
        e^{-\mathrm{i}t\big(
                x\hat{\varepsilon}_{ij}^{(a)}
                +y\hat{\varepsilon}_{kl}^{(a)}
                \big)}
        \ket*{\varepsilon_i^{(a)}}
        \ket*{\varepsilon_j^{(a)}}
        \ket*{\varepsilon_k^{(a)}}
        \ket*{\varepsilon_l^{(a)}}
        \\
         & \qquad\qquad\qquad\qquad\qquad\qquad\qquad+
        \ket{1}_{\mathcal B}
        \sum_{i j k l}
        c_i^{(b)} c_j^{(b)} d_k^{(b)} d_l^{(b)}
        e^{-\mathrm{i}t\big(
                x\hat{\varepsilon}_{ij}^{(b)}
                +y\hat{\varepsilon}_{kl}^{(b)}
                \big)}
        \ket*{\varepsilon_i^{(b)}}
        \ket*{\varepsilon_j^{(b)}}
        \ket*{\varepsilon_k^{(b)}}
        \ket*{\varepsilon_l^{(b)}}
        \Big]                                          \\[0.5ex]
         & =
        \frac{1}{N_m^2\sqrt{2}}
        \sum_{q,r=0}^{N_m-1}
        \sum_{x,y=0}^{N_m-1}
        e^{\,\mathrm{i}\frac{2\pi}{N_m}(qx+ry)}
        \ket{q}
        \otimes
        \ket{r}
        \otimes
        \ket{\Omega(x,y)}=:\ket{\Psi_{\mathrm{out}}},
    \end{aligned}
\end{equation}
where $\ket{\Omega(x,y)}$ denotes the term in square brackets in the previous line and $\hat{\varepsilon}_{ij}^{(s)}=
(\varepsilon_i^{(s)}-\varepsilon_j^{(s)})$ for $s\in\{a,b\}$.

Calculating
\begin{equation}
    \begin{aligned}
         & \braket{\Omega(x',y')}{\Omega(x,y)}
        =                                   \\
         & \sum_{i j k l}\sum_{i' j' k' l'}
        \big(c_{i'}^{(a)} c_{j'}^{(a)} d_{k'}^{(a)} d_{l'}^{(a)}\big)^{*}
        \big(c_{i}^{(a)} c_{j}^{(a)} d_{k}^{(a)} d_{l}^{(a)}\big)
        e^{\,\mathrm{i} t\big(
                x'\hat{\varepsilon}_{i'j'}^{(a)}
                +y'\hat{\varepsilon}_{k'l'}^{(a)}
                \big)}
        e^{-\mathrm{i} t\big(
                x\hat{\varepsilon}_{ij}^{(a)}
                +y\hat{\varepsilon}_{kl}^{(a)}
                \big)}
        \braket*{\varepsilon_{i'}^{(a)}}{\varepsilon_i^{(a)}}
        \braket*{\varepsilon_{j'}^{(a)}}{\varepsilon_j^{(a)}}
        \braket*{\varepsilon_{k'}^{(a)}}{\varepsilon_k^{(a)}}
        \braket*{\varepsilon_{l'}^{(a)}}{\varepsilon_l^{(a)}}
        \\
         & +
        \sum_{i j k l}\sum_{i' j' k' l'}
        \big(c_{i'}^{(b)} c_{j'}^{(b)} d_{k'}^{(b)} d_{l'}^{(b)}\big)^{*}
        \big(c_{i}^{(b)} c_{j}^{(b)} d_{k}^{(b)} d_{l}^{(b)}\big)
        e^{\,\mathrm{i} t\big(
                x'\hat{\varepsilon}_{i'j'}^{(b)}
                +y'\hat{\varepsilon}_{k'l'}^{(b)}
                \big)}
        e^{-\mathrm{i} t\big(
                x\hat{\varepsilon}_{ij}^{(b)}
                +y\hat{\varepsilon}_{kl}^{(b)}
                \big)}
        \braket*{\varepsilon_{i'}^{(b)}}{\varepsilon_i^{(b)}}
        \braket*{\varepsilon_{j'}^{(b)}}{\varepsilon_j^{(b)}}
        \braket*{\varepsilon_{k'}^{(b)}}{\varepsilon_k^{(b)}}
        \braket*{\varepsilon_{l'}^{(b)}}{\varepsilon_l^{(b)}}
        \\[0.5ex]
         & =
        \sum_{s\in\{a,b\}}
        \sum_{i j k l}
        \abs{c_{i}^{(s)}}^{2}\abs{c_{j}^{(s)}}^{2}
        \abs{d_{k}^{(s)}}^{2}\abs{d_{l}^{(s)}}^{2}\,
        e^{\,\mathrm{i} t\big(
                (x'-x)\hat{\varepsilon}_{ij}^{(s)}
                +(y'-y)\hat{\varepsilon}_{kl}^{(s)}
                \big)},
    \end{aligned}
\end{equation}
and defining
\begin{equation}
    \begin{aligned}
        F_m(\omega)
         & :=
        \frac{1}{N_m^2}
        \left|
        \sum_{x=0}^{N_m-1}
        e^{-\mathrm{i}xt\omega}
        \right|^2
        =
        \begin{cases}
            1, & \omega=0,\\[1ex]
            \displaystyle
            \frac{1-\cos(N_m t\omega)}
            {N_m^2\left[1-\cos(t\omega)\right]},
            & \omega\neq 0,
        \end{cases}
    \end{aligned}
\end{equation}
where the second equality follows from the use of the geometric series, gives us
\begin{equation}
    \begin{aligned}
        P^\mathcal{A}_{0,0}
         & =
        \bra{\Psi_{\mathrm{out}}}
        \big(
        \ketbra{0}{0}^{\otimes m}
        \otimes
        \ketbra{0}{0}^{\otimes m}
        \otimes
        \mathbb{1}
        \big)
        \ket{\Psi_{\mathrm{out}}}
        \\
         & =
        \frac{1}{2N_m^4}
        \sum_{q,r,q',r'=0}^{N_m-1}
        \sum_{x,y,x',y'=0}^{N_m-1}
        e^{-\,\mathrm{i}\frac{2\pi}{N_m}(q'x'+r'y')}
        e^{\,\mathrm{i}\frac{2\pi}{N_m}(qx+ry)}
        \braket{q'}{0}\braket{0}{q}
        \braket{r'}{0}\braket{0}{r}
        \braket{\Omega(x',y')}{\Omega(x,y)}
        \\
         & =
        \frac{1}{2N_m^4}
        \sum_{x,y,x',y'=0}^{N_m-1}
        \braket{\Omega(x',y')}{\Omega(x,y)}
        \\
         & =
        \frac{1}{2N_m^4}
        \sum_{s\in\{a,b\}}
        \sum_{i j k l}
        \abs{c_{i}^{(s)}}^{2}
        \abs{c_{j}^{(s)}}^{2}
        \abs{d_{k}^{(s)}}^{2}
        \abs{d_{l}^{(s)}}^{2}
        \sum_{x,x'=0}^{N_m-1}
        e^{\,\mathrm{i}t(x'-x)\hat{\varepsilon}_{ij}^{(s)}}
        \sum_{y,y'=0}^{N_m-1}
        e^{\,\mathrm{i}t(y'-y)\hat{\varepsilon}_{kl}^{(s)}}
        \\
         & =
        \frac{1}{2}
        \sum_{s\in\{a,b\}}
        \sum_{i j k l}
        \abs{c_{i}^{(s)}}^{2}
        \abs{c_{j}^{(s)}}^{2}
        \abs{d_{k}^{(s)}}^{2}
        \abs{d_{l}^{(s)}}^{2}
        F_m\!\left(\hat{\varepsilon}_{ij}^{(s)}\right)
        F_m\!\left(\hat{\varepsilon}_{kl}^{(s)}\right)
        \\
         & =
        \frac{1}{2}
        \sum_{s\in\{a,b\}}
        \Bigg[
        \sum_{i j}
        \abs{c_{i}^{(s)}}^{2}
        \abs{c_{j}^{(s)}}^{2}
        F_m\!\left(\hat{\varepsilon}_{ij}^{(s)}\right)
        \Bigg]
        \Bigg[
        \sum_{k l}
        \abs{d_{k}^{(s)}}^{2}
        \abs{d_{l}^{(s)}}^{2}
        F_m\!\left(\hat{\varepsilon}_{kl}^{(s)}\right)
        \Bigg].
    \end{aligned}
\end{equation}

For a non-degenerate Hamiltonian,
$\hat{\varepsilon}_{ij}^{(s)}=0$ implies $i=j$. Using $F_m(0)=1$, we have
\begin{equation}
    \begin{aligned}
        \sum_{ij}
        \abs{c_i^{(s)}}^{2}
        \abs{c_j^{(s)}}^{2}
        F_m\!\left(\hat{\varepsilon}_{ij}^{(s)}\right)
        & =
        \sum_i \abs{c_i^{(s)}}^{4}
        +
        \sum_{i\neq j}
        \abs{c_i^{(s)}}^{2}
        \abs{c_j^{(s)}}^{2}
        F_m\!\left(\hat{\varepsilon}_{ij}^{(s)}\right) 
        =
        I^{(s)}(\psi)
        +
        \epsilon_{\psi}^{(s)},
    \end{aligned}
\end{equation}
and analogously
\begin{equation}
    \begin{aligned}
        \sum_{kl}
        \abs{d_k^{(s)}}^{2}
        \abs{d_l^{(s)}}^{2}
        F_m\!\left(\hat{\varepsilon}_{kl}^{(s)}\right)
        & =
        \sum_k \abs{d_k^{(s)}}^{4}
        +
        \sum_{k\neq l}
        \abs{d_k^{(s)}}^{2}
        \abs{d_l^{(s)}}^{2}
        F_m\!\left(\hat{\varepsilon}_{kl}^{(s)}\right) 
        =
        I^{(s)}(\phi)
        +
        \epsilon_{\phi}^{(s)}.
    \end{aligned}
\end{equation}
Thus,
\begin{equation}
    \begin{aligned}
        P^\mathcal{A}_{0,0}
         & =
        \frac{1}{2}
        \sum_{s\in\{a,b\}}
        \left(
        I^{(s)}(\psi)+\epsilon_{\psi}^{(s)}
        \right)
        \left(
        I^{(s)}(\phi)+\epsilon_{\phi}^{(s)}
        \right)
        \\
         & =
        \frac{1}{2}
        \Bigg[
        \sum_{s\in\{a,b\}}
        I^{(s)}(\psi)I^{(s)}(\phi)
        +
        \sum_{s\in\{a,b\}}
        \Big(
        I^{(s)}(\psi)\epsilon_{\phi}^{(s)}
        +
        I^{(s)}(\phi)\epsilon_{\psi}^{(s)}
        +
        \epsilon_{\psi}^{(s)}\epsilon_{\phi}^{(s)}
        \Big)
        \Bigg]
        \\
         & =
        \frac{1}{2}
        \Big(
        K_{\mathrm{IPR}}(\psi,\phi)
        +
        \epsilon_r
        \Big),
    \end{aligned}
\end{equation}
where the finite-resolution error $\epsilon_r$ decays as $\mathcal{O}(4^{-m})$. The derivation follows along the lines of \cite{liu_quantum_2025}, with the difference that two independent phase-estimation registers are required here to resolve the two energy differences separately.

For the finite-resolution errors we have
\begin{equation}
    \begin{aligned}
        \epsilon_{\psi}^{(s)}
        =
        \sum_{i\neq j}
        \abs{c_i^{(s)}}^{2}
        \abs{c_j^{(s)}}^{2}
        F_m\!\left(\hat{\varepsilon}_{ij}^{(s)}\right)
        =
        \frac{1}{N_m^2}
        \sum_{i\neq j}
        \abs{c_i^{(s)}}^{2}
        \abs{c_j^{(s)}}^{2}
        \frac{
        1-\cos\!\left(N_m t\hat{\varepsilon}_{ij}^{(s)}\right)
        }{
        1-\cos\!\left(t\hat{\varepsilon}_{ij}^{(s)}\right)
        },
    \end{aligned}
\end{equation}
and analogously,
\begin{equation}
    \begin{aligned}
        \epsilon_{\phi}^{(s)}
        =
        \sum_{k\neq l}
        \abs{d_k^{(s)}}^{2}
        \abs{d_l^{(s)}}^{2}
        F_m\!\left(\hat{\varepsilon}_{kl}^{(s)}\right)
        =
        \frac{1}{N_m^2}
        \sum_{k\neq l}
        \abs{d_k^{(s)}}^{2}
        \abs{d_l^{(s)}}^{2}
        \frac{
        1-\cos\!\left(N_m t\hat{\varepsilon}_{kl}^{(s)}\right)
        }{
        1-\cos\!\left(t\hat{\varepsilon}_{kl}^{(s)}\right)
        }.
    \end{aligned}
\end{equation}
A further bound can be obtained when
$0<\abs{t\hat{\varepsilon}_{ij}^{(s)}}\leq\pi$
for all $i\neq j$ and $s\in\{a,b\}$. Since $1-\cos\theta\geq\frac{2\theta^2}{\pi^2}$ for $\theta\in[-\pi,\pi]$
and $1-\cos\theta\leq2$, we obtain
\begin{equation}
    \begin{aligned}
        F_m\!\left(\hat{\varepsilon}_{ij}^{(s)}\right)
         & \leq
        \frac{1}{N_m^2}
        \frac{
        2
        }{
        1-\cos\!\left(t\hat{\varepsilon}_{ij}^{(s)}\right)
        }
        \\
         & \leq
        \frac{1}{N_m^2}
        \frac{
        2
        }{
        \frac{2}{\pi^2}
        \left(t\hat{\varepsilon}_{ij}^{(s)}\right)^2
        }
        \\
         & =
        \frac{\pi^2}{
        N_m^2t^2
        \left(\hat{\varepsilon}_{ij}^{(s)}\right)^2
        }.
    \end{aligned}
\end{equation}
Defining the minimum nonzero energy gap of $H_s$ as $\Delta_{\min}^{(s)}:=\min_{i\neq j}\abs{\varepsilon_i^{(s)}-\varepsilon_j^{(s)}}$, we therefore have
\begin{equation}
    \begin{aligned}
        \epsilon_{\psi}^{(s)}
         & \leq
        \frac{\pi^2}{
        N_m^2t^2
        \left(\Delta_{\min}^{(s)}\right)^2
        }
        \sum_{i\neq j}
        \abs{c_i^{(s)}}^{2}
        \abs{c_j^{(s)}}^{2}
        \\
         & =
        \frac{\pi^2}{
        N_m^2t^2
        \left(\Delta_{\min}^{(s)}\right)^2
        }
        \left(
        1-I^{(s)}(\psi)
        \right)
        \\
         & \leq
        \frac{\pi^2}{
        N_m^2t^2
        \left(\Delta_{\min}^{(s)}\right)^2
        },
    \end{aligned}
\end{equation}
and similarly $\epsilon_{\phi}^{(s)}\leq\frac{\pi^2}{N_m^2t^2\left(\Delta_{\min}^{(s)}\right)^2}$. For convenience, define $A_s:=\frac{\pi^2}{N_m^2t^2\left(\Delta_{\min}^{(s)}\right)^2}$.
Using
\begin{equation}
    \epsilon_r
    =
    \sum_{s\in\{a,b\}}
    \Big(
    I^{(s)}(\psi)\epsilon_{\phi}^{(s)}
    +
    I^{(s)}(\phi)\epsilon_{\psi}^{(s)}
    +
    \epsilon_{\psi}^{(s)}\epsilon_{\phi}^{(s)}
    \Big),
\end{equation}
together with $2^{-L}\leq I^{(s)}(\psi), I^{(s)}(\phi)\leq1$, we obtain
\begin{equation}
    \begin{aligned}
        \epsilon_r
         & \leq
        \sum_{s\in\{a,b\}}
        \Big(
        I^{(s)}(\psi)A_s
        +
        I^{(s)}(\phi)A_s
        +
        A_s^2
        \Big)
        \\
         & \leq
        \sum_{s\in\{a,b\}}
        \left(
        2A_s+A_s^2
        \right).
    \end{aligned}
\end{equation}
Defining $\Delta_{\min}:=\min\left\{\Delta_{\min}^{(a)},\Delta_{\min}^{(b)}\right\}$ and $A:=\frac{\pi^2}{N_m^2t^2\Delta_{\min}^2}$, such that $A_s\leq A$, gives
\begin{equation}
    \begin{aligned}
        \epsilon_r
         & \leq
        4A+2A^2
        \\
         & =
        \frac{4\pi^2}{
        N_m^2t^2\Delta_{\min}^2
        }
        +
        \frac{2\pi^4}{
        N_m^4t^4\Delta_{\min}^4
        }
        \\
         & =
        \frac{4\pi^2}{
        4^m t^2\Delta_{\min}^2
        }
        +
        \frac{2\pi^4}{
        16^m t^4\Delta_{\min}^4
        }
        \\
         & =
        \mathcal{O}\!\left(4^{-m}\right).
    \end{aligned}
\end{equation}

\end{document}